\documentclass[journal,onecolumn]{IEEEtran}
\pdfoutput=1

\newenvironment{EnumerateRoman}{\begin{enumerate}%
\setlength{\parsep}{0pt}\setlength{\topsep}{0.1\baselineskip}\setlength{\itemsep}{0.1\baselineskip}}{\end{enumerate}}

\usepackage{amsmath,amsfonts,amssymb}
\allowdisplaybreaks
\usepackage{mathrsfs} %Font used for sets
\usepackage{units}
\usepackage{multirow}
\newcommand{\FigScaleMultipleBigger}{0.43}
\newcommand{\FigScaleMultipleBiggest}{0.50}

\usepackage{graphicx} %Use option \usepackage[draft]{graphicx} to plot placeholders instead
\usepackage[tight,footnotesize]{subfigure} %For subfigures, use subfigure package, or newer version, subfig
\usepackage{array} %Extension of environments tabular and array
\usepackage[footnotesize]{caption} %Small captions [small], even smaller, [footnotesize], default is normalsize [normal]
\usepackage{tabularx}
\usepackage{ctable}

\usepackage{url}

\usepackage[noadjust]{cite} %To sort and range citation numbers. [noadjust] eliminates leading space.

\def\hyph{-\penalty0\hskip0pt\relax}

\DeclareMathOperator{\oD}{D}                                            %Relative entropy
\DeclareMathOperator{\oClo}{cl}                                           %Closure
\newcommand{\cL}{\mathcal{L}}                                           %Lagrangian cost
\newcommand{\cR}{\mathcal{R}}                                           %Privacy risk
\newcommand{\cP}{\mathcal{P}}                                           %Privacy (probability)
\newcommand{\cCP}{\bar{P}}                                           %Complementary cumulative function of distribution P
\newcommand{\cCQ}{\bar{Q}}                                           %Complementary cumulative function of distribution Q
\newcommand{\rhocrit}{\rho_\textnormal{crit}}
\newcommand{\cC}{\mathscr{C}}
\newtheorem{proposition}{Proposition}

\newtheorem{theorem}[proposition]{Theorem}
\newtheorem{lemma}[proposition]{Lemma}
\newtheorem{corollary}[proposition]{Corollary}
\newcommand{\BeginProof}{\indent{\itshape Proof: }}

\newcommand{\ProofSquare}{\blacksquare}
\newcommand{\EndProof}{\hspace*{\fill}~$\ProofSquare$\par\unskip}
\newcommand{\SpaceAfterPropositionEndedWithFormula}{\vspace{\belowdisplayskip}}

\begin{document}

%%%%%%%%%%%%%%%%%%%%%%%%%%%%%%%%%%%%%%%%%%%%%%%%%%%%%%%%%%%%%%%%%%%%%%%%%%%%%%%%%%%%%%%%%%%%%%%%%%%%%%%%%%%%%%%%%%%%%%%%%%%%%%%%%%
%%%%%%%%%%%%%%%%%%%%%%%%%%%%%%%%%%%%%%%%%%%%%%%%%%%%%%%%%%%%%%%%%%%%%%%%%%%%%%%%%%%%%%%%%%%%%%%%%%%%%%%%%%%%%%%%%%%%%%%%%%%%%%%%%%
%TITLE AND AUTHORS
%%%%%%%%%%%%%%%%%%%%%%%%%%%%%%%%%%%%%%%%%%%%%%%%%%%%%%%%%%%%%%%%%%%%%%%%%%%%%%%%%%%%%%%%%%%%%%%%%%%%%%%%%%%%%%%%%%%%%%%%%%%%%%%%%%
%%%%%%%%%%%%%%%%%%%%%%%%%%%%%%%%%%%%%%%%%%%%%%%%%%%%%%%%%%%%%%%%%%%%%%%%%%%%%%%%%%%%%%%%%%%%%%%%%%%%%%%%%%%%%%%%%%%%%%%%%%%%%%%%%%
\title{Optimal Forgery and Suppression of Ratings for\\ Privacy Enhancement in Recommendation Systems}
\author{Javier Parra\hyph Arnau, David Rebollo\hyph Monedero and Jordi Forn\'e%<- Prevents space before \thanks
\thanks{Some parts of this paper (a reduced version of Secs.~\ref{sec:Introduction} and~\ref{sec:StateOfTheArt})
were presented at the International Workshop on Data Privacy Management, Leuven, Belgium, Sep. 2011~\cite{Parra11DPM}.
The formulation of the trade\hyph off between privacy and utility (Sec.~\ref{sec:FSTechnique}),
the theoretical analysis (Sec.~\ref{sec:Theory}), the experiments (Sec.~\ref{sec:Experiments}) and the conclusions (Sec.~\ref{sec:Conclusion}) are all new work.}
\thanks{The authors are with the Department of Telematics Engineering, Universitat Polit\`ecnica de Catalunya (UPC),
E-08034 Barcelona, Spain (e-mail: javier.parra@entel.upc.edu; david.rebollo@entel.upc.edu; jforne@entel.upc.edu).}}

\maketitle

%%%%%%%%%%%%%%%%%%%%%%%%%%%%%%%%%%%%%%%%%%%%%%%%%%%%%%%%%%%%%%%%%%%%%%%%%%%%%%%%%%%%%%%%%%%%%%%%%%%%%%%%%%%%%%%%%%%%%%%%%%%%%%%%%%
%%%%%%%%%%%%%%%%%%%%%%%%%%%%%%%%%%%%%%%%%%%%%%%%%%%%%%%%%%%%%%%%%%%%%%%%%%%%%%%%%%%%%%%%%%%%%%%%%%%%%%%%%%%%%%%%%%%%%%%%%%%%%%%%%%
%ABSTRACT & KEYWORDS
%%%%%%%%%%%%%%%%%%%%%%%%%%%%%%%%%%%%%%%%%%%%%%%%%%%%%%%%%%%%%%%%%%%%%%%%%%%%%%%%%%%%%%%%%%%%%%%%%%%%%%%%%%%%%%%%%%%%%%%%%%%%%%%%%%
%%%%%%%%%%%%%%%%%%%%%%%%%%%%%%%%%%%%%%%%%%%%%%%%%%%%%%%%%%%%%%%%%%%%%%%%%%%%%%%%%%%%%%%%%%%%%%%%%%%%%%%%%%%%%%%%%%%%%%%%%%%%%%%%%%

\begin{abstract}
Recommendation systems are information\hyph filtering systems that tailor information to users on the basis of knowledge about their preferences. The ability of these systems to profile users is what enables such intelligent functionality, but at the same time, it is the source of serious privacy concerns. In this paper we investigate a privacy\hyph enhancing technology that aims at hindering an attacker in its efforts to accurately profile users based on the items they rate. Our approach capitalizes on the combination of two perturbative mechanisms---the forgery and the suppression of ratings. While this technique enhances user privacy to a certain extent, it inevitably comes at the cost of a loss in data utility, namely a degradation of the recommendation's accuracy. In short, it poses a trade\hyph off between privacy and utility.

The theoretical analysis of said trade\hyph off is the object of this work. We measure privacy as the Kullback\hyph Leibler divergence between the user's and the population's item distributions, and quantify utility as the proportion of ratings users consent to forge and eliminate. Equipped with these quantitative measures, we find a closed\hyph form solution to the problem of optimal forgery and suppression of ratings, and characterize the optimal trade\hyph off surface among privacy, forgery rate and suppression rate.
Experimental results on a popular recommendation system show how our approach may contribute to privacy enhancement.
\end{abstract}

\begin{IEEEkeywords}
Information privacy, Kullback\hyph Leibler divergence, user profiling, privacy\hyph enhancing technologies, data perturbation, recommendation systems.
\end{IEEEkeywords}

%FOOTNOTES
\setcounter{footnote}{0} %Reset footnote counter unless same format is used
\renewcommand{\thefootnote}{(\alph{footnote})}  %If \alph,\arabic number style prererred to default

%%%%%%%%%%%%%%%%%%%%%%%%%%%%%%%%%%%%%%%%%%%%%%%%%%%%%%%%%%%%%%%%%%%%%%%%%%%%%%%%%%%%%%%%%%%%%%%%%%%%%%%%%%%%%%%%%%%%%%%%%%%%%%%%%%
%%%%%%%%%%%%%%%%%%%%%%%%%%%%%%%%%%%%%%%%%%%%%%%%%%%%%%%%%%%%%%%%%%%%%%%%%%%%%%%%%%%%%%%%%%%%%%%%%%%%%%%%%%%%%%%%%%%%%%%%%%%%%%%%%%
%%%%%%%%%%%%%%%%%%%%%%%%%%%%%%%%%%%%%%%%%%%%%%%%%%%%%%%%%%%%%%%%%%%%%%%%%%%%%%%%%%%%%%%%%%%%%%%%%%%%%%%%%%%%%%%%%%%%%%%%%%%%%%%%%%
%1 INTRODUCTION
%%%%%%%%%%%%%%%%%%%%%%%%%%%%%%%%%%%%%%%%%%%%%%%%%%%%%%%%%%%%%%%%%%%%%%%%%%%%%%%%%%%%%%%%%%%%%%%%%%%%%%%%%%%%%%%%%%%%%%%%%%%%%%%%%%
%%%%%%%%%%%%%%%%%%%%%%%%%%%%%%%%%%%%%%%%%%%%%%%%%%%%%%%%%%%%%%%%%%%%%%%%%%%%%%%%%%%%%%%%%%%%%%%%%%%%%%%%%%%%%%%%%%%%%%%%%%%%%%%%%%
%%%%%%%%%%%%%%%%%%%%%%%%%%%%%%%%%%%%%%%%%%%%%%%%%%%%%%%%%%%%%%%%%%%%%%%%%%%%%%%%%%%%%%%%%%%%%%%%%%%%%%%%%%%%%%%%%%%%%%%%%%%%%%%%%%
\section{Introduction}
\label{sec:Introduction}
\noindent
%----------------------------------------------------- Recommendation Systems - Introduction -------------------------------------------------
From the advent of the Internet and the World Wide Web, the amount of information available to users has grown exponentially.
As a result, the ability to find information relevant for their interests has become a central issue in recent years.
In this context of information overload, \emph{recommendation systems} arise to provide information tailored to users on the basis of knowledge about their preferences~\cite{Hanani01UMUAI}.
In essence, a recommendation system may be regarded as a type of information\hyph filtering system that suggests information items users may be interested in.
% Examples of Recommendation Systems
Examples of such systems include recommending music at \emph{Last.fm} %~\cite{Lastfm}
and \emph{Pandora Radio}, %~\cite{Pandora},
movies by \emph{MovieLens} %~\cite{Movielens}
and \emph{Netflix}, %~\cite{Netflix},
videos at \emph{YouTube}, %~\cite{YouTube},
news at \emph{Digg} %~\cite{Digg}
and \emph{Google News}, %~\cite{GoogleNews}
and books and other products at \emph{Amazon}. %~\cite{Amazon}.

%----------------------------------------------------- Interaction between the user and the system -------------------------------------------------
Most of these systems capitalize on the creation of \emph{profiles} that represent interests and preferences of users.
Such profiles are the result of the collection and analysis of the data that users communicate to those systems.
A distinction is frequently made between \emph{explicit} and \emph{implicit} forms of data collection.
The most popular form of explicit data collection is that users communicate their preferences by rating items.
This is the case of many of the applications mentioned above,
where users assign \emph{ratings} to songs, movies or news they have already listened, watched or read.
Other strategies to capture users' interests include asking them to sort a number of items by order of predilection, or suggesting that they mark the
items they like.
On the other hand, recommendation systems may collect data from users without requiring them to explicitly convey their preferences~\cite{Oard98RS}.
These practices comprise observing the items clicked by users in an online store,
analyzing the time it takes users to examine an item, or simply keeping a record of the purchased items.

The prolonged collection of these personal data allows the system to extract an accurate snapshot of user interests, i.e., their profiles.
With this invaluable source of information, the recommendation system applies some technique~\cite{Adomavicius05TKDE}
to generate a prediction of users' interests for those items they have not yet considered.
For example, \emph{Movielens} and \emph{Digg} use collaborative\hyph filtering techniques to predict the rating that a user would give to a movie and to create a personalized list of recommended news, respectively.
In a nutshell, the ability of profiling users based on such personal information is precisely what enables the intelligent functionality of those systems.

Despite the many advantages recommendation systems are bringing to users, the information collected, processed and stored by these
systems prompts serious privacy concerns.
One of the main privacy risks perceived by users is that of a computer ``figuring things out'' about them~\cite{Cranor03WPES}.
Many users are worried about the idea that their profiles may reveal sensitive information such as health\hyph related issues,
political preferences, salary or religion.
Such privacy risk is exacerbated especially when these profiles are combined across several information services or enriched with data from social networks.
An illustrative example is~\cite{Narayanan08SP}, which demonstrates that it is possible to unveil sensitive information about a person from their movie rating history
by cross\hyph referencing data from other sources.
The authors analyzed the \emph{Netflix Prize} data set~\cite{NetflixPrizeDataSet}, which contained anonymous movie ratings of around half a million users of \emph{Netflix},
and were able to uncover the identity, political leaning and even sexual orientation of some of those users,
by simply correlating their ratings with reviews they posted on the popular movie Web site \emph{IMDb}.
Apart from the risk of cross\hyph referencing,
users are also concerned that the system's predictions may be totally erroneous and be later used to defame them.
This latter situation is examined in~\cite{Zaslow02TWSJ}, where the accuracy of the predictions provided by \emph{TiVo} digital video recorder
and \emph{Amazon} is questioned.
Lastly, other privacy risks embrace unsolicited marketing, information leaked to other users of the same computer, court subpoenas,
and government surveillance~\cite{Cranor03WPES}.
\begin{figure}
\centering
\includegraphics[scale=0.49]{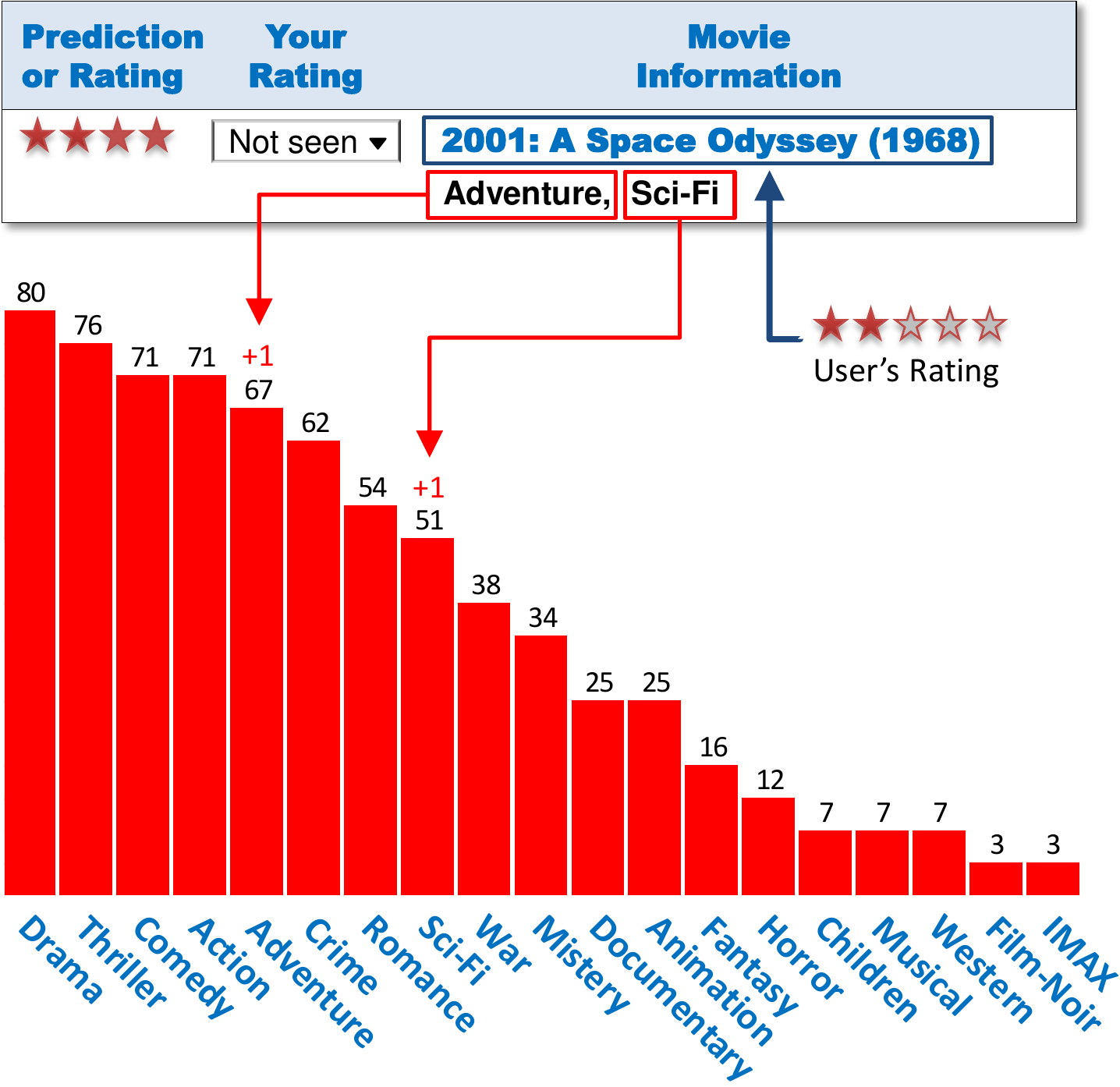}
\caption{The profile of a user is modeled in \emph{Movielens} as a histogram of absolute frequencies of ratings within a set of movie genres (bottom).
Based on this profile, the recommender predicts the rating that the user would probably give to a movie (top). After having watched the movie, the user rates it and their profile is updated.}
\label{fig:user_profiles}
\end{figure}

As a result of all this, it is not surprising that some users are reticent to reveal their interests.
In fact, \cite{Fox00PIALP}~reports that the 24\% of Internet users surveyed provided false information in order to
avoid giving private information to a Web site.
Alternatively, another study~\cite{Hoffman99ACMCOM} finds that 95\% of the respondents refused, at some point, to provide personal
information when requested by a Web site.
In closing, these studies seem to indicate that submitting false information and refusing to give private information are
strategies accepted by users concerned with their privacy.
%%%%%%%%%%%%%%%%%%%%%%%%%%%%%%%%%%%%%%%%%%%%%%%%%%%%%%%%%%%%%%%%%%%%%%%%%%%%%%%%%%%%%%%%%%%%%%%%%%%%%%%%%%%%%%%%%%%%%%%%%%%%%%%%%%
%%%%%%%%%%%%%%%%%%%%%%%%%%%%%%%%%%%%%%%%%%%%%%%%%%%%%%%%%%%%%%%%%%%%%%%%%%%%%%%%%%%%%%%%%%%%%%%%%%%%%%%%%%%%%%%%%%%%%%%%%%%%%%%%%%
% CONTRIBUTION AND PLAN OF THIS PAPER
%%%%%%%%%%%%%%%%%%%%%%%%%%%%%%%%%%%%%%%%%%%%%%%%%%%%%%%%%%%%%%%%%%%%%%%%%%%%%%%%%%%%%%%%%%%%%%%%%%%%%%%%%%%%%%%%%%%%%%%%%%%%%%%%%%
%%%%%%%%%%%%%%%%%%%%%%%%%%%%%%%%%%%%%%%%%%%%%%%%%%%%%%%%%%%%%%%%%%%%%%%%%%%%%%%%%%%%%%%%%%%%%%%%%%%%%%%%%%%%%%%%%%%%%%%%%%%%%%%%%%
\subsection{Contribution and Plan of this Paper}
\label{sec:Introduction:Contribution}
\noindent
%----------------------------------------------------------------- CONTRIBUTION --------------------------------------------------------
In this paper we approach the problem of protecting user privacy in those recommendation systems that profile users on the
basis of the items they rate.
Given the willingness of users to provide fake information and elude disclosing private data,
we investigate a privacy\hyph enhancing technology (PET) that combines these two forms of data perturbation,
namely the forgery and the suppression of ratings.
Concordantly, in our scenario users rate those items they have an opinion on.
However, in order to avoid being accurately profiled by the recommender or, in general, by any privacy attacker capable of collecting this information,
users may wish to refrain from rating some of those items and/or rate items that do not reflect their actual preferences.
Our approach thus protects user privacy to a certain degree, without having to trust the recommendation system
or the network operator, but at the cost a loss in utility, a degradation of the quality of the recommendation.
In other words, our PET poses a trade\hyph off between privacy and utility.

The theoretical analysis of the trade\hyph off between these two contrasting aspects is the object of this work.
We tackle the issue in a systematic fashion,
drawing upon the methodology of multiobjective optimization.
Before proceeding, though, we adopt a quantifiable measure of user privacy---the Kullback\hyph Leibler (KL) divergence between
the probability distribution of the user's items and the population's distribution,
a criterion that we introduced in previous work~\cite{Rebollo10IT} and justified and
interpreted in~\cite{Rebollo11SECTECH, Parra13FGCS} by leveraging on the rationale behind entropy\hyph maximization methods.
Equipped with a measure of both privacy and utility,
we formulate an optimization problem modeling the trade\hyph off between privacy on the one hand, and on the other forgery rate and suppression rate as utility metrics.
Our extensive theoretical analysis finds a closed\hyph form solution to the problem of optimal forgery and suppression of ratings,
and characterizes the optimal trade\hyph off between the aspects of privacy and utility.

In addition, we provide an empirical evaluation of our data\hyph perturbative approach.
Specifically, we apply the forgery and the suppression of ratings in the popular movie recommendation system \emph{Movielens},
and show how these two strategies may preserve the privacy of its users.

%%----------------------------------------------------------------- ORGANIZATION --------------------------------------------------------
Sec.~\ref{sec:StateOfTheArt} reviews several data\hyph perturbative approaches aimed at enhancing user privacy in the context of recommender systems.
Sec.~\ref{sec:FSTechnique} introduces our privacy\hyph enhancing technology, proposes a quantitative measure of the privacy of user profiles,
and formulates the trade\hyph off between privacy and utility.
Sec.~\ref{sec:Theory} presents a theoretical analysis of the optimization problem
characterizing the privacy\hyph forgery\hyph suppression trade\hyph off.
In this same section we also provide a numerical example that illustrates our formulation and theoretical results.
Sec.~\ref{sec:Experiments} evaluates our privacy\hyph protecting mechanism in a real recommendation system.
Finally, conclusions are drawn in Sec.~\ref{sec:Conclusion}.

%%%%%%%%%%%%%%%%%%%%%%%%%%%%%%%%%%%%%%%%%%%%%%%%%%%%%%%%%%%%%%%%%%%%%%%%%%%%%%%%%%%%%%%%%%%%%%%%%%%%%%%%%%%%%%%%%%%%%%%%%%%%%%%%%%
%%%%%%%%%%%%%%%%%%%%%%%%%%%%%%%%%%%%%%%%%%%%%%%%%%%%%%%%%%%%%%%%%%%%%%%%%%%%%%%%%%%%%%%%%%%%%%%%%%%%%%%%%%%%%%%%%%%%%%%%%%%%%%%%%%
%%%%%%%%%%%%%%%%%%%%%%%%%%%%%%%%%%%%%%%%%%%%%%%%%%%%%%%%%%%%%%%%%%%%%%%%%%%%%%%%%%%%%%%%%%%%%%%%%%%%%%%%%%%%%%%%%%%%%%%%%%%%%%%%%%
%2 STATE OF THE ART
%%%%%%%%%%%%%%%%%%%%%%%%%%%%%%%%%%%%%%%%%%%%%%%%%%%%%%%%%%%%%%%%%%%%%%%%%%%%%%%%%%%%%%%%%%%%%%%%%%%%%%%%%%%%%%%%%%%%%%%%%%%%%%%%%%
%%%%%%%%%%%%%%%%%%%%%%%%%%%%%%%%%%%%%%%%%%%%%%%%%%%%%%%%%%%%%%%%%%%%%%%%%%%%%%%%%%%%%%%%%%%%%%%%%%%%%%%%%%%%%%%%%%%%%%%%%%%%%%%%%%
%%%%%%%%%%%%%%%%%%%%%%%%%%%%%%%%%%%%%%%%%%%%%%%%%%%%%%%%%%%%%%%%%%%%%%%%%%%%%%%%%%%%%%%%%%%%%%%%%%%%%%%%%%%%%%%%%%%%%%%%%%%%%%%%%%
\section{State of the Art}
\label{sec:StateOfTheArt}
\noindent
% Overview
Numerous approaches have been proposed to protect user privacy in the context of recommendation systems.
These approaches fundamentally suggest either perturbing the information provided by users or using cryptographic techniques.
%, and distributing the information stored by recommenders.

%-------------------------------------------------------------------------------------------------------------------------------
%------------------------------------------------------------ Perturbing -------------------------------------------------------
%-------------------------------------------------------------------------------------------------------------------------------
In the case of perturbative methods for recommendation systems,
\cite{Polat03SDM}~proposes that users add random values to their ratings and then submit these perturbed ratings to the recommender.
After receiving these ratings, the system executes an algorithm and sends the users some information that allows them to compute the prediction.
When the number of participating users is sufficiently large, the authors find that user privacy is protected to a certain extent and the
system reaches a decent level of accuracy.
However, even though a user disguises all their ratings, it is evident that the items themselves may uncover sensitive information.
Simply put, the mere fact of showing interest in a certain item may be more revealing than the rating assigned to that item.
For instance, a user rating a book called ``How to Overcome Depression'' indicates a clear interest in depression, regardless of the score assigned to this book.
Apart from this critique, other works~\cite{Kargupta03ICDM,Huang05SIGMOD} stress that the use of \emph{randomized} data distortion techniques
might not be able to preserve privacy.

In line with this work, \cite{Polat05SAC}~applies the same data\hyph perturbative technique to collaborative\hyph filtering algorithms
based on singular\hyph value decomposition.
Specifically, the authors focus on the impact that their technique has on privacy.
For this purpose, they use the privacy metric proposed by~\cite{Agrawal01SIGMOD}, which is essentially equivalent to differential entropy,
and conduct some experiments with data sets from \emph{Movielens} and \emph{Jester}.
The results show the trade\hyph off curve between accuracy in recommendations and privacy.
In particular, they measure accuracy as the mean absolute error between the predicted values from the original ratings and the predictions obtained from the perturbed ratings.

At this point, we would like to remark that the use of perturbative techniques is by no means new in other scenarios such as private information retrieval
and the semantic Web.
In the former scenario, users send general\hyph purpose queries to an information service provider.
A perturbative approach to protect user profiles in this context consists in combining genuine with false queries.
Precisely, \cite{Rebollo10IT}~proposes a \emph{nonrandomized} method for query forgery and investigates
the trade\hyph off between privacy and the additional traffic overhead.
In the semantic Web scenario, users annotate resources with the purpose of classifying them.
In this application domain, the perturbation of user profiles for privacy preservation may be carried out by dropping certain annotations or \emph{tags}.
An example of this kind of perturbation may be found in~\cite{Parra10TB,Parra12DKE,Parra12TKDE}, where the authors propose the elimination of tags as a privacy\hyph enhancing strategy.

Regarding the use of cryptographic techniques,
\cite{Canny02SIGIR,Canny02SP}~propose a method that enables a community of users to calculate a public aggregate of their profiles
without revealing them on an individual basis.
In particular, the authors use a homomorphic encryption scheme and a peer\hyph to\hyph peer communication protocol
for the recommender to perform this calculation.
Once the aggregated profile is computed, the system sends it to users, who finally use local computation to obtain personalized recommendations.
This proposal prevents the system or any external attacker from ascertaining the individual user profiles.
However, its main handicap is assuming that an acceptable number of users is online and willing to participate in the protocol.
In line with this, \cite{Ahmad07ISIAS}~uses a variant of Pailliers' homomorphic cryptosystem which improves the efficiency in the
communication protocol.
Another solution~\cite{Zhan10SMC} presents an algorithm aimed at providing more efficiency by using the scalar product protocol.

%%%%%%%%%%%%%%%%%%%%%%%%%%%%%%%%%%%%%%%%%%%%%%%%%%%%%%%%%%%%%%%%%%%%%%%%%%%%%%%%%%%%%%%%%%%%%%%%%%%%%%%%%%%%%%%%%%%%%%%%%%%%%%%%%%
%%%%%%%%%%%%%%%%%%%%%%%%%%%%%%%%%%%%%%%%%%%%%%%%%%%%%%%%%%%%%%%%%%%%%%%%%%%%%%%%%%%%%%%%%%%%%%%%%%%%%%%%%%%%%%%%%%%%%%%%%%%%%%%%%%
%%%%%%%%%%%%%%%%%%%%%%%%%%%%%%%%%%%%%%%%%%%%%%%%%%%%%%%%%%%%%%%%%%%%%%%%%%%%%%%%%%%%%%%%%%%%%%%%%%%%%%%%%%%%%%%%%%%%%%%%%%%%%%%%%%
%3 Privacy Protection via Forgery and Suppression of Ratings
%%%%%%%%%%%%%%%%%%%%%%%%%%%%%%%%%%%%%%%%%%%%%%%%%%%%%%%%%%%%%%%%%%%%%%%%%%%%%%%%%%%%%%%%%%%%%%%%%%%%%%%%%%%%%%%%%%%%%%%%%%%%%%%%%%
%%%%%%%%%%%%%%%%%%%%%%%%%%%%%%%%%%%%%%%%%%%%%%%%%%%%%%%%%%%%%%%%%%%%%%%%%%%%%%%%%%%%%%%%%%%%%%%%%%%%%%%%%%%%%%%%%%%%%%%%%%%%%%%%%
%%%%%%%%%%%%%%%%%%%%%%%%%%%%%%%%%%%%%%%%%%%%%%%%%%%%%%%%%%%%%%%%%%%%%%%%%%%%%%%%%%%%%%%%%%%%%%%%%%%%%%%%%%%%%%%%%%%%%%%%%%%%%%%%%%
\section{Privacy Protection via Forgery and Suppression of Ratings}
\label{sec:FSTechnique}
\noindent
In this section, first we present the forgery and the suppression of ratings as a privacy\hyph enhancing technology.
The description of our approach is prefaced by a brief introduction of the concepts of soft privacy and hard privacy.
Secondly, we propose a model of user profile and set forth our assumptions about the adversary capabilities.
Finally, we provide a quantitative measure of both privacy and utility,
 and present a formulation of the trade\hyph off between these two contrasting aspects.
%%%%%%%%%%%%%%%%%%%%%%%%%%%%%%%%%%%%%%%%%%%%%%%%%%%%%%%%%%%%%%%%%%%%%%%%%%%%%%%%%%%%%%%%%%%%%%%%%%%%%%%%%%%%%%%%%%%%%%%%%%%%%%%%%%
%%%%%%%%%%%%%%%%%%%%%%%%%%%%%%%%%%%%%%%%%%%%%%%%%%%%%%%%%%%%%%%%%%%%%%%%%%%%%%%%%%%%%%%%%%%%%%%%%%%%%%%%%%%%%%%%%%%%%%%%%%%%%%%%%%
%3.1 Soft Privacy vs. Hard Privacy
%%%%%%%%%%%%%%%%%%%%%%%%%%%%%%%%%%%%%%%%%%%%%%%%%%%%%%%%%%%%%%%%%%%%%%%%%%%%%%%%%%%%%%%%%%%%%%%%%%%%%%%%%%%%%%%%%%%%%%%%%%%%%%%%%
%%%%%%%%%%%%%%%%%%%%%%%%%%%%%%%%%%%%%%%%%%%%%%%%%%%%%%%%%%%%%%%%%%%%%%%%%%%%%%%%%%%%%%%%%%%%%%%%%%%%%%%%%%%%%%%%%%%%%%%%%%%%%%%%%%
\subsection{Soft Privacy vs. Hard Privacy}
\label{sec:FSTechnique:HardSoft}
\noindent
The privacy research literature~\cite{Mina10PHD} recognizes the distinction between the concepts of~\emph{soft privacy} and \emph{hard privacy}.
A privacy\hyph enhancing mechanism providing \emph{soft privacy} assumes that users entrust their private data to an entity,
which is thereafter responsible for the protection of their data.
In the literature, numerous attempts to protect privacy have followed the traditional method of
anonymous communications~\cite{Chaum81CACM,Reed96CSAC,Goldschlag99CACM,Dingledine04SSYM},
which is fundamentally based on the suppositions of soft privacy.
Unfortunately, anonymous\hyph communication systems are not completely
effective~\cite{Levine04FINCRYP,Bauer07TECH,Murdoch05SEPR,Pfitzmann90EURO},
they normally come at the cost of infrastructure, and assume that users are willing to trust other parties.

Our privacy\hyph protecting technique, per contra, leverages on the principle of \emph{hard privacy},
which assumes that users mistrust communicating entities and therefore strive to reveal as little private information as possible.
In the motivating scenario of this work, hard privacy means that users need not trust an external entity such as the recommender or the network operator.
Consequently, because users just trust themselves, it is their own responsibility to protect their privacy.
In this state of affairs, the forgery and the suppression of ratings appear as a technique that may hinder privacy attackers in their efforts to accurately profile users on the basis of the items they rate.
Specifically, when users are adhered to this technique, they have the possibility to submit ratings to items that do not reflect their genuine preferences, and/or
refrain from rating some items of their interest---this is what we refer to as the \emph{forgery} and the \emph{suppression} of ratings, respectively.

%%%%%%%%%%%%%%%%%%%%%%%%%%%%%%%%%%%%%%%%%%%%%%%%%%%%%%%%%%%%%%%%%%%%%%%%%%%%%%%%%%%%%%%%%%%%%%%%%%%%%%%%%%%%%%%%%%%%%%%%%%%%%%%%%%
%%%%%%%%%%%%%%%%%%%%%%%%%%%%%%%%%%%%%%%%%%%%%%%%%%%%%%%%%%%%%%%%%%%%%%%%%%%%%%%%%%%%%%%%%%%%%%%%%%%%%%%%%%%%%%%%%%%%%%%%%%%%%%%%%%
%3.2 User Profile and Adversary Model
%%%%%%%%%%%%%%%%%%%%%%%%%%%%%%%%%%%%%%%%%%%%%%%%%%%%%%%%%%%%%%%%%%%%%%%%%%%%%%%%%%%%%%%%%%%%%%%%%%%%%%%%%%%%%%%%%%%%%%%%%%%%%%%%%%
%%%%%%%%%%%%%%%%%%%%%%%%%%%%%%%%%%%%%%%%%%%%%%%%%%%%%%%%%%%%%%%%%%%%%%%%%%%%%%%%%%%%%%%%%%%%%%%%%%%%%%%%%%%%%%%%%%%%%%%%%%%%%%%%%%
\subsection{User Profile and Adversary Model}
\label{sec:FSTechnique:UserModel}
\noindent
In the scenario of recommendation systems,
users rate items of a very different nature, e.g., music, pictures, videos or news,
according to their personal preferences.
The information conveyed allows those systems to extract a profile of interests or
\emph{user profile}, which turns to be essential in the provision of personalized recommendations.

We mentioned in Sec.~\ref{sec:Introduction} that \emph{Movielens} represents user profiles by using some kind of histogram.
Other systems such as \emph{Jinni} and \emph{Last.fm} show this information by means of a tag cloud,
which in essence may be regarded as another kind of histogram.
In this same spirit, recent privacy\hyph protecting approaches in the scenario of recommendation systems also
propose using histograms of absolute frequencies for modeling user profiles~\cite{Toubiana10SNDSS,Fredrikson11SP}.

According to these examples and inspired by other works in the field~\cite{Rebollo10IT,Parra10TB,Parra11DPM,Rebollo11DSC,Parra12DKE,Parra12TKDE},
we model the \emph{items} rated by users as random variables (r.v.'s) taking on values in a common finite alphabet of categories,
namely the set $\{1,\dots,n\}$ for some integer $n\geqslant2$.
Concordantly, we model the profile of a user as a probability mass function (PMF) $q=(q_{1},\dots,q_{n})$, that is,
a histogram of relative frequencies of %(rated)
items within a predefined set of categories of interest.

We would like to emphasize that, under this model, user profiles do not capture the particular scores given to items,
but what we consider to be more sensitive: the categories these items belong to.
This is exactly the case of \emph{Movielens} and numerous content\hyph based recommendation systems.
Fig.~\ref{fig:user_profiles} provides an example that illustrates how user profiles are constructed in \emph{Movielens}.
In this particular example, a user assigns two stars to a movie, meaning that they consider it to be ``fairly bad''.
However, the recommender updates their profile based only on the categories this movie belongs to.

According to this model, a privacy attacker supposedly observes a perturbed version of this profile,
resulting from the forgery and the suppression of certain ratings, %according to a forgery strategy and a suppression strategy,
and is unaware or ignores the fact that the observed user profile, also in the form of a histogram,
does not reflect the actual profile of interests of the user in question.
In principle, our passive attacker could be the recommender itself or the network operator.
However, the set of potential attackers is not restricted merely to these two entities.
Since ratings are often publicly available to other users of the recommendation system,
any other attacker able to crawl through this information is taken into consideration in our adversary model.

When users adhere to the forgery and the suppression of ratings,
they specify a \emph{forgery rate}~$\rho \in[0,\infty)$ and a \emph{suppression rate}~$\sigma \in[0,1)$.
The former is the ratio of forged ratings to total genuine ratings that a user consents to submit.
The latter ratio is the fraction of genuine ratings that the user agrees to eliminate~\footnote{%
The description of an architecture implementing this data\hyph perturbative approach may be found in~\cite{Parra11DPM}.}.
Note that, in our approach, the number of false ratings submitted by the user can exceed the number of genuine ratings, that is, $\rho$ can be greater than 1.
Nevertheless, the number of suppressed ratings is always lower than the number of genuine ratings.

By forging and suppressing ratings, the \emph{actual} profile of interests $q$ is then perceived from the outside as the \emph{apparent} PMF $t=\frac{q+r-s}{1+\rho-\sigma}$,
according to a \emph{forgery strategy} $r=(r_1,\dots,r_n)$ and a \emph{suppression strategy} $s=(s_1,\dots,s_n)$.
Such strategies represent the proportion of ratings that the user should forge and eliminate in each of the $n$~categories.
Naturally, these strategies must satisfy, on the one hand, that $r_i \geqslant 0$, $s_i \geqslant 0$ and $q_i + r_i - s_i\geqslant 0$ for $i=1,\dots,n$,
and on the other, that $\sum_{i=1}^n r_i = \rho$ and $\sum_{i=1}^n s_i = \sigma$.
In conclusion, the apparent profile is the result of the addition and the substraction of certain items to/from the actual profile, and the posterior normalization by $\frac{1}{1+\rho-\sigma}$ so that $\sum_{i=1}^n t_i =1$.
%%%%%%%%%%%%%%%%%%%%%%%%%%%%%%%%%%%%%%%%%%%%%%%%%%%%%%%%%%%%%%%%%%%%%%%%%%%%%%%%%%%%%%%%%%%%%%%%%%%%%%%%%%%%%%%%%%%%%%%%%%%%%%%%%%
%%%%%%%%%%%%%%%%%%%%%%%%%%%%%%%%%%%%%%%%%%%%%%%%%%%%%%%%%%%%%%%%%%%%%%%%%%%%%%%%%%%%%%%%%%%%%%%%%%%%%%%%%%%%%%%%%%%%%%%%%%%%%%%%%%
%3.3 Measuring the Privacy User Profiles
%%%%%%%%%%%%%%%%%%%%%%%%%%%%%%%%%%%%%%%%%%%%%%%%%%%%%%%%%%%%%%%%%%%%%%%%%%%%%%%%%%%%%%%%%%%%%%%%%%%%%%%%%%%%%%%%%%%%%%%%%%%%%%%%%%
%%%%%%%%%%%%%%%%%%%%%%%%%%%%%%%%%%%%%%%%%%%%%%%%%%%%%%%%%%%%%%%%%%%%%%%%%%%%%%%%%%%%%%%%%%%%%%%%%%%%%%%%%%%%%%%%%%%%%%%%%%%%%%%%%%
\subsection{Measuring the Privacy of User Profiles}
\label{sec:FSTechnique:Metric}
\noindent
Inspired by the privacy measures proposed in~\cite{Li07ICDE,Rebollo10IT,Parra10TB,Rebollo11SECTECH,Parra13FGCS},
and according to the model of user profile assumed in Sec.~\ref{sec:FSTechnique:UserModel},
we define \emph{initial privacy risk} as the KL divergence~\cite{Cover06B} between the user's genuine profile and
the population's distribution, that is,
$$\cR_0 = \oD(q\,\|\,p).$$

Similarly, we define \emph{(final) privacy risk}~$\mathcal{R}$ as the KL divergence between the user's apparent profile and
the population's distribution,
$$\cR=\oD(t\,\|\,p)=\oD\left(\left. \frac{q+r-s}{1+\rho-\sigma}\,\right\|\,p\right).$$

An intuitive justification of our privacy metric stems from the observation that, whenever the user's apparent item distribution diverges too much from the population's,
a privacy attacker will have actually gained some information about the user, in contrast to the statistics of the general population.

A richer argument may be found in~\cite{Rebollo11SECTECH,Parra13FGCS}, where we establish some riveting connections between
Jaynes' rationale on entropy\hyph maximization methods and the use of entropies and divergences as measures of privacy.
The leading idea is that the method of types from information theory establishes an approximate monotonic relationship
between the likelihood of a PMF in a stochastic system and its Shannon's entropy.
Loosely speaking and in our context, the higher the entropy of a profile, the more likely it is,
the more users behave similarly.
This is in absence of a probability distribution model for the PMFs, viewed abstractly as r.v.'s themselves.
Under this interpretation, Shannon's entropy is a measure of anonymity, \emph{not} in the sense that the user's identity remains unknown,
but only in the sense that higher likelihood of an apparent profile, believed by an external observer to be the actual profile,
makes that profile more common, helping the user go unnoticed,
less interesting to an attacker assumed to strive to target peculiar users.

If an aggregated histogram of the population were available as a reference profile, as we assume in this work,
the extension of Jaynes' argument to relative entropy also gives an acceptable measure of privacy (or anonymity).
Recall~\cite{Cover06B} that KL divergence is a measure of discrepancy between probability distributions,
which includes Shannon's entropy as the special case when the reference distribution is uniform.
Conceptually, a lower KL divergence hides discrepancies with respect to a reference profile, say the population's,
and there also exists a monotonic relationship between the likelihood of a distribution and
its divergence with respect to the reference distribution of choice,
which enables us to regard KL divergence as a measure of anonymity in a sense entirely analogous to the above mentioned.
%%%%%%%%%%%%%%%%%%%%%%%%%%%%%%%%%%%%%%%%%%%%%%%%%%%%%%%%%%%%%%%%%%%%%%%%%%%%%%%%%%%%%%%%%%%%%%%%%%%%%%%%%%%%%%%%%%%%%%%%%%%%%%%%%%
%%%%%%%%%%%%%%%%%%%%%%%%%%%%%%%%%%%%%%%%%%%%%%%%%%%%%%%%%%%%%%%%%%%%%%%%%%%%%%%%%%%%%%%%%%%%%%%%%%%%%%%%%%%%%%%%%%%%%%%%%%%%%%%%%
%3.4. Formulation of the Trade\hyph Off among Privacy, Forgery and Suppression
%%%%%%%%%%%%%%%%%%%%%%%%%%%%%%%%%%%%%%%%%%%%%%%%%%%%%%%%%%%%%%%%%%%%%%%%%%%%%%%%%%%%%%%%%%%%%%%%%%%%%%%%%%%%%%%%%%%%%%%%%%%%%%%%%
%%%%%%%%%%%%%%%%%%%%%%%%%%%%%%%%%%%%%%%%%%%%%%%%%%%%%%%%%%%%%%%%%%%%%%%%%%%%%%%%%%%%%%%%%%%%%%%%%%%%%%%%%%%%%%%%%%%%%%%%%%%%%%%%%%
%----------------------------------------------------------Introduction-----------------------------------------------------------
 \subsection{Formulation of the Trade\hyph Off among Privacy, Forgery and Suppression}
 \label{sec:FSTechnique:Formulation}
 \noindent
Our data\hyph perturbative mechanism allows users to enhance their privacy to a certain extent, since the resulting profile, as observed from the outside, no longer captures their actual interests.
The price to be paid, however, is a loss in data utility, in particular in the accuracy of the recommender's predictions.

For the sake of tractability, in this work we consider as utility metrics the forgery rate and the suppression rate.
This consideration enables us to formulate the problem of choosing a forgery strategy and a suppression strategy as a multiobjective optimization problem that takes into account privacy, forgery rate and suppression rate.
Specifically, under the assumption that the population of users is large enough to neglect the impact of the choice of $r$ and $s$ on~$p$,
we define the \emph{privacy\hyph forgery\hyph suppression} function
\begin{equation}
\label{eqn:PrivacyUtilityFunction}
\cR(\rho,\sigma)=  \min_{\substack{r,s\\r_i \geqslant 0, \, s_i \geqslant 0,\\ q_i + r_i - s_i\geqslant 0,\\
\sum r_i = \rho, \,\sum s_i = \sigma}} \oD\left(\left. \frac{q+r-s}{1+\rho-\sigma}\,\right\|\,p\right),
\end{equation}
which characterizes the optimal trade\hyph off among privacy, forgery rate and suppression rate.

Conceptually,
the result of this optimization are two strategies $r$ and $s$ that contain information about which ratings should be forged and which ones should be suppressed,
in order to achieve the minimum privacy risk. More precisely, the component $r_i$ is the percentage of items that the user should forge in the category $i$.
The component $s_i$ is defined analogously for suppression.

%%%%%%%%%%%%%%%%%%%%%%%%%%%%%%%%%%%%%%%%%%%%%%%%%%%%%%%%%%%%%%%%%%%%%%%%%%%%%%%%%%%%%%%%%%%%%%%%%%%%%%%%%%%%%%%%%%%%%%%%%%%%%%%%%%
%%%%%%%%%%%%%%%%%%%%%%%%%%%%%%%%%%%%%%%%%%%%%%%%%%%%%%%%%%%%%%%%%%%%%%%%%%%%%%%%%%%%%%%%%%%%%%%%%%%%%%%%%%%%%%%%%%%%%%%%%%%%%%%%%
%%%%%%%%%%%%%%%%%%%%%%%%%%%%%%%%%%%%%%%%%%%%%%%%%%%%%%%%%%%%%%%%%%%%%%%%%%%%%%%%%%%%%%%%%%%%%%%%%%%%%%%%%%%%%%%%%%%%%%%%%%%%%%%%%%
%4 THEORETICAL ANALYSIS
%%%%%%%%%%%%%%%%%%%%%%%%%%%%%%%%%%%%%%%%%%%%%%%%%%%%%%%%%%%%%%%%%%%%%%%%%%%%%%%%%%%%%%%%%%%%%%%%%%%%%%%%%%%%%%%%%%%%%%%%%%%%%%%%%%
%%%%%%%%%%%%%%%%%%%%%%%%%%%%%%%%%%%%%%%%%%%%%%%%%%%%%%%%%%%%%%%%%%%%%%%%%%%%%%%%%%%%%%%%%%%%%%%%%%%%%%%%%%%%%%%%%%%%%%%%%%%%%%%%%%
%%%%%%%%%%%%%%%%%%%%%%%%%%%%%%%%%%%%%%%%%%%%%%%%%%%%%%%%%%%%%%%%%%%%%%%%%%%%%%%%%%%%%%%%%%%%%%%%%%%%%%%%%%%%%%%%%%%%%%%%%%%%%%%%%%

\section{Optimal Forgery and Suppression of Ratings}
\label{sec:Theory}
\noindent
This section is entirely devoted to the theoretical analysis of the privacy\hyph forgery\hyph suppression function~\eqref{eqn:PrivacyUtilityFunction} defined in Sec.~\ref{sec:FSTechnique:Formulation}.
In our attempt to characterize the trade\hyph off among privacy risk, forgery rate and suppression rate,
we shall present a closed\hyph form solution to the optimization problem inherent in the definition of this function.
Afterwards, we shall analyze some fundamental properties of said trade\hyph off.
For the sake of brevity, our theoretical analysis only contemplates the case when all given probabilities are strictly positive:
%------------------------------------------------------------ Positivity assumption ------------------------------------------------------------
\begin{equation}
\label{eqn:PositivityAssumption}
q_i, p_i >0 \textnormal{ for all }i=1,\dots,n.
\end{equation}
Additionally, we suppose without loss of generality that
%------------------------------------------------------------ Labeling assumption ------------------------------------------------------------
\begin{equation}\label{eqn:LabelingAssumption}
\frac{q_1}{p_1}\leqslant \cdots\leqslant\frac{q_n}{p_n}.
\end{equation}

Before diving into the mathematical analysis, it is immediate from the definition of the privacy\hyph forgery\hyph suppression function that its initial value is
$\cR(0,0)=\oD(q\,\|\,p)$.
The characterization of the optimal trade\hyph off surface modeled by $\cR(\rho,\sigma)$ at any other values of $\rho$ and $\sigma$ is the focus of this section.

%%%%%%%%%%%%%%%%%%%%%%%%%%%%%%%%%%%%%%%%%%%%%%%%%%%%%%%%%%%%%%%%%%%%%%%%%%%%%%%%%%%%%%%%%%%%%%%%%%%%%%%%%%%%%%%%%%%%%%%%%%%%%%%%%%
%%%%%%%%%%%%%%%%%%%%%%%%%%%%%%%%%%%%%%%%%%%%%%%%%%%%%%%%%%%%%%%%%%%%%%%%%%%%%%%%%%%%%%%%%%%%%%%%%%%%%%%%%%%%%%%%%%%%%%%%%%%%%%%%%%
%4.1 CLOSE-FORM SOLUTION
%%%%%%%%%%%%%%%%%%%%%%%%%%%%%%%%%%%%%%%%%%%%%%%%%%%%%%%%%%%%%%%%%%%%%%%%%%%%%%%%%%%%%%%%%%%%%%%%%%%%%%%%%%%%%%%%%%%%%%%%%%%%%%%%%%
%%%%%%%%%%%%%%%%%%%%%%%%%%%%%%%%%%%%%%%%%%%%%%%%%%%%%%%%%%%%%%%%%%%%%%%%%%%%%%%%%%%%%%%%%%%%%%%%%%%%%%%%%%%%%%%%%%%%%%%%%%%%%%%%%%
\subsection{Closed\hyph Form Solution}
\label{sec:Theory:ClosedFormSolution}
\noindent
Our first theorem, Theorem~\ref{Theorem:ClosedFormSolution}, will present a closed\hyph form solution to the minimization problem involved in the definition of
function~\eqref{eqn:PrivacyUtilityFunction}.
The solution will be derived from Lemma~\ref{Lemma:ResourceAllocation},
which addresses a resource allocation problem.
This a theoretical problem encountered in many fields, from load distribution and production planning to communication networks, computer scheduling and portfolio selection~\cite{Ibaraki88MIT}.
Although this lemma provides a parametric\hyph form solution, we shall be able to proceed towards an explicit closed\hyph form solution, albeit piecewise.

%%%%%%%%%%%%%%%%%%%%%%%%%%%%%%%%%%%%%%%%%%%%%%%%%%%%%%%%%%%%%%%%%%%%%%%%%%%%%%%%%%%%%%%%%%%%%%%%%%%%%%%%%%%%%%%%%%%%%%%%%%%%%%%%%%
%RESOURCE ALLOCATION LEMMA
%%%%%%%%%%%%%%%%%%%%%%%%%%%%%%%%%%%%%%%%%%%%%%%%%%%%%%%%%%%%%%%%%%%%%%%%%%%%%%%%%%%%%%%%%%%%%%%%%%%%%%%%%%%%%%%%%%%%%%%%%%%%%%%%%%
\begin{lemma}[Resource Allocation]
\label{Lemma:ResourceAllocation}
For all $k=1,\ldots,n$, let $f_k$ be a real\hyph valued function on $\{(x_k,y_k)\in \mathbb{R}^2 \colon \kappa_k + x_k - y_k \geqslant 0\},$
twice differentiable in the interior of its domain.
Assume that $\frac{\partial f_k}{\partial x_k} = - \frac{\partial f_k}{\partial y_k}$, that $\frac{\partial^2 f _k}{\partial x_k^2} = \frac{\partial^2 f _k}{\partial y_k^2}>0$ and that the Hessian $H(f_k)$ is positive semidefinite.
Define $h_k = \frac{\partial f_k}{\partial x_k}$.
Because $\frac{\partial h_k}{\partial x_k}>0$ and $\frac{\partial h_k}{\partial y_k}<0$,
it follows that $h_k$ is strictly increasing in $x_k$ and strictly decreasing in $y_k$.
Consequently, for a fixed~$y_k$, $h_k(x_k,y_k)$ is an invertible function of $x_k$. Denote by $h_k^{-1}$ the inverse of $h_k(x_k,0)$.
Suppose further that $h_k(x_k,y_k) = h_k(x_k-y_k, 0)$ and finally that $\lim \limits_{x_k \downarrow y_k - \kappa_k} h_k(x_k,y_k) = -\infty$.
Now consider the following optimization problem in the variables $x_1,\ldots, x_n$ and $y_1,\ldots, y_n$:
\begin{equation*}
\begin{aligned}
&\textnormal{minimize }    && \sum_{k=1}^n f_k(x_k,y_k) \\
&\textnormal{subject to }    && x_k,\, y_k \geqslant 0,\\
                                                    &&& \kappa_k + x_k - y_k \geqslant 0 \textnormal{ for } k =1,\ldots,n,\\
                                                   &&& \textnormal{and } \sum_{k=1}^n x_k = \eta,\,\sum_{k=1}^n y_k = \theta \textnormal{ for some } \eta,\theta \geqslant 0.
\end{aligned}
\end{equation*}

\begin{EnumerateRoman}
\item The solution to the problem $(x^*_k, y^*_k)$ depends on two real numbers $\psi,\omega$ that satisfy the equality constraints $\sum_k x_k^*=\eta$ and $\sum_k y_k^*=\theta$.
The solution exists provided that $\psi\leqslant\omega$. If $\psi<\omega$, then the solution is unique and yields
$$(x^*_k, y^*_k)= \left( \max\left\{0,h_k^{-1}(\psi)\right\},\max\left\{0,-h_k^{-1}(\omega)\right\}\right).$$
If $\psi=\omega$, then there exists an infinite number of solutions of the form $(x^*_k +\alpha_k, y^*_k+\alpha_k)$ for all $\alpha_k \in\mathbb{R}_+$ meeting the two aforementioned equality constraints.
\end{EnumerateRoman}
\noindent
Without loss of generality, suppose that $h_1(0,0)\leqslant\dots\leqslant h_n(0,0)$.
\begin{EnumerateRoman}
\item [(ii)] For $\psi<\omega$, consider the following cases:
\begin{EnumerateRoman}
\item[(a)] $h_i(0,0) < \psi \leqslant h_{i+1}(0,0)$ for some $i=1,\ldots,j-1$ and $h_{j-1}(0,0) \leqslant \omega < h_j(0,0)$ for some $j=2,\ldots,n$.
\item[(b)] $h_{j-1}(0,0) \leqslant \omega$ for $j=n+1$ and, either $h_i(0,0) < \psi \leqslant h_{i+1}(0,0)$ for some $i=1,\ldots,n-1$ or $h_i(0,0) < \psi$ for $i=n$.
\item[(c)] $\psi \leqslant h_{i+1}(0,0)$ for $i=0$ and, either $h_{j-1}(0,0) \leqslant \omega < h_j(0,0)$ for some $j=2,\ldots,n$ or $\omega < h_j(0,0)$ for $j=1$.
\item[(d)] $h_{j-1}(0,0) \leqslant \omega$ for $j=n+1$ and $\psi \leqslant h_{i+1}(0,0)$ for $i=0$.
\end{EnumerateRoman}
In each case, and for the corresponding indexes $i$ and $j$,
%\vspace{0.07mm}
\begin{equation*}
\begin{array}{ll}
x^*_k & =\left\{ \begin{array}{l@{,\quad}l}
                                            h_k^{-1}(\psi) \quad \quad & k=1,\dots,i \\
                                            0 \quad \quad & k=i+1,\dots,n
                \end{array}\right.,\\
\rule{0pt}{4ex}
y^*_k & =\left\{ \begin{array}{l@{,\quad}l}
                                            0 \,\,\, & k=1,\dots,j-1\\
                                            -h_k^{-1}(\omega) \,\,\,  & k=j,\dots,n
                \end{array}\right..
\end{array}
\end{equation*}
%\vspace{0.07mm}
\end{EnumerateRoman}
\begin{EnumerateRoman}
\item [(iii)] For $\psi=\omega$, consider the following cases:
\begin{EnumerateRoman}
\item[(a)]either $h_i(0,0)<\psi<h_j(0,0)$ for some $j=2,\ldots,n$ and $i=j-1$, or $h_i(0,0) < \psi = h_{i+1}(0,0) = \cdots = h_{j-1}(0,0) < h_j(0,0)$ for some $i=1,\ldots,j-2$ and some $j=3,\ldots,n$.
\item[(b)]for $j=n+1$, either $h_i(0,0)<h_{i+1}(0,0)=\cdots=h_{j-1}(0,0)=\omega$ for some $i=1,\ldots,j-2$ or $h_{j-1}(0,0)<\omega$ with $i=n$.
\item[(c)] for $i=0$, either $\psi = h_{i+1}(0,0)=\cdots=h_{j-1}(0,0)<h_j(0,0)$ for some $j=2,\ldots,n$ or $\psi<h_{i+1}(0,0)$ with $j=1$.
\end{EnumerateRoman}
In each case, and for the corresponding indexes $i$ and $j$,
%\vspace{0.07mm}
\begin{equation*}
\begin{array}{rl}
x^*_k &=\left\{ \begin{array}{l@{,\quad}l}
                                            h_k^{-1}(\psi) + \alpha_k \quad \quad & k=1,\dots,i \\
                                            \alpha_k \quad \quad & k=i+1,\dots,n
                \end{array}\right.,\\
\rule{0pt}{4ex}
y^*_k &=\left\{ \begin{array}{l@{,\quad}l}
                                            \alpha_k \,\,\, & k=1,\dots,j-1\\
                                            -h_k^{-1}(\omega)  + \alpha_k \,\,\,  & k=j,\dots,n
                \end{array}\right..
\end{array}
\end{equation*}
%\vspace{0.07mm}
\end{EnumerateRoman}
\end{lemma}
\SpaceAfterPropositionEndedWithFormula
\BeginProof
The proof of statement~(i) consists of two steps.
In the first step, we show that the optimization problem stated in the lemma is convex; %in the lemma poses the structure of a convex problem.
then we apply Karush\hyph Kuhn\hyph Tucker (KKT) conditions to said problem, and finally reformulate these conditions into a reduced number of equations.
The bulk of this proof comes later, in the second step, where we proceed to solve the system of equations for the two cases considered in the lemma,
$\psi <\omega$ and $\psi=\omega$.
Lastly, statements~(ii) and~(iii) follow from~(i).

To see that the problem is convex, simply observe that the objective function is convex on account of $H(f_k)\succeq 0$, and
that the inequality and equality constraint functions are affine.
Since the objective and constraint functions are also differentiable and Slater's constraint qualification holds, KKT conditions are necessary and sufficient conditions for optimality~\cite{Boyd04B}.
Systematic application of these optimality conditions leads to the Lagrangian cost,
\begin{multline*}
\cL =\sum f_k(x_k,y_k) - \sum \lambda_k x_k -\sum \mu_k y_k +\sum \nu_k (y_k - \kappa_k - x_k) - \psi\left( \sum x_k - \eta \right) + \omega \left( \sum y_k - \theta\right),
\end{multline*}
and finally to the conditions
%\vspace{0.15mm}
\begin{equation*}
\begin{array}{lrr}
\rule{0pt}{1ex}
 x_k \geqslant 0,\, y_k \geqslant 0,\, \kappa_k + x_k - y_k\geqslant 0,&\\
\sum x_k=\eta,\, \sum y_k=\theta, &\textnormal{(primal feasibility)}\\
\rule{0pt}{3ex}
\lambda_k\geqslant 0,\, \mu_k \geqslant 0,\,\nu_k \geqslant 0,& \textnormal{(dual feasibility)}\\
\rule{0pt}{3ex}
\lambda_k\, x_k=0, \,\mu_k\, y_k=0,&\\
\nu_k\, (y_k - \kappa_k - x_k)=0, &\textnormal{(complementary slackness)}\\
\rule{0pt}{3ex}
\frac{\partial \cL}{\partial x_k} = h_k(x_k,y_k) - \lambda_k -\nu_k - \psi = 0,&\\
\frac{\partial \cL}{\partial y_k} = h_k(x_k,y_k) + \mu_k - \nu_k - \omega = 0.  &\textnormal{(dual optimality)}
\end{array}
\end{equation*}
%\vspace{0.10mm}

Because $\lim \limits_{x_k \downarrow y_k - \kappa_k} h_k(x_k,y_k) = -\infty$, it follows from the dual optimality conditions that $\kappa_k + x_k - y_k>0$,
which implies, by complementary slackness, that $\nu_k=0$.
Subsequently, we may rewrite the dual optimality conditions as $\lambda_k = h_k(x_k,y_k) - \psi$ and $\mu_k =  \omega - h_k(x_k,y_k)$.
By eliminating the slack variables $\lambda_k,\,\mu_k$, we obtain the simplified conditions $h_k(x_k,y_k) \geqslant \psi$ and $h_k(x_k,y_k) \leqslant \omega$.
Lastly, we substitute the above expressions of $\lambda_k$ and $\mu_k$ into the complementary slackness conditions, so that
we can formulate the dual optimality and complementary slackness conditions equivalently as
\begin{align}
        &h_k(x_k,y_k) \geqslant\,\psi, \label{eqn:Lemma1}\\
        &h_k(x_k,y_k) \leqslant\,\omega, \label{eqn:Lemma2}\\
        &(h_k(x_k,y_k) - \psi)\,x_k=0, \label{eqn:Lemma3}\\
        &(h_k(x_k,y_k) - \omega)\,y_k\,=0. \label{eqn:Lemma4}
\end{align}

In the following, we shall proceed to solve these equations which, together with the primal and dual feasibility conditions, are necessary and sufficient conditions for optimality.
To this end, first note that, if $\psi>\omega$, then there exists no $(x_k,y_k)$ that satisfies equations~\eqref{eqn:Lemma1} and~\eqref{eqn:Lemma2} at the same time, and consequently,
as stated in part (i) of the lemma, there is no solution.
Concordantly, next we shall study the case when $\psi<\omega$; afterwards we shall tackle the other case when $\psi=\omega$.

Before plunging into the analysis of the former case, recall that the function $h_k$ is strictly increasing in $x_k$ and strictly decreasing in~$y_k$.
Having said this, observe that, under the assumption $\psi<\omega$,
the variables $x_k$ and $y_k$ cannot be positive simultaneously by virtue of equations~\eqref{eqn:Lemma3} and~\eqref{eqn:Lemma4}.
Bearing this in mind, consider these three possibilities for each~$k$: $h_k(0,0)<\psi$, $\psi\leqslant h_k(0,0)\leqslant \omega$ and $\omega < h_k(0,0)$.

% Case 1. V < W and H_k(0,0) < V < W
When $h_k(0,0)<\psi$, the only conclusion consistent with~\eqref{eqn:Lemma1} and with the fact that $h_k$ is strictly increasing in $x_k$ is that $x_k>0$.
Since $x_k$ must be positive, the complementary slackness condition~\eqref{eqn:Lemma3} implies that $h_k(x_k,y_k) = \psi$ and, because of~\eqref{eqn:Lemma4}, that $y_k=0$.
As a result, $x_k$ must satisfy $h_k(x_k,0) = \psi$, or equivalently, $x_k= h_k^{-1}(\psi)$.
% Case 1. The solution is unique.
Next, we show that the solution $(x_k,0)$ is unique.
For this purpose, suppose that $y_k>0$ and, in consequence, that $x_k=0$.
It follows from~\eqref{eqn:Lemma4}, however, that $h_k(0,y_k) = \omega$,
which contradicts the fact that $h_k$ is a strictly decreasing function of~$y_k$.
In the end, we verify that $x_k=y_k=0$ does not satisfy~\eqref{eqn:Lemma1} and thus prove that $(x_k,y_k)=(h_k^{-1}(\psi),0)$ is the unique minimizer of the objective function when $h_k(0,0)<\psi$.

% Case 2. V < W and V <= H_k(0,0) <= W
Now consider the case when $\psi\leqslant h_k(0,0)\leqslant\omega$.
First, suppose that $x_k>0$, and therefore that $y_k=0$.
By complementary slackness, it follows that $h_k(x_k,0)=\psi$, which is not consistent with the fact that $h_k$ is strictly increasing in $x_k$. Consequently, $x_k$ cannot be positive.
Secondly, assume that $x_k$ is zero and $y_k$ positive.
Under this assumption, equation~\eqref{eqn:Lemma4} implies that $h_k(0,y_k)=\omega$,
a contradiction since $h_k$ is a strictly decreasing function of $y_k$.
Accordingly, $y_k$ cannot be positive either.
Finally, check that $x_k=y_k=0$ satisfies the optimality conditions and hence
it is the unique solution.

% Case 3. V < W and  W < H_k(0,0)
The last possibility corresponds to the case when $\omega < h_k(0,0)$.
Note that, in this case, the only conclusion consistent with~\eqref{eqn:Lemma2} and with the fact that $h_k$ is strictly decreasing in $y_k$ is that $y_k>0$. %, and thus $x_k=0$.
Thus, because of~\eqref{eqn:Lemma4}, $y_k$~must satisfy $h_k(0,y_k) = \omega$.
Recalling from the lemma that $h_k(x_k,y_k)=h_k(x_k-y_k,0)$,
we may express the condition $h_k(0,y_k) = \omega$ equivalently as $y_k=-h_k^{-1}(\omega)$.
Lastly, we check that this solution is unique in the case under study.
To this end, note that a solution such that $x_k>0$ and $y_k=0$ contradicts the fact that $h_k$ is strictly increasing in $x_k$.
As a result, $x_k$ cannot be positive.
Finally, we confirm that equation~\eqref{eqn:Lemma2} does not hold for $x_k=y_k=0$ and
therefore prove that $(x_k,y_k)=(0,-h_k^{-1}(\omega))$ is the unique solution when $\omega<h_k(0,0)$.

In summary, $x_k=h_k^{-1}(\psi)$ if $h_k(0,0)<\psi$, or equivalently, $h_k^{-1}(\psi)>0$; otherwise $x_k=0$.
Further, $y_k=-h_k^{-1}(\omega)$ if $h_k(0,0)>\omega$, or equivalently, $h_k^{-1}(\omega)<0$; otherwise $y_k=0$.
Accordingly, we may write the solution compactly as
$$(x_k, y_k)= \left(\max\left\{0,h_k^{-1}(\psi)\right\},\max\left\{0,-h_k^{-1}(\omega)\right\}\right),$$
where $\psi,\omega$ must satisfy the primal equality constraints $\sum_k x_k =\eta$ and $\sum_k y_k =\theta$.

% Case 4. V = W. Presentation
Having examined the case when $\psi <\omega$, next we proceed to solve the optimality conditions at hand for~$\psi=\omega$.
Observe that, in this new case, \eqref{eqn:Lemma1}~and~\eqref{eqn:Lemma2} transform into the equation
\begin{equation}\label{eqn:Lemma5}
h_k(x_k,y_k)=\psi.
\end{equation}
Moreover, note that any pair $(x_k,y_k)$ satisfying~\eqref{eqn:Lemma5} also
meets the complementary slackness conditions~\eqref{eqn:Lemma3} and~\eqref{eqn:Lemma4}.
However, notice that this does not mean that all those pairs are optimal.
To elaborate on this point, consider the following three possibilities for each $k$: $h_k(0,0)<\psi$, $h_k(0,0)=\psi$ and $\psi < h_k(0,0)$.

% Case 4.1 V = W and  H_k(0,0) < V
In the case when $h_k(0,0)<\psi$, the only condition consistent with~\eqref{eqn:Lemma5} and with the fact that $h_k$ is strictly increasing in $x_k$ is that $x_k>0$.
From the lemma, it is immediate that $\frac{\partial h_k}{\partial x_k} = - \frac{\partial h_k}{\partial y_k}$, which implies that $x_k$ must also be greater than $y_k$.
Hence, the set of solutions is
$$\{(x_k,y_k)\colon h_k(x_k,y_k) = \psi,\, x_k>y_k\},$$
where every pair in this set must also fulfill the primal equality conditions.
Let $x'_k$ satisfy $h_k(x'_k,0) = \psi$, or equivalently, $x'_k=h_k^{-1}(\psi)$.
Then, because $h_k(x'_k+\alpha_k,\alpha_k)=\psi$ for any $\alpha\geqslant0$,
this set may be recast equivalently as
$$\{(x_k,y_k)\colon x_k = x'_k + \alpha_k,\, y_k=\alpha_k\}.$$

% Case 4.2 V = W and  H_k(0,0) = V
For the two remaining cases, i.e., $h_k(0,0)=\psi$ and $\psi < h_k(0,0)$,
the set of solutions is obtained in a completely analogous way as above.
In the former case, the pairs $(x_k,y_k)$ must satisfy $x_k=y_k$, and the set of solutions may be expressed as
$$\{(x_k,y_k)\colon x_k =\alpha_k,\, y_k=\alpha_k\}.$$
% Case 4.3 V = W and  V < H_k(0,0)
In the latter case, it follows that $y_k>x_k$ and, consequently, that the set of solutions is
$$\{(x_k,y_k)\colon x_k = \alpha_k,\, y_k=y'_k + \alpha_k\},$$
where $y'_k$ must satisfy $h_k(0,y'_k)=\psi$.

To sum up, the case $\psi=\omega$ leads to the following solutions:
$x_k=h_k^{-1}(\psi) + \alpha_k$ if $h_k(0,0)<\psi$, or equivalently, $h_k^{-1}(\psi)>0$; otherwise $x_k=\alpha_k$.
In addition, $y_k=-h_k^{-1}(\omega)+\alpha_k$ if $h_k(0,0)>\omega$, or equivalently, $h_k^{-1}(\omega)<0$; otherwise $y_k=\alpha_k$.
Accordingly, the solutions $(x_k, y_k)$ yield
\begin{equation}\label{eqn:Lemma6}
\left(\max\left\{0,h_k^{-1}(\psi)\right\}+\alpha_k,\max\left\{0,-h_k^{-1}(\omega)\right\}+\alpha_k\right),
\end{equation}
for some $\psi,\omega$ and nonnegative sequence $\alpha_1,\ldots,\alpha_n$ such that $\sum_k x_k =\eta$ and $\sum_k y_k =\theta$.
Note that, although $\psi=\omega$, we intentionally write $\omega$ instead of $\psi$ to highlight that the solutions for $\psi<\omega$ and for $\psi=\omega$ just differ in the term $\alpha_k$, as we claimed in part (i) of the lemma.

To complete the proof of statement~(i), it suffices to show that the number of solutions is infinite when $\psi=\omega$.
To this end, simply observe that there exists an infinite number of sequences $\alpha_1,\ldots,\alpha_n$ such that
$$\sum_k x_k = \sum_k h_k^{-1}(\psi) + \sum_k \alpha_k = \eta\textnormal{\,\,\,\, and}$$
$$\sum_k y_k = -\sum_k h_k^{-1}(\psi) + \sum_k \alpha_k = \theta,$$
which results in an infinite number of solutions of the form given in~\eqref{eqn:Lemma6}.

Now we proceed to prove~(ii), which is an immediate consequence of~(i).
For this purpose, observe that if $\psi \leqslant h_{i+1}(0,0)\leqslant\dots\leqslant h_n(0,0)$ holds for some $i=0,\ldots,n-1$, then
$h_{i+1}^{-1}(\psi),\ldots,h_n^{-1}(\psi) \leqslant 0$, and accordingly $x_{i+1}=\cdots=x_n=0$.
Similarly, if $h_1(0,0)\leqslant\dots\leqslant h_{j-1}(0,0)\leqslant\omega$ is satisfied for some $j=2,\ldots, n+1$, then
$h_1^{-1}(\omega),\ldots,h_{j-1}^{-1}(\omega) \geqslant 0$, and thus $y_1=\cdots=y_{j-1}=0$.

Note that the particular case when the index $i$ ranges from $1$ to $j-1$ and the index $j$ goes from $2$ to $n$ is the case described in (ii)~(a),
which corresponds to $\eta,\theta>0$.
Further, observe that the case assumed in (ii)~(b), i.e., when $j=n+1$, implies that $\theta=0$.
Here, the index $i$ starts at $1$, therefore excluding $\eta=0$, and ends at $n$, including the possibility that $x_i > 0$ for all $i$.
In part (ii)~(c), we consider $i=0$, which is equivalent to the condition $\eta=0$.
In this case, the index $j$ starts at $1$, permitting $y_j>0$ for all $j$, and ends at $n$, avoiding $\theta=0$.
Finally, the case described in (ii)~(d), namely when $j=n+1$ and $i=0$, is precisely the trivial case $x=y=0$.

In order to verify statement~(iii), we proceed analogously by noting that if $\psi=h_{i+1}(0,0)=\cdots=h_{j-1}(0,0)$ holds for some $i=1,\ldots,j-2$ and some $j=3,\ldots,n$,
then $h_{i+1}^{-1}(\psi)=\cdots =h_{j-1}^{-1}(\psi)=0$, and consequently $x_k=y_k=\alpha_k$ for $k=i+1,\ldots,j-1$.
\EndProof

%%%%%%%%%%%%%%%%%%%%%%%%%%%%%%%%%%%%%%%%%%%%%%%%%%%%%%%%%%%%%%%%%%%%%%%%%%%%%%%%%%%%%%%%%%%%%%%%%%%%%%%%%%%%%%%%%%%%%%%%%%%%%%%%%%
% CRITICAL-PRIVACY REGION
%%%%%%%%%%%%%%%%%%%%%%%%%%%%%%%%%%%%%%%%%%%%%%%%%%%%%%%%%%%%%%%%%%%%%%%%%%%%%%%%%%%%%%%%%%%%%%%%%%%%%%%%%%%%%%%%%%%%%%%%%%%%%%%%%%
The previous lemma presented the solution to a resource allocation problem that minimizes a rather general but convex objective function, subject to affine constraints.
Our next theorem, Theorem~\ref{Theorem:ClosedFormSolution}, applies the results of this lemma to the special case of the objective function of problem~\eqref{eqn:PrivacyUtilityFunction}.
In doing so, we shall confirm the intuition that there must exist a set of ordered pairs $(\rho,\sigma)$
where the privacy risk vanishes
and another set where it does not.
We shall refer to the former set as the \emph{critical\hyph privacy region} and formally define it as
$$\cC =\{(\rho,\sigma) \colon \cR(\rho,\sigma)=0\}.$$
The latter set will be the complementary set~$\bar{\cC}$ and we shall refer to it as the \emph{noncritical\hyph privacy region}.

%%%%%%%%%%%%%%%%%%%%%%%%%%%%%%%%%%%%%%%%%%%%%%%%%%%%%%%%%%%%%%%%%%%%%%%%%%%%%%%%%%%%%%%%%%%%%%%%%%%%%%%%%%%%%%%%%%%%%%%%%%%%%%%%%%
% SUPPRESSION & FORGERY THRESHOLDS
%%%%%%%%%%%%%%%%%%%%%%%%%%%%%%%%%%%%%%%%%%%%%%%%%%%%%%%%%%%%%%%%%%%%%%%%%%%%%%%%%%%%%%%%%%%%%%%%%%%%%%%%%%%%%%%%%%%%%%%%%%%%%%%%%%
Before proceeding with Theorem~\ref{Theorem:ClosedFormSolution},
first we shall introduce what we term \emph{forgery} and \emph{suppression thresholds},
two sequences of rates that will play a fundamental role in the characterization of the solution to the minimization problem defining the privacy\hyph forgery\hyph suppression function.
Secondly, we shall investigate certain properties of these thresholds in Proposition~\ref{Proposition:Monotonicity}.
And thereafter, we shall introduce some definitions that will facilitate the exposition of the aforementioned theorem.

Let $Q_i=\sum_{k=1}^i q_k$ and $P_i=\sum_{k=1}^i p_k$ be the cumulative distribution functions corresponding to $q$ and~$p$.
Denote by $\cCQ_i=\sum_{k=i}^n q_k$ and $\cCP_i=\sum_{k=i}^n p_k$ the complementary cumulative distribution functions of $q$ and~$p$.
Define the \emph{forgery thresholds} $\rho_i$ as
$$\rho_i=\left\{ \begin{array}{l@{,\quad}l}
                                            P_i \frac{q_i}{p_i} - Q_i \quad & i=1,\dots,j-1 \\
                                            \frac{P_{j-1}}{\cCP_j}(\cCQ_j - \sigma) - Q_{j-1} \quad & i=j \\
                                            \infty \quad & i=j+1
                \end{array}\right.,$$
for $j=2,\ldots,n$.
Additionally, define the \emph{suppression thresholds} $\sigma_j$ as
$$\sigma_j = \cCQ_j - \cCP_j \frac{q_j}{p_j}$$
%\vspace{0.10mm}
for $j=1,\ldots,n$, and $\sigma_0=1$.
Observe that $\rho_1=\sigma_n=0$ and that the forgery threshold $\rho_j$ is a linear function of~$\sigma$.
We shall refer to this latter threshold as the \emph{critical forgery\hyph suppression threshold} and denote it also by~$\rhocrit(\sigma)$.
The reason is that said threshold will determine the boundary of the critical\hyph privacy region, as we shall see later. %$\rho_j (\sigma)$
The following result, Proposition~\ref{Proposition:Monotonicity}, characterizes the monotonicity of the forgery and the suppression thresholds.
%%%%%%%%%%%%%%%%%%%%%%%%%%%%%%%%%%%%%%%%%%%%%%%%%%%%%%%%%%%%%%%%%%%%%%%%%%%%%%%%%%%%%%%%%%%%%%%%%%%%%%%%%%%%%%%%%%%%%%%%%%%%%%%%%%
% MONOTONICITY OF SUPPRESSION & FORGERY THRESHOLDS
%%%%%%%%%%%%%%%%%%%%%%%%%%%%%%%%%%%%%%%%%%%%%%%%%%%%%%%%%%%%%%%%%%%%%%%%%%%%%%%%%%%%%%%%%%%%%%%%%%%%%%%%%%%%%%%%%%%%%%%%%%%%%%%%%%
\begin{proposition}[Monotonicity of Thresholds]
\label{Proposition:Monotonicity}
\leavevmode
\begin{EnumerateRoman}
\item For $j=3,\ldots,n$ and $i=1,\ldots,j-2$, the forgery thresholds satisfy $\rho_i \leqslant \rho_{i+1}$, with equality if, and only if, $\frac{q_i}{p_i}=\frac{q_{i+1}}{p_{i+1}}.$
\item For $j=2,\ldots,n$, the suppression thresholds satisfy $\sigma_j \leqslant \sigma_{j-1}$, with equality if, and only if, $\frac{q_j}{p_j}=\frac{q_{j-1}}{p_{j-1}}.$
\item Further, for any $j=2,\ldots,n$ and any $\sigma \in (\sigma_j,\sigma_{j-1}]$, the critical forgery\hyph suppression threshold satisfies $\rho_j (\sigma) \geqslant \rho_{j-1}$,
with equality if, and only if, $\sigma=\sigma_{j-1}$.
\end{EnumerateRoman}
\end{proposition}
\BeginProof
The first statement can be shown from the definition of the forgery thresholds by routine algebraic manipulation and under the labeling assumption~\eqref{eqn:LabelingAssumption}.
To this end, it is helpful to note that
$$P_i \frac{q_{i+1}}{p_{i+1}} - Q_i = P_{i+1} \frac{q_{i+1}}{p_{i+1}} - Q_{i+1}.$$
The second statement can be shown analogously, observing that
$$\cCQ_j - \cCP_j \frac{q_{j-1}}{p_{j-1}} = \cCQ_{j-1} - \cCP_{j-1} \frac{q_{j-1}}{p_{j-1}}.$$
For the last statement, use the definitions of the forgery and the suppression thresholds to note that the condition $\rho_j (\sigma) \geqslant \rho_{j-1}$ is equivalent to $\sigma \leqslant \sigma_{j-1}$.
\EndProof

%%%%%%%%%%%%%%%%%%%%%%%%%%%%%%%%%%%%%%%%%%%%%%%%%%%%%%%%%%%%%%%%%%%%%%%%%%%%%%%%%%%%%%%%%%%%%%%%%%%%%%%%%%%%%%%%%%%%%%%%%%%%%%%%%%
% DEFINITIONS of DISTRIBUTIONS AND TUPLES
%%%%%%%%%%%%%%%%%%%%%%%%%%%%%%%%%%%%%%%%%%%%%%%%%%%%%%%%%%%%%%%%%%%%%%%%%%%%%%%%%%%%%%%%%%%%%%%%%%%%%%%%%%%%%%%%%%%%%%%%%%%%%%%%%%
Prior to investigate a closed\hyph form solution to the problem~\eqref{eqn:PrivacyUtilityFunction},
we introduce some definitions for ease of presentation.
For $i=1,\ldots,j-1$ and $j=2,\ldots,n$, define
\begin{equation*}
\setlength\arraycolsep{0.11em}
 \begin{array}{lcccccccccccccccr}
\vspace{0.1cm}
\tilde{q}=\big(&Q_i&,& q_{i+1}&, &\ldots&,&q_{j-1}&,&\cCQ_j&\big),\\
\vspace{0.1cm}
\tilde{r}=\big(&\rho&,& 0 &,&\ldots&,& 0 &,&0&\big),\\
\vspace{0.1cm}
\tilde{s}=\big(&0&,& 0 &,&\ldots&,& 0 &,&\sigma&\big),\\
\vspace{0.1cm}
\tilde{p}=\big(&P_i&,& p_{i+1}&, &\ldots&,&p_{j-1}&,&\cCP_j&\big),
\end{array}
\end{equation*}
where $\tilde{q}$ and $\tilde{p}$ are distributions in the probability simplex of $j-i+1$ dimensions,
and $\tilde{r}$ and $\tilde{s}$ are tuples of the same dimension that represent a forgery strategy and a suppression strategy, respectively.
Particularly, note that the indexes $i=1$ and $j=n$ lead to $\tilde{q}=q$ and~$\tilde{p}=p$.

%%%%%%%%%%%%%%%%%%%%%%%%%%%%%%%%%%%%%%%%%%%%%%%%%%%%%%%%%%%%%%%%%%%%%%%%%%%%%%%%%%%%%%%%%%%%%%%%%%%%%%%%%%%%%%%%%%%%%%%%%%%%%%%%%%
%CLOSED-FORM SOLUTION THEOREM
%%%%%%%%%%%%%%%%%%%%%%%%%%%%%%%%%%%%%%%%%%%%%%%%%%%%%%%%%%%%%%%%%%%%%%%%%%%%%%%%%%%%%%%%%%%%%%%%%%%%%%%%%%%%%%%%%%%%%%%%%%%%%%%%%%
\begin{theorem}\label{Theorem:ClosedFormSolution}
Let $\partial{\cC}$ be the boundary of $\cC$, and $\oClo\bar{\cC}$ the closure of $\bar{\cC}$.
\begin{EnumerateRoman}
\item  $\partial \cC \subset \cC$ and
$$\partial \cC = \{(\rho,\sigma) \colon \rho = \rho_j(\sigma), \sigma\in [\sigma_{j}, \sigma_{j-1}], \textnormal{ for } j=2,\ldots,n\}.$$
\item For any $(\rho,\sigma) \in \oClo\bar{\cC}$,
either $\rho \in [\rho_i,\rho_{i+1}]$ for $i=1$ or $\rho \in (\rho_i,\rho_{i+1}]$ for some $i=2,\ldots,j-1$,
and either $\sigma \in [\sigma_j,\sigma_{j-1}]$ for $j=n$ or $\sigma\in (\sigma_j,\sigma_{j-1}]$ for some $j=2,\ldots,n-1$.
Then, for the corresponding indexes $i,j$, the optimal forgery and suppression strategies are
%\vspace{0.10mm}
\begin{equation*}
\begin{array}{ll}
r^*_k & =\left\{ \begin{array}{l@{,\quad}l}
                                           \frac{p_k}{P_i}(Q_i + \rho) - q_k \,\, & k=1,\dots,i \\
                                            0 \,\, & k=i+1,\dots,n
               \end{array}\right.,\\
\rule{0pt}{5ex}
s^*_k & =\left\{ \begin{array}{l@{,\quad}l}
                                            0 & k=1,\dots,j-1\\
                                            q_k - \frac{p_k}{\cCP_j} (\cCQ_j - \sigma) & k=j,\dots,n
                \end{array}\right.,
\end{array}
\end{equation*}
and the corresponding, minimum KL divergence yields the privacy\hyph forgery\hyph suppression function
%\vspace{0.24mm}
$$\cR(\rho,\sigma) = \oD\left(\left.\frac{\tilde{q} + \tilde{r} - \tilde{s}}{1+\rho-\sigma}\right\|\tilde{p}\right).$$
\end{EnumerateRoman}
\end{theorem}
\SpaceAfterPropositionEndedWithFormula
\BeginProof
The proof is structured as follows.
We begin by showing that the optimization problem~\eqref{eqn:PrivacyUtilityFunction} may be construed as a particular case of that stated in Lemma~\ref{Lemma:ResourceAllocation}.
Accordingly, we apply this lemma, namely the cases~(ii) and~(iii), to obtain the optimal forgery and suppression strategies.
The application of the former case allows us to derive the solution for $(\rho,\sigma) \in \bar{\cC}$.
The latter case enables us, first, to confirm that this solution is also valid on~$\partial{\bar{\cC}}$,
and secondly, to prove statement~(i).
Lastly, we complete the proof of~(ii) by expressing function~\eqref{eqn:PrivacyUtilityFunction} in terms of the optimal apparent distribution.

Use the definition of KL divergence to write the objective function of the optimization problem as
$\oD(t\,\|\,p)=\sum_k t_k \log \frac{t_k}{p_k}$, with $t=\frac{q+r-s}{1+\rho - \sigma}$.
Observe that the functions $f_k(r_k,s_k)=t_k \log \frac{t_k}{p_k}$ are twice differentiable on $\{(r_k,s_k)\colon q_k + r_k - s_k>0\}$.
Denote by $h_k$ the derivative of $f_k$ with respect to $r_k$,
\begin{equation}\label{eqn:FunctionH}
h_k(r_k,s_k)=\frac{1}{1+\rho-\sigma}\left(\log \frac{q_k + r_k - s_k}{(1+\rho-\sigma)p_k} + 1 \right).
\end{equation}
Then, note that the functions $f_k$ and $h_k$ satisfy the assumptions of Lemma~\ref{Lemma:ResourceAllocation},
and that the inequality and equality constraints of function~\eqref{eqn:PrivacyUtilityFunction} coincide with those in the lemma.
This exposes the structure of the optimization problem as a special case of the resource allocation lemma.

Before proceeding any further,
notice from~\eqref{eqn:FunctionH} that $h_k(r_k,0)$ is a strictly increasing function of $r_k$ and hence invertible.
Note also that, according to the lemma, the solutions are completely determined by the inverse of this function,
which is denoted by $h_k^{-1}$ and yields
$$h_k^{-1}(\phi) = p_k (1 + \rho - \sigma) 2^{(1 + \rho - \sigma)\phi - 1} - q_k.$$
Finally, observe that the assumption $h_1(0,0)\leqslant \cdots \leqslant h_n(0,0)$ in the lemma
is equivalent to the labeling assumption~\eqref{eqn:LabelingAssumption},
as $h_k(0,0)$ is a strictly increasing function of $\frac{q_k}{p_k}$.

%%%%%%%%%%%%%%%%% case \psi < \omega
Next we apply Lemma~\ref{Lemma:ResourceAllocation}~(ii), where it is assumed the condition~$\psi<\omega$.
We start with case~(ii)~(a).
On account of part~(i) of the lemma, the optimal forgery strategy must satisfy
\begin{equation*}
\rho=\sum_{k=1}^i h_k^{-1}(\psi) = P_i (1 + \rho - \sigma) 2^{(1 + \rho - \sigma)\psi - 1} - Q_i,
\end{equation*}
or equivalently,
\begin{equation*}
\psi = \frac{1}{1 + \rho - \sigma}\left(\log \frac{Q_i + \rho}{(1 + \rho - \sigma)P_i } + 1 \right).
\end{equation*}
Analogously for the suppression strategy,
\begin{equation*}
\sigma = - \sum_{k=j}^n h_k^{-1}(\omega) = \cCQ_j - \cCP_j (1 + \rho - \sigma) 2^{(1 + \rho - \sigma)\omega - 1},
\end{equation*}
and therefore
\begin{equation*}
\omega = \frac{1}{1 + \rho - \sigma}\left(\log \frac{\cCQ_j - \sigma}{(1 + \rho - \sigma)\cCP_j} + 1 \right).
\end{equation*}
Then it suffices to substitute the expressions of $\psi$ and $\omega$ into the function $h_k^{-1}$, % it suffices to
to obtain the nonzero optimal solutions claimed in assertion (ii) of the theorem.

Now we proceed to confirm the interval of values of $\rho$ and $\sigma$ where these solutions are defined.
In the case under study, $\psi$ and $\omega$ satisfy
$h_i(0,0) < \psi \leqslant h_{i+1}(0,0)$ for some $i=1,\ldots,j-1$ and $h_{j-1}(0,0) \leqslant \omega < h_j(0,0)$ for some $j=2,\ldots,n$.
We split the discussion into two cases, namely $i<j-1$ and $i=j-1$.

Assume the former case. Observe that the condition $h_i(0,0) < \psi$ is equivalent to
\begin{equation*}
\frac{1}{1+\rho-\sigma}\left(\log \frac{q_i }{(1+\rho-\sigma)p_i} + 1 \right) < \frac{1}{1 + \rho - \sigma}\left(\log \frac{Q_i + \rho}{(1 + \rho - \sigma)P_i } + 1 \right)
\end{equation*}
and finally, after routine algebraic manipulation, to
$$\rho>P_i \frac{q_i}{p_i} - Q_i.$$
Similarly, the upper\hyph bound condition $\psi \leqslant h_{i+1}(0,0)$ leads to
$$\rho\leqslant P_i \frac{q_{i+1}}{p_{i+1}} - Q_i.$$
Hence, the intervals resulting from imposing $h_i(0,0) < \psi \leqslant h_{i+1}(0,0)$ are of the form $(\rho_i,\rho_{i+1}]$.
The monotonicity of the thresholds $\rho_i$, demonstrated in Proposition~\ref{Proposition:Monotonicity}, guarantees that these intervals are contiguous and nonoverlapping.
In an analogous manner, it can be shown that the condition $h_{j-1}(0,0) \leqslant \omega < h_j(0,0)$ leads to intervals of the form $(\sigma_j,\sigma_{j-1}]$,
also contiguous and nonoverlapping by virtue of Proposition~\ref{Proposition:Monotonicity}.

Now assume the latter case, where $h_{i}(0,0) < \psi < \omega < h_j(0,0)$ with $i=j-1$.
On the one hand, the assumption $h_{j-1}(0,0) < \psi$ is, as shown above, equivalent to the condition $\rho > \rho_{j-1}$.
On the other hand, straightforward manipulation allows us to write the inequality $\psi < \omega$ as
$$\rho < \frac{P_{j-1}}{\cCP_j}(\cCQ_j - \sigma) - Q_{j-1}.$$
Combining these two bounds on $\psi$, we obtain the interval $\left(\rho_{j-1},\rhocrit(\sigma)\right)$.
With this last interval, we complete the range of validity of the solution for the case~(ii)~(a) in the lemma.
Ultimately, it is easy to verify that, in those intervals of $\rho$ and $\sigma$,
the optimal apparent profile $t=\frac{q+r-s}{1+\rho-\sigma}$ does not coincide with the population's profile $p$.
In consequence, $\oD(t\,\|\,p)>0$. %the privacy risk function does not vanish.

Next, we turn to case (ii)~(b) of the lemma.
Here, the assumption $h_n(0,0)\leqslant\omega$ leads to $\sigma=0$, or equivalently, to the solution $s=0$. %$s_k =0$ for $k=1,\ldots,n$. % all $k$.
Note that, precisely, this is the solution given in the theorem for $\sigma = \sigma_j$ with $j=n$.
On the other hand, the application of the condition $\sum_{k=1}^i r_k =\rho$ results in the same optimal forgery strategy obtained in case (ii)~(a).
Proceeding analogously as in this case, from the assumptions on $\psi$ we derive the intervals of values of $\rho$ where the solution is defined:
$(\rho_i,\rho_{i+1}]$ for $i=1,\ldots,n-1$ and $(\rho_i,\rho_{i+1})$ for $i=n$.
Given these intervals, it is then straightforward to check that $\cR(\rho,0) = 0$ if, and only if, $\rho \geqslant \rho_n$.
This provides us with the pairs $(\rho,0)$ that belong to~$\oClo\bar{\cC}$.

% Case (ii) c)
In case (ii)~(c), the condition $\psi \leqslant h_1(0,0)$ means that $\rho=0$, or equivalently, $r=0$. %$r_k =0$ for $k=1,\ldots,n$.
Observe that this is the solution stated in the theorem for $\rho = \rho_i$ with $i=1$.
Then again, the condition $\sum_{k=j}^n s_k =\sigma$ leads to the same optimal suppression strategy found in case (ii)~(a).
From the assumptions in the lemma on $\omega$, we obtain the intervals $(\sigma_j,\sigma_{j-1}]$ for $j=2,\ldots,n$ and $(\sigma_j,\sigma_{j-1})$ for $j=1$.
Then, we verify that $\cR(0,\sigma) = 0$ if, and only if, $\sigma \geqslant \sigma_1$,
from which it follows the pairs $(0,\sigma)$ that belong to~$\oClo\bar{\cC}$.

% Case (ii) d)
Finally, the case (ii)~(d) in the lemma, in which $h_n(0,0)\leqslant \omega$ and $\psi \leqslant h_1(0,0)$,
corresponds to the trivial case $\sigma=\sigma_j$ for $j=n$ and $\rho=\rho_i$ for $i=1$, that is, the solution $r=s=0$.

% Summary
After having applied Lemma~\ref{Lemma:ResourceAllocation}~(ii) to function~\eqref{eqn:PrivacyUtilityFunction},
now we proceed with case (iii)~(a).
In applying it, we shall show that the solution claimed in the theorem is also valid for the extreme values of the intervals in case (ii)~(a),
specifically the set
\begin{equation*}
\{(\rho,\sigma)\colon \rho=\rhocrit(\sigma), \sigma\in (\sigma_{j},\sigma_{j-1}] \textnormal{ for } j=3,\ldots,n, \textnormal{ and } \sigma \in (\sigma_j,\sigma_{j-1}) \textnormal{ for } j=2\}.
\end{equation*}

Assume the case (iii)~(a) in which $h_i(0,0) < \psi = \omega < h_j(0,0)$ for some $j=2,\ldots,n$ and $i=j-1$.
Under this assumption, the equality constraint $\sum_{k=1}^i r_k = \rho$ in the lemma is equivalent,
after simple algebraic manipulation, to
\begin{equation}\label{eqn:psi}
\psi = \frac{1}{1 + \rho - \sigma}\left(\log \frac{Q_{j-1} + \rho - \zeta}{(1 + \rho - \sigma)P_{j-1} } + 1 \right),
\end{equation}
where we define $\zeta=\sum_{k=1}^n \alpha_k$.
Similarly, the equality constraint $\sum_{k=j}^n s_k = \sigma$ becomes
$$\omega = \frac{1}{1 + \rho - \sigma}\left(\log \frac{\cCQ_j - \sigma + \zeta}{(1 + \rho - \sigma)\cCP_j} + 1 \right).$$
But $\psi=\omega$, therefore
$$\frac{Q_{j-1} + \rho - \zeta}{P_{j-1}} = \frac{\cCQ_j - \sigma + \zeta}{\cCP_j},$$
or equivalently,
$$\rho=\rhocrit(\sigma) + \frac{\zeta}{\cCP_j}.$$
In short, the assumption $\psi=\omega$ imposes the condition $(\rho,\sigma) \succeq (\rhocrit(\sigma),\sigma)$
for some nonnegative sequence $\alpha_1,\ldots,\alpha_n$ satisfying the above equality.
Next we examine, for a given $\sigma$, these two possibilities, $\rho=\rhocrit(\sigma)$ and $\rho>\rhocrit(\sigma)$.

Consider the former possibility and observe that $\rho=\rhocrit(\sigma)$ if, and only if, $\alpha_k=0$ for $k=1,\ldots,n$.
According to the lemma, the nonzero optimal solutions yield
\begin{equation*}
\begin{aligned}
r_k = h_k^{-1}(\psi) &= p_k \frac{Q_{j-1} + \rhocrit(\sigma)}{P_{j-1}} - q_k\\
                                           &= p_k (1 + \rhocrit(\sigma) - \sigma) - q_k
\end{aligned}
\end{equation*}
for $k=1,\ldots, j-1$, and
\begin{equation*}
s_k = - h_k^{-1}(\psi) = q_k - p_k (1 + \rhocrit(\sigma) - \sigma)
\end{equation*}
for $k=j,\ldots,n$, that is,
the solutions obtained after applying case (ii)~(a), but evaluated at $\rho=\rhocrit(\sigma)$.
From these expression for $r$ and $s$, it is immediate to verify then that $t=p$ and thus $\cR(\rho,\sigma)=0$.

%%%%%%%%%%%%%%%%%%%%%%%%%%%%%%%%%%%%%%%%%%%%%%%%%%%%%%%%%%%%%%%%%%%%%%%%%%%%%%%%%%%%%%%%%%%%
%%%%%%%%%%%%%%%%%%%%%%%%%%%%%%%%%%%%%%% FIGURE - THEOREM CONCEPTUAL %%%%%%%%%%%%%%%%%%%%%%%%%%%%%%%%
%%%%%%%%%%%%%%%%%%%%%%%%%%%%%%%%%%%%%%%%%%%%%%%%%%%%%%%%%%%%%%%%%%%%%%%%%%%%%%%%%%%%%%%%%%%%
\begin{figure*}
\centering
\includegraphics[width=\columnwidth]{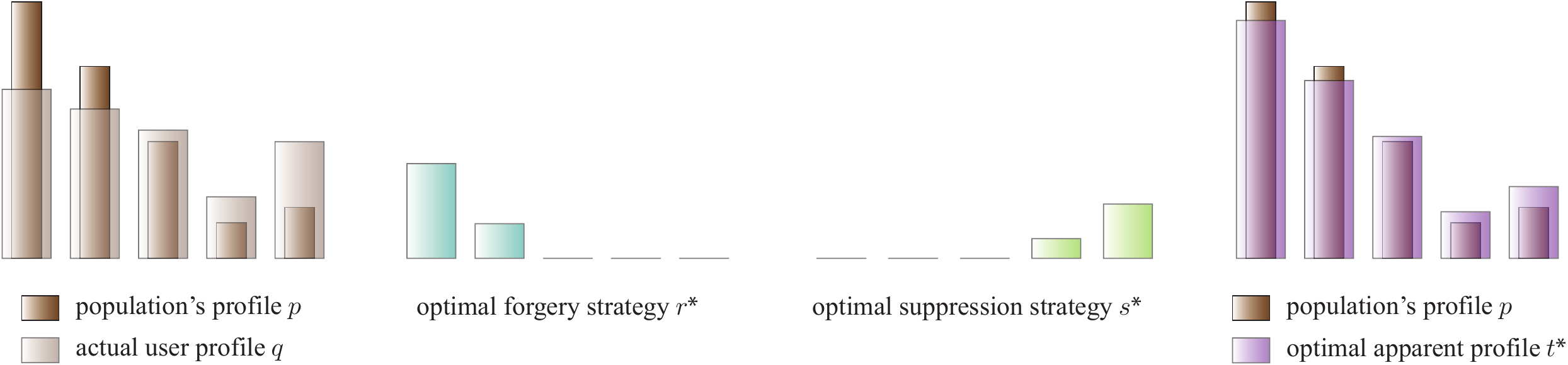}
\caption{A user's item distribution is perturbed according to two optimal forgery and suppression strategies, in order for the resulting profile to minimize the KL divergence
with respect to the population's distribution.}
\label{fig:theorem_conceptual}
\end{figure*}
%%%%%%%%%%%%%%%%%%%%%%%%%%%%%%%%%%%%%%%%%%%%%%%%%%%%%%%%%%%%%%%%%%%%%%%%%%%%%%%%%%%%%%%%%%%%
%%%%%%%%%%%%%%%%%%%%%%%%%%%%%%%%%%%%%%%%%%%%%%%%%%%%%%%%%%%%%%%%%%%%%%%%%%%%%%%%%%%%%%%%%%%%
%%%%%%%%%%%%%%%%%%%%%%%%%%%%%%%%%%%%%%%%%%%%%%%%%%%%%%%%%%%%%%%%%%%%%%%%%%%%%%%%%%%%%%%%%%%%

Now we assume the latter possibility, i.e., $(\rho,\sigma)\succ (\rhocrit(\sigma),\sigma)$, to show that the privacy\hyph risk function also vanishes for these values of $\rho$ and~$\sigma$.
On account of part (iii)~(a) of the lemma and~\eqref{eqn:psi}, we derive the optimal forgery and suppression strategies
$$r_k = p_k (1+\rhocrit(\sigma) - \sigma) + \frac{p_k\, \zeta}{\cCP_j} - q_k + \alpha_k$$
and $s_k=\alpha_k$ for $k=1,\ldots,j-1$, and
$$s_k = q_k - p_k (1+\rhocrit(\sigma) - \sigma) - \frac{p_k\, \zeta}{\cCP_j} + \alpha_k$$
and $r_k=\alpha_k$ for $k=j,\ldots,n$.
Then, we substitute $r$ and $s$ back into the apparent profile $t$ and check that $\oD(t\,\|\,p)=0$.
In doing so, we determine the pairs $(\rho,\sigma) \succ 0$ that belong to~$\oClo\bar{\cC}$, and
finally obtain the expression for the boundary of the critical\hyph privacy region claimed in statement~(i) of the theorem. %

To conclude the proof, it remains only to write the privacy\hyph risk function $\cR(\rho,\sigma)=\sum_{k=1}^n t_k \log \frac{t_k}{p_k}$ in terms of the optimal apparent distribution.
With this aim, we split the summation into three parts.
The first part, corresponding to $t_k=\frac{p_k (Q_i + \rho)}{P_i (1+\rho-\sigma)}$, is
$$\sum_{k=1}^i t_k \log \frac{t_k}{p_k} = \frac{Q_i + \rho}{1 + \rho - \sigma} \log \frac{Q_i + \rho}{(1 + \rho - \sigma)P_i},$$
where we leverage on the fact that $\frac{t_k}{p_k}$ does not depend on $k$.
The second part of the sum, corresponding to $t_k = \frac{q_k}{1 +\rho - \sigma}$, yields
$$\sum_{k=i+1}^{j-1} t_k \log \frac{t_k}{p_k}=\sum_{k=i+1}^{j-1}\frac{q_k}{1 +\rho - \sigma} \log \frac{q_k}{(1 +\rho - \sigma)p_k}.$$
The last part, corresponding to $t_k=\frac{p_k (\cCQ_j - \sigma)}{\cCP_j (1+\rho-\sigma)}$, is
$$\sum_{k=j}^n t_k \log \frac{t_k}{p_k}=\frac{\cCQ_j - \sigma}{1 + \rho - \sigma} \log \frac{\cCQ_j - \sigma}{(1 + \rho - \sigma)\cCP_j},$$
where we also note that $\frac{t_k}{p_k}$ does not depend on $k$ either.
Now, it is straightforward to identify the terms of $\cR(\rho,\sigma)$ as the KL divergence between the distributions
\begin{equation*}
\left(\frac{Q_i+\rho}{1+\rho-\sigma}, \frac{q_{i+1}}{1+\rho-\sigma},\ldots,\frac{q_{j-1}}{1+\rho-\sigma},\frac{\cCQ_j - \sigma}{1+\rho-\sigma}\right)
\end{equation*}
and
\begin{equation*}
\left(P_i,p_{i+1},\ldots,p_{j-1},\cCP_j\right),
\end{equation*}
precisely the distributions stated in the theorem.
\EndProof

In light of Theorem~\ref{Theorem:ClosedFormSolution},
we would like to remark the intuitive principle that both the optimal forgery and suppression strategies follow.
On the one hand, the forgery strategy suggests adding ratings to those categories with a low ratio $\tfrac{q_k}{p_k}$,
that is, to those in which the user's interest is considerably lower than the population's.
On the other hand, the suppression strategy recommends eliminating ratings from those categories where the ratio $\tfrac{q_k}{p_k}$ is high,
i.e., where the interest of the user exceeds that of the population.

Another straightforward consequence of Theorem~\ref{Theorem:ClosedFormSolution} is the role of the forgery and the suppression thresholds.
In particular, we identify $\rho_i$ as the forgery rate beyond which the components of $r_k$ for $k=1,\ldots,i$ become positive.
A similar reasoning applies to $\sigma_j$, which indicates the suppression rate beyond which the components of $s_k$ for $k=j,\ldots,n$ are positive. %reasoning
In a nutshell, these thresholds determine the number of nonzero components of the optimal strategies.

Also, from this theorem we deduce that the perturbation of the user profile does not \emph{only} affect those categories where either $r_k>0$ or $s_k>0$.
In fact, since we are dealing with relative frequencies,
the components of the apparent distribution $t_k$ belonging to the categories $k=i+1,\ldots, j-1$ are normalized by $\tfrac{1}{1+\rho - \sigma}$.
Fig.~\ref{fig:theorem_conceptual} illustrates these three conclusions by means of a simple example with $n=5$ categories of interest.

In this example we consider a user who is disposed to submit a percentage of false ratings $\rho \in (\rho_2, \rho_3]$,
and to refrain from sending a fraction of genuine ratings $\sigma \in (\sigma_4, \sigma_3]$.
Given these rates, the optimal forgery strategy recommends that the user forge ratings belonging to the categories 1 and 2,
where clearly there is a lack of interest, compared to the reference distribution.
On the contrary, the suppression strategy specifies that the user eliminate ratings from the categories 4 and 5,
that is, from those categories where they show too much interest, again compared to the population's profile.
In adopting these two strategies, the apparent user profile approaches the population's distribution,
especially in those components where the ratio $\tfrac{q_k}{p_k}$ deviates significantly from 1.
Finally, the component of the apparent profile $t_3$, which is not directly affected by the forgery and the suppression strategies,
gets closer to $p_3$ as a result of the aforementioned normalization.

In the following subsections, we shall analyze a number of important consequences of Theorem~\ref{Theorem:ClosedFormSolution}.

%%%%%%%%%%%%%%%%%%%%%%%%%%%%%%%%%%%%%%%%%%%%%%%%%%%%%%%%%%%%%%%%%%%%%%%%%%%%%%%%%%%%%%%%%%%%%%%%%%%%%%%%%%%%%%%%%%%%%%%%%%%%%%%%%%
%%%%%%%%%%%%%%%%%%%%%%%%%%%%%%%%%%%%%%%%%%%%%%%%%%%%%%%%%%%%%%%%%%%%%%%%%%%%%%%%%%%%%%%%%%%%%%%%%%%%%%%%%%%%%%%%%%%%%%%%%%%%%%%%%%
%4.1 Orthogonality, Continuity and Proportionality
%%%%%%%%%%%%%%%%%%%%%%%%%%%%%%%%%%%%%%%%%%%%%%%%%%%%%%%%%%%%%%%%%%%%%%%%%%%%%%%%%%%%%%%%%%%%%%%%%%%%%%%%%%%%%%%%%%%%%%%%%%%%%%%%%%
%%%%%%%%%%%%%%%%%%%%%%%%%%%%%%%%%%%%%%%%%%%%%%%%%%%%%%%%%%%%%%%%%%%%%%%%%%%%%%%%%%%%%%%%%%%%%%%%%%%%%%%%%%%%%%%%%%%%%%%%%%%%%%%%%%
\subsection{Orthogonality, Continuity and Proportionality}
\label{sec:Theory:OrthContProp}
\noindent
In this subsection we study some interesting properties of the closed\hyph form solution obtained in Sec.~\ref{sec:Theory:ClosedFormSolution}.
Specifically, we investigate the orthogonality and continuity of the optimal forgery and suppression strategies,
and then establish a proportionality relationship between the optimal apparent user profile and the population's distribution.
%%%%%%%%%%%%%%%%%%%%%%%%%%%%%%%%%%%%%%%%%%%%%%%%%%%%%%%%%%%%%%%%%%%%%%%%%%%%%%%%%%%%%%%%%%%%%%%%%%%%%%%%%%%%%%%%%%%%%%%%%%%%%%%%%%%
%%COROLLARY: ORTHOGONALITY AND CONTINUITY
%%%%%%%%%%%%%%%%%%%%%%%%%%%%%%%%%%%%%%%%%%%%%%%%%%%%%%%%%%%%%%%%%%%%%%%%%%%%%%%%%%%%%%%%%%%%%%%%%%%%%%%%%%%%%%%%%%%%%%%%%%%%%%%%%%%
\begin{corollary}[Orthogonality and Continuity]
\label{Corollary:OrthoCont}
\leavevmode
\begin{EnumerateRoman}
\item For any $(\rho,\sigma) \in \oClo\bar{\cC}$, the optimal forgery and suppression strategies satisfy $r_k^*\,s_k^* =0$ for $k=1,\ldots,n$.
\item The components of $r^*$ and $s^*$, interpreted as functions of $\rho$ and $\sigma$ respectively, are continuous on $\oClo\bar{\cC}$.
\end{EnumerateRoman}
\end{corollary}

\BeginProof
The proof of~(i) is trivial from Theorem~\ref{Theorem:ClosedFormSolution}.
To prove statement~(ii) we also resort to this theorem.
According to it, each component $r_k^*$ may be regarded as a piecewise function of $\rho$ defined on the contiguous, nonoverlapping intervals
$[\rho_i, \rho_{i+1}]$ for $i=1$ and $(\rho_i,\rho_{i+1}]$ for $i=2,\ldots,j-1$.
A direct verification shows that, for any $k=j,\ldots,n$, the component $r_k^*$ is identically zero on the whole interval $[\rho_1,\rho_j]$ and hence continuous.
For any $k=1,\ldots,j-1$, we immediately check the continuity of $r_k^*$ on the interior of each of the intervals parameterized by $i$.
Now we examine the endpoints of such intervals.
The continuity at the extreme points $\rho_1$ and $\rho_j$ is verified straightforwardly as the intervals are closed at these points.
Then, we check that the limit at the remaining endpoints $\rho_i$ exists, since
\begin{align*}
\lim \limits_{\rho \rightarrow \rho_i^{-}}  r_k^*(\rho) & =  \frac{p_k}{P_{i-1}}(Q_{i-1} + \rho_i) - q_k \\
                           & = \frac{p_k}{P_{i}}(Q_{i} + \rho_i) - q_k  = \lim \limits_{\rho \rightarrow \rho_i^{+}} r_k^*(\rho),
\end{align*}
for $i=2,\ldots,j-1$.
Because each limit coincides with the corresponding value $r_k^*(\rho_i)$, we prove the continuity of the components $r_1,\ldots,r_{j-1}$.
The proof of the continuity of the components of $s^*$ is analogous to that of $r^*$.
\EndProof

The orthogonality of the optimal forgery and suppression strategies, in the sense indicated by Corollary~\ref{Corollary:OrthoCont}~(i),
conforms to intuition---it would not make any sense to
submit false ratings to items of a particular category and, at the same time, eliminate genuine ratings from this category.
This intuitive result is illustrated in Fig.~\ref{fig:theorem_conceptual}.
The second part of Corollary~\ref{Corollary:OrthoCont} is applied to show our next result, Proposition~\ref{Proposition:Proportionality}.

%%%%%%%%%%%%%%%%%%%%%%%%%%%%%%%%%%%%%%%%%%%%%%%%%%%%%%%%%%%%%%%%%%%%%%%%%%%%%%%%%%%%%%%%%%%%%%%%%%%%%%%%%%%%%%%%%%%%%%%%%%%%%%%%%%%
%%PROPOSITION: PROPORTIONALITY
%%%%%%%%%%%%%%%%%%%%%%%%%%%%%%%%%%%%%%%%%%%%%%%%%%%%%%%%%%%%%%%%%%%%%%%%%%%%%%%%%%%%%%%%%%%%%%%%%%%%%%%%%%%%%%%%%%%%%%%%%%%%%%%%%%%
\begin{proposition}[Proportionality] % of Apparent's and Population's Distributions
\label{Proposition:Proportionality}
Define the piecewise functions $\phi(\rho,\sigma) = \frac{Q_i + \rho}{(1 + \rho - \sigma) P_i}$ and $\chi(\rho,\sigma) = \frac{\cCQ_j - \sigma}{(1 + \rho - \sigma) \cCP_j}$
on the intervals $[\sigma_j,\sigma_{j-1}]$ for $j=2,\ldots,n$ and $[\rho_i,\rho_{i+1}]$ for $i=1,\ldots,j-1$.
\begin{EnumerateRoman}
%%%%%%%%%%%%%%%%% FIRST STATEMENT
\item For any $j=2,\ldots,n$ and $i=1,\ldots,j-1$, and for any $\sigma\in [\sigma_j,\sigma_{j-1}]$ and $\rho \in [\rho_i,\rho_{i+1}]$,
the optimal apparent profile $t^*$ and the population's distribution $p$ satisfy
\begin{equation*}
\frac{t^*_1}{p_1} = \cdots = \frac{t^*_i}{p_i} = \phi(\rho,\sigma),
\end{equation*}
\begin{equation*}
\frac{t^*_j}{p_j} = \cdots = \frac{t^*_n}{p_n} = \chi(\rho,\sigma),
\end{equation*}
and
\begin{equation*}
\phi(\rho,\sigma)\leqslant \frac{t_{i+1}^*}{p_{i+1}}\leqslant \cdots\leqslant \frac{t_{j-1}^*}{p_{j-1}} \leqslant\chi(\rho,\sigma).
\end{equation*}
%%%%%%%%%%%%%%%%% SECOND STATEMENT
\item %For any fixed $\sigma$,
The function $\phi$ is continuous and strictly \emph{increasing} in each of its arguments, and satisfies $\phi(\rho,\sigma) \leqslant 1$, %on its domain,
with equality if, and only if, $(\rho,\sigma)=(\rho_j(\sigma),\sigma)$.
%%%%%%%%%%%%%%%%% THIRD STATEMENT
\item %For any fixed $\rho$,
The function $\chi$ is continuous and strictly \emph{decreasing} in each of its arguments, and satisfies $\chi(\rho,\sigma) \geqslant 1$, % on its domain,
with equality if, and only if, $(\rho,\sigma)=(\rho_j(\sigma),\sigma)$.
\end{EnumerateRoman}
\end{proposition}
\BeginProof
The continuity of the components of $t^*$ on $\oClo\bar{\cC}$ follows from Corollary~\ref{Corollary:OrthoCont}~(ii).
This allows us to write the intervals in Theorem~\ref{Theorem:ClosedFormSolution}
as $[\rho_i,\rho_{i+1}]$ and $[\sigma_j,\sigma_{j-1}]$, in lieu of $(\rho_i,\rho_{i+1}]$ and $(\sigma_j,\sigma_{j-1}]$, respectively.
From the expressions of $r_k^*$ and $s_k^*$ in the theorem, it is immediate to identify the ratios $\frac{t^*_k}{p_k}$ as either $\phi(\rho,\sigma)$ or $\chi(\rho,\sigma)$.
The inner inequalities in statement~(i) of this proposition also follow immediately from the labeling assumption~\eqref{eqn:LabelingAssumption}.
Direct manipulation shows that the outer inequalities $\frac{t^*_i}{p_i} \leqslant \frac{t_{i+1}^*}{p_{i+1}}$ and $\frac{t_{j-1}^*}{p_{j-1}} \leqslant \frac{t^*_j}{p_j}$
are equivalent to $\rho \leqslant \rho_{i+1}$ and $\sigma \leqslant \sigma_{j-1}$, respectively.
This proves~(i).

Next, we proceed to demonstrate the strict monotonicity of~$\phi$.
A simple calculation shows that
$$\frac{\partial \phi}{\partial \rho} = \frac{\cCQ_{i+1} - \sigma}{(1 + \rho - \sigma)^2 P_i}.$$
To prove that $\frac{\partial \phi}{\partial \rho}>0$,
it is sufficient to verify that $\cCQ_j > \sigma_{j-1}$,
or equivalently, that $\cCP_j \frac{q_{j-1}}{p_{j-1}}>0.$
Then, by the positivity assumption~\eqref{eqn:PositivityAssumption},
we immediately see that this latter inequality holds for any $j=2,\ldots,n$.
The strict monotonicity of $\phi$ in $\sigma$ also follows from assumption~\eqref{eqn:PositivityAssumption}.

To complete~(ii),
we write the condition $\phi(\rho,\sigma) \leqslant 1$ as
$$\rho \leqslant \frac{(1 - \sigma) P_i - Q_i}{\cCP_{i+1}}.$$
A routine computation shows that the equality holds for $\rho_j(\sigma)$
and any $\sigma\in [\sigma_j,\sigma_{j-1}]$ with $j=2,\ldots,n$.
Therefore,
for any fixed $\sigma$, the inequality holds strictly for any other $\rho$.
The converse, that is, $\phi(\rho,\sigma) =1$ implies $(\rho,\sigma)=(\rho_j(\sigma),\sigma)$, is
immediate from the strict monotonicity of $\phi$.
The proof of statement~(iii) proceeds along the same lines of that of~(ii) and is omitted.
\EndProof

%%%%%%%%%%%%%%%%%%%%%%%%%%%%%%%%% FIGURE - PROPORTIONALITY CONCEPTUAL %%%%%%%%%%%%%%%%%%%%%%%%%%%%%%%%
\begin{figure}
\centering
\includegraphics[scale= 0.80]{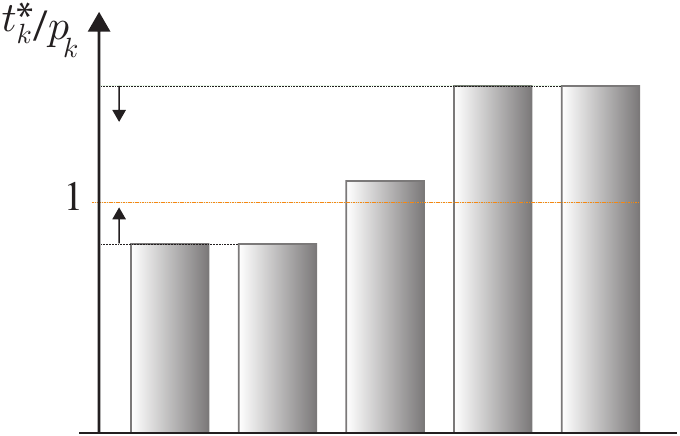}
\caption{Proportionality relationship between the optimal user's apparent item distribution and the population's profile.
In this figure we show the ratios $\frac{t_k^*}{p_k}$ of the example illustrated in Fig.~\ref{fig:theorem_conceptual},
where the number of categories is $n=5$, $\rho \in [\rho_2, \rho_3]$ and $\sigma \in [\sigma_4, \sigma_3]$.}
\label{fig:proportionality}
\end{figure}
%%%%%%%%%%%%%%%%%%%%%%%%%%%%%%%%%%%%%%%%%%%%%%%%%%%%%%%%%%%%%%%%%%%%%%%%%%%%%%%%%%%%%%%%%%%%

Our previous result tells us how perturbation operates.
According to Proposition~\ref{Proposition:Proportionality},
the optimal strategies perturb the user profile in such a manner that, in those categories with the lowest and highest ratios $\tfrac{q_k}{p_k}$,
the apparent profile becomes proportional to the population's distribution.
More precisely, the common ratio $\tfrac{t_k^*}{p_k}$ increases with both $\rho$ and $\sigma$ in those categories affected by forgery, that is, $k=1,\ldots,i$.
Exactly the opposite happens in those categories affected by suppression, where the common ratio $\tfrac{t_j^*}{p_j}$ decreases with both rates.
This tendency
continues until $\rho=\rhocrit(\sigma)$, at which point $t^*=p$.
Fig.~\ref{fig:proportionality} illustrates this proportionality property in the case of the example depicted in Fig.~\ref{fig:theorem_conceptual}.

%%%%%%%%%%%%%%%%%%%%%%%%%%%%%%%%%%%%%%%%%%%%%%%%%%%%%%%%%%%%%%%%%%%%%%%%%%%%%%%%%%%%%%%%%%%%%%%%%%%%%%%%%%%%%%%%%%%%%%%%%%%%%%%%%%
%%%%%%%%%%%%%%%%%%%%%%%%%%%%%%%%%%%%%%%%%%%%%%%%%%%%%%%%%%%%%%%%%%%%%%%%%%%%%%%%%%%%%%%%%%%%%%%%%%%%%%%%%%%%%%%%%%%%%%%%%%%%%%%%%%
%4.3 CRITICAL-PRIVACY REGION
%%%%%%%%%%%%%%%%%%%%%%%%%%%%%%%%%%%%%%%%%%%%%%%%%%%%%%%%%%%%%%%%%%%%%%%%%%%%%%%%%%%%%%%%%%%%%%%%%%%%%%%%%%%%%%%%%%%%%%%%%%%%%%%%%%
%%%%%%%%%%%%%%%%%%%%%%%%%%%%%%%%%%%%%%%%%%%%%%%%%%%%%%%%%%%%%%%%%%%%%%%%%%%%%%%%%%%%%%%%%%%%%%%%%%%%%%%%%%%%%%%%%%%%%%%%%%%%%%%%%%
\subsection{Critical\hyph Privacy Region}
\label{sec:Theory:CriticalRegion}
\noindent
One of the results of Theorem~\ref{Theorem:ClosedFormSolution} is that the boundary of the critical\hyph privacy region
is determined by the critical forgery\hyph suppression threshold~$\rho_j(\sigma)$, which we also denote by~$\rhocrit(\sigma)$
to highlight this fact.
The following proposition leverages on this result and characterizes said region.
In particular, Proposition~\ref{Proposition:CriticalRegion} first examines some properties of this threshold and then investigates the convexity of the critical\hyph privacy region.
\begin{proposition}[Convexity of the Critical\hyph Privacy Region]
\label{Proposition:CriticalRegion}
\leavevmode
\begin{EnumerateRoman}
\item $\rho_j$ is a convex, piecewise linear function of $\sigma\in [\sigma_{j}, \sigma_{j-1}]$ for $j=2,\ldots,n$.
\item $\cC$ is convex.
\end{EnumerateRoman}
\end{proposition}

\BeginProof
From Theorem~\ref{Theorem:ClosedFormSolution}, it is routine to check the continuity of $\rho_j$ on $[\sigma_n,\sigma_1]$.
To show its convexity, we conveniently write this function as $\rho_j(\sigma) = m_j\, \sigma + b_j$, where $m_j =- \frac{P_{j-1}}{\cCP_j}$
and $b_j= \frac{P_{j-1} - Q_{j-1}}{\cCP_j}$.
Next, we prove that the slopes satisfy $m_j < m_{j-1}$ for all $j=3,\ldots,n$.
We proceed by contradiction, assuming that $m_j \geqslant m_{j-1}$.
Note that this inequality is equivalent to $P_{j-1} \cCP_{j-1} \leqslant \cCP_j - \cCP_j \cCP_{j-1}$
and, after algebraic simplification, to $p_{j-1} \leqslant 0$.
This contradicts the positivity assumption~\eqref{eqn:PositivityAssumption}, which, in turn, implies that $m_j < 0$ for all $j=2,\ldots,n$.
Therefore, since $\rho_j$ is a piecewise linear function defined by the strictly increasing sequence of negative slopes $\{m_n,\ldots,m_2\}$,
we can conclude that $\rho_j$ is convex.
This proves statement~(i).
The second statement follows from the first one.
As $\rho_j$ is convex, so is its epigraph, i.e., the critical\hyph privacy region.
\EndProof

\begin{figure}
\centering
\includegraphics[scale=0.50]{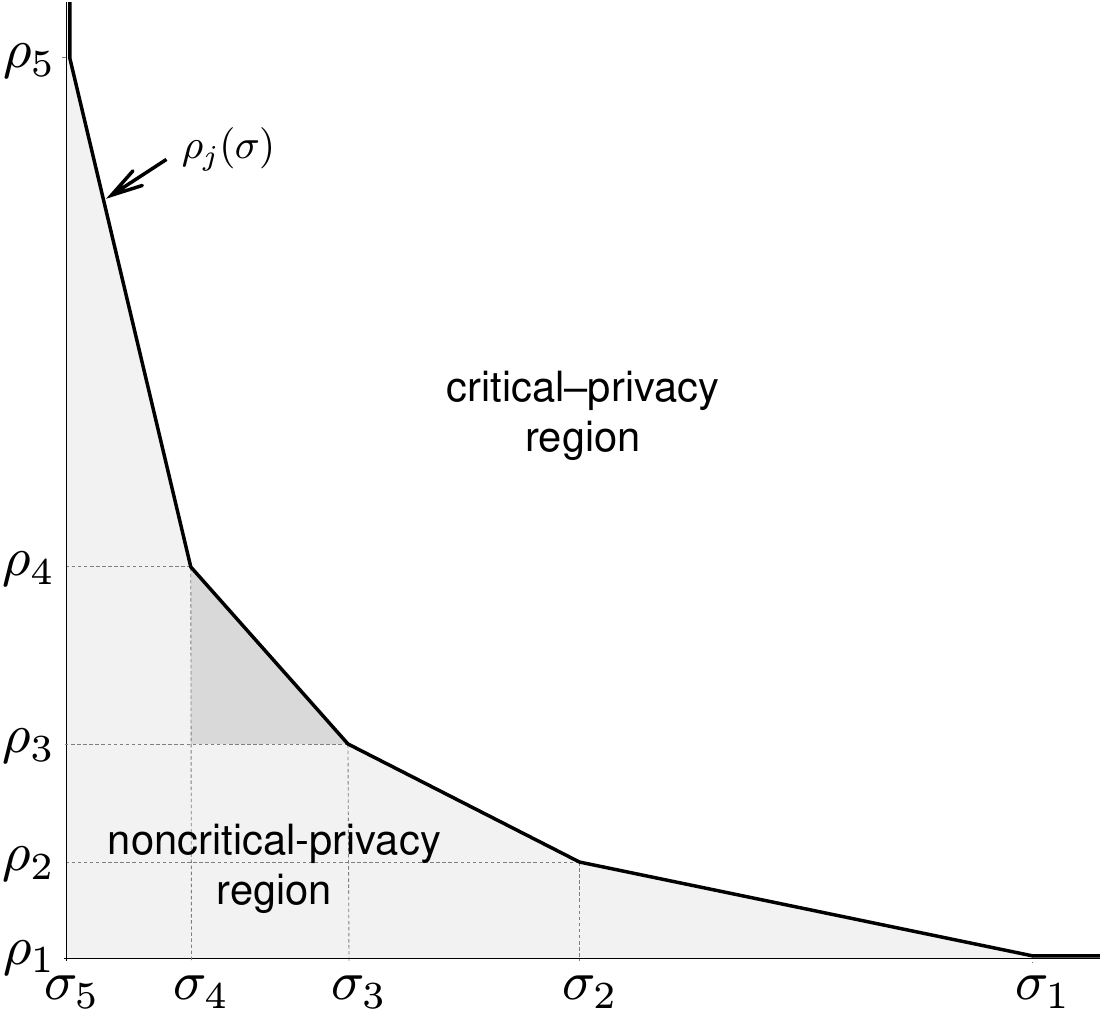}
\caption{Conceptual plot of the critical and noncritical privacy regions for $n=5$ categories.}
\label{fig:convexity}
\end{figure}
The conclusions drawn from Proposition~\ref{Proposition:CriticalRegion} are illustrated in Fig.~\ref{fig:convexity}.
In this figure we represent the critical and noncritical\hyph privacy regions for $n=5$ categories of interest;
the distributions $q$ and $p$ assumed in this conceptual example are different from those considered in Figs.~\ref{fig:theorem_conceptual} and~\ref{fig:proportionality}.
That said, the figure in question shows a straightforward consequence of our previous proposition---the noncritical\hyph privacy region is nonconvex.

In this illustrative example, the sequences of forgery thresholds $\{\rho_1\, \ldots,\rho_5\}$ and suppression thresholds $\{\sigma_5,\ldots, \sigma_1\}$ are strictly increasing.
By Proposition~\ref{Proposition:Monotonicity}, we can conclude then that the inequalities of the labeling assumption~\eqref{eqn:LabelingAssumption} hold strictly.
Related to these thresholds is also the number of nonzero components of the optimal strategies, as follows from Theorem~\ref{Theorem:ClosedFormSolution}.
Fig.~\ref{fig:convexity} shows the sets of pairs $(\rho,\sigma)$ where the number of nonzero components of $r^*$ and $s^*$ is fixed.
Thus, in the triangular area shown darker, corresponding to the Cartesian product of the intervals $[\rho_3,\rho_4]$ and $[\sigma_4,\sigma_3]$,
the solutions $r^*$ and $s^*$ have $i=3$ and $n-j+1=2$ nonzero components, respectively.

%%%%%%%%%%%%%%%%%%%%%%%%%%%%%%%%%%%%%%%%%%%%%%%%%%%%%%%%%%%%%%%%%%%%%%%%%%%%%%%%%%%%%%%%%%%%%%%%%%%%%%%%%%%%%%%%%%%%%%%%%%%%%%%%%%
%%%%%%%%%%%%%%%%%%%%%%%%%%%%%%%%%%%%%%%%%%%%%%%%%%%%%%%%%%%%%%%%%%%%%%%%%%%%%%%%%%%%%%%%%%%%%%%%%%%%%%%%%%%%%%%%%%%%%%%%%%%%%%%%%%
%4.3 CASE OF LOW FORGERY AND SUPPRESSION
%%%%%%%%%%%%%%%%%%%%%%%%%%%%%%%%%%%%%%%%%%%%%%%%%%%%%%%%%%%%%%%%%%%%%%%%%%%%%%%%%%%%%%%%%%%%%%%%%%%%%%%%%%%%%%%%%%%%%%%%%%%%%%%%%%
%%%%%%%%%%%%%%%%%%%%%%%%%%%%%%%%%%%%%%%%%%%%%%%%%%%%%%%%%%%%%%%%%%%%%%%%%%%%%%%%%%%%%%%%%%%%%%%%%%%%%%%%%%%%%%%%%%%%%%%%%%%%%%%%%%
\subsection{Case of Low Forgery and Suppression}
\label{sec:Theory:LowRates}
\noindent
This subsection characterizes the privacy\hyph forgery\hyph suppression function
in the special case when $\rho,\sigma \simeq 0$.

%%%%%%%%%%%%%%%%%%%%%%%%%%%%%%%%%%%%%%%%%%%%%%%%%%%%%%%%%%%%%%%%%%%%%%%%%%%%%%%%%%%%%%%%%%%%%%%%%%%%%%%%%%%%%%%%%%%%%%%%%%%%%%%%%%%
%%PROPOSITION: LOW FORGERY AND SUPPRESSION RATES
%%%%%%%%%%%%%%%%%%%%%%%%%%%%%%%%%%%%%%%%%%%%%%%%%%%%%%%%%%%%%%%%%%%%%%%%%%%%%%%%%%%%%%%%%%%%%%%%%%%%%%%%%%%%%%%%%%%%%%%%%%%%%%%%%%%
\begin{proposition}[Low Rates of Forgery and Suppression]
\label{Proposition:LowCase}
Assume the nontrivial case in which $q\neq p$.
Then, there exist two indexes $i,j$ such that $0=\rho_1=\cdots = \rho_i < \rho_{i+1}$
and $0=\sigma_n=\cdots = \sigma_j < \sigma_{j-1}$.
For any $\rho \in [0,\rho_{i+1}]$ and $\sigma \in [0,\sigma_{j-1}]$,
the number of nonzero components of the optimal forgery and suppression strategies is $i$ and $n-j+1$, respectively.
Further, the gradient of the privacy\hyph forgery\hyph suppression function at the origin is
\renewcommand{\arraystretch}{1.5}
\begin{equation*}
\nabla \cR(0,0) =
\left(\begin{array}{c}
\frac{\partial \cR(0,0)}{\partial \rho}\\
\frac{\partial \cR(0,0)}{\partial \sigma}
\end{array}\right) =
\left(\begin{array}{c}
\log \frac{q_1}{p_1} - \oD(q\,\|\,p)\\
\oD(q\,\|\,p) - \log \frac{q_n}{p_n}
\end{array}\right).
\end{equation*}
\end{proposition}
\SpaceAfterPropositionEndedWithFormula
\BeginProof
The existence of the indexes $i$ and $j$ is guaranteed by the assumption that $q\neq p$.
The number of nonzero components of $r^*$ and $s^*$ is trivial from Theorem~\ref{Theorem:ClosedFormSolution}.
In view of this theorem, for any $\rho \in [0,\rho_{i+1}]$ and $\sigma \in [0,\sigma_{j-1}]$, we have
$$\cR(\rho,\sigma) = \oD\left(\left.\frac{\tilde{q} + \rho (1,0,\ldots,0) - \sigma (0,\ldots,0,1)}{1+\rho-\sigma}\right\|\tilde{p}\right).$$

The continuity of the components of $r^*$ and $s^*$ proven in Corollary~\ref{Corollary:OrthoCont}~(ii) ensures the continuity of the privacy\hyph forgery\hyph suppression function on $\bar{\cC}$.
It is routine to check its differentiability in this region and to obtain its derivative with respect to $\sigma$ at the origin,
$$\frac{\partial \cR(0,0)}{\partial \sigma} = Q_i \log \frac{Q_i\, \cCP_j}{P_i\, \cCQ_j} + \sum_{k=i+1}^{j-1} q_k \log \frac{\cCP_j\,q_k}{\cCQ_j\,p_k}.$$
On account of Proposition~\ref{Proposition:Monotonicity}, the conditions $\rho_1=\cdots = \rho_i $ and $\sigma_j=\cdots = \sigma_n$
imply
$$\frac{q_1}{p_1} = \cdots = \frac{q_i}{p_i} = \frac{Q_i}{P_i}$$
and
$$\frac{q_j}{p_j} =\cdots = \frac{q_n}{p_n} = \frac{\cCQ_j}{\cCP_j}.$$
Therefore,
\begin{equation*}
\begin{aligned}
\frac{\partial \cR(0,0)}{\partial \sigma} &= \sum_{k=1}^{j-1} q_k \log \frac{q_k}{p_k} - Q_{j-1} \log \frac{q_n}{p_n}               \\
                                                                         &= \oD(q\,\|\,p) - \log \frac{q_n}{p_n}.
\end{aligned}
\end{equation*}
The derivative of $\cR$ with respect to $\rho$ at $\rho=\sigma=0$ follows analogously.
\EndProof
Next, we shall derive an expression for the relative decrement of the privacy\hyph risk function at $\rho,\sigma \simeq 0$.
To this end, define the \emph{forgery relative decrement factor}
$$\delta_{\rho} = -  \frac{\frac{\partial \cR(0,0)}{\partial \rho}}{\cR(0,0)} = 1 - \frac{\log\frac{q_1}{p_1}}{\oD(q\,\|\,p)}, $$
and the \emph{suppression relative decrement factor}
$$\delta_{\sigma} = -  \frac{\frac{\partial \cR(0,0)}{\partial \sigma}}{\cR(0,0)} = \frac{\log\frac{q_n}{p_n}}{\oD(q\,\|\,p)} - 1. $$
By dint of Proposition~\ref{Proposition:LowCase},
the first\hyph order Taylor approximation of function~\eqref{eqn:PrivacyUtilityFunction} around $\rho=\sigma=0$ yields
\begin{equation*}
\cR(\rho,\sigma) \simeq \oD(q\,\|\,p) + \rho\left(\log \frac{q_1}{p_1} - \oD(q\,\|\,p)\right) + \sigma\left(\oD(q\,\|\,p) - \log \frac{q_n}{p_n}\right),
\end{equation*}
or more compactly, in terms of the decrement factors,
$$\frac{\oD(q\,\|\,p) - \cR(\rho,\sigma)}{\oD(q\,\|\,p)} \simeq \delta_{\rho}\,\rho + \delta_{\sigma}\,\sigma.$$
In words, the minimum and maximum ratios $\frac{q_k}{p_k}$ characterize the relative reduction in privacy risk.
The following result, Proposition~\ref{Proposition:DecrementFactors}, establishes a bound on these relative decrement factors.

%%%%%%%%%%%%%%%%%%%%%%%%%%%%%%%%%%%%%%%%%%%%%%%%%%%%%%%%%%%%%%%%%%%%%%%%%%%%%%%%%%%%%%%%%%%%%%%%%%%%%%%%%%%%%%%%%%%%%%%%%%%%%%%%%%%
%%PROPOSITION: LOW FORGERY AND SUPPRESSION RATES
%%%%%%%%%%%%%%%%%%%%%%%%%%%%%%%%%%%%%%%%%%%%%%%%%%%%%%%%%%%%%%%%%%%%%%%%%%%%%%%%%%%%%%%%%%%%%%%%%%%%%%%%%%%%%%%%%%%%%%%%%%%%%%%%%%%
\begin{proposition}[Relative Decrement Factors]
\label{Proposition:DecrementFactors}
In the nontrivial case when $q\neq p$, the relative decrement factors satisfy $\delta_{\rho} > 1$ and $\delta_{\sigma} > 0$.
\end{proposition}

\BeginProof
Observe that the statement $\delta_{\rho} > 1$ is equivalent to the condition $q_1 < p_1$.
We prove this by contradiction.
Suppose that $q_1 > p_1$.
By the labeling assumption~\eqref{eqn:LabelingAssumption},
it follows that $q_k>p_k$ for all $k$,
what leads to the contradiction that $1 = \sum q_k  > \sum p_k = 1$.
Now assume that $q_1 = p_1$.
Since $q \neq p$, there must exist an index $i$ such that
$$\frac{q_1}{p_1} = \cdots = \frac{q_{i-1}}{p_{i-1}} < \frac{q_i}{p_i} \leqslant \cdots \leqslant \frac{q_n}{p_n}.$$
But this implies that
$$1 - \sum_{k=1}^{i-1} q_k =  \sum_{k=i}^{n} q_k > \sum_{k=i}^{n} p_k = 1 - \sum_{k=1}^{i-1} q_k,$$
a contradiction.
This proves the first part of the proposition.

For the second part, note that the statement $\delta_{\sigma} > 0$ is equivalent to
$$q_1 \log \frac{q_1}{p_1} + \cdots + q_n \log \frac{q_n}{p_n} < \log \frac{q_n}{p_n},$$
and, after algebraic manipulation, to
$$q_1 \log \frac{q_1}{ p_1} \frac{p_n}{q_n} + \cdots + q_{n-1} \log \frac{q_{n-1}}{p_{n-1}}\frac{p_n}{q_n} < 0.$$
The positivity and labeling assumptions~\eqref{eqn:PositivityAssumption},~\eqref{eqn:LabelingAssumption}
ensure that all terms in the sum are nonpositive.
However, the additional assumption $q \neq p$ implies that $\frac{q_1}{p_1} < \frac{q_n}{p_n}$,
which in turn implies that the first term is negative and so is, consequently, the entire summation.
\EndProof

Conceptually, the bound on $\delta_{\rho}$ tells us that the relative decrement in privacy risk is greater than the forgery rate introduced.
This is under the assumption that $q \neq p$ and at low rates of forgery and suppression.
The bound on $\delta_{\sigma}$, however, is looser than the previous one and just ensures that an increase in the suppression rate always leads to a decrease in privacy risk,
as one would expect.

%%%%%%%%%%%%%%%%%%%%%%%%%%%%%%%%%%%%%%%%%%%%%%%%%%%%%%%%%%%%%%%%%%%%%%%%%%%%%%%%%%%%%%%%%%%%%%%%%%%%%%%%%%%%%%%%%%%%%%%%%%%%%%%%%%
%%%%%%%%%%%%%%%%%%%%%%%%%%%%%%%%%%%%%%%%%%%%%%%%%%%%%%%%%%%%%%%%%%%%%%%%%%%%%%%%%%%%%%%%%%%%%%%%%%%%%%%%%%%%%%%%%%%%%%%%%%%%%%%%%%
%4.4 PURE STRATEGIES
%%%%%%%%%%%%%%%%%%%%%%%%%%%%%%%%%%%%%%%%%%%%%%%%%%%%%%%%%%%%%%%%%%%%%%%%%%%%%%%%%%%%%%%%%%%%%%%%%%%%%%%%%%%%%%%%%%%%%%%%%%%%%%%%%%
%%%%%%%%%%%%%%%%%%%%%%%%%%%%%%%%%%%%%%%%%%%%%%%%%%%%%%%%%%%%%%%%%%%%%%%%%%%%%%%%%%%%%%%%%%%%%%%%%%%%%%%%%%%%%%%%%%%%%%%%%%%%%%%%%%
\subsection{Pure Strategies}
\label{sec:Theory:PureStrategies}
\noindent
In the previous subsections we investigated the forgery and the suppression of ratings as a \emph{mixed} strategy that users
may adopt to enhance their privacy.
In this subsection we contemplate the case in which users may be reluctant to use these two mechanisms in conjunction; %certain
and as a consequence, they may opt for a \emph{pure} strategy consisting in the application of either forgery or suppression.
In this case, it would be useful to determine which is the most appropriate technique in terms of the privacy\hyph utility trade\hyph off posed.
Our next result, Corollary~\ref{Corollary:PureStrategies}, provides some insight on this,
under the assumption that, from the user's perspective, the impact on utility due to forgery is equivalent to that caused by the effect of suppression.

Before showing this result, observe from Theorem~\ref{Theorem:ClosedFormSolution} that $\rho_n = \frac{q_n}{p_n} - 1$ is the minimum forgery rate such that
$\cR(\rho,0)=0$.
Analogously, $\sigma_1 = 1 - \frac{q_1}{p_1}$ is the minimum suppression rate satisfying $\cR(0,\sigma)=0$.
In other words, $\rho_n$ and $\sigma_1$ are the \emph{critical rates} of the pure forgery and suppression strategies, respectively.
Further, note that $\sigma_1 < \sigma_0 = 1$, on account of the positivity assumption~\eqref{eqn:PositivityAssumption}.
However, $\rho_n > 1$ if, and only if, $\frac{q_n}{p_n}>2$.
%%%%%%%%%%%%%%%%%%%%%%%%%%%%%%%%%%%%%%%%%%%%%%%%%%%%%%%%%%%%%%%%%%%%%%%%%%%%%%%%%%%%%%%%%%%%%%%%%%%%%%%%%%%%%%%%%%%%%%%%%%%%%%%%%%%
%%COROLLARY: PURE STRATEGIES
%%%%%%%%%%%%%%%%%%%%%%%%%%%%%%%%%%%%%%%%%%%%%%%%%%%%%%%%%%%%%%%%%%%%%%%%%%%%%%%%%%%%%%%%%%%%%%%%%%%%%%%%%%%%%%%%%%%%%%%%%%%%%%%%%%%
\begin{corollary}[Pure Strategies]
\label{Corollary:PureStrategies}
%\leavevmode
Consider the nontrivial case when $q\neq p$.
\begin{EnumerateRoman}
\item The critical rates of the pure forgery and suppression strategies satisfy $\rho_n < \sigma_1$ if, and only if,
$$\frac{\nicefrac{q_1}{p_1} + \nicefrac{q_n}{p_n}}{2} < 1.$$
\item The forgery and the suppression relative decrement factors satisfy $\delta_{\rho} > \delta_{\sigma}$ if, and only if,
$$\sqrt{\frac{q_1}{p_1}\, \frac{q_n}{p_n}} < 2^{\oD(q\,\|\,p)}.$$
\end{EnumerateRoman}
\end{corollary}
\SpaceAfterPropositionEndedWithFormula
\BeginProof
Both statements are immediate from the definitions of $\rho_n$ and $\sigma_1$ on the one hand, and $\delta_{\rho}$ and $\delta_{\sigma}$ on the other.
\EndProof

In conceptual terms, the condition $\rho_n < \sigma_1$ means that the pure forgery strategy is the most appropriate mechanism
in terms of causing the minimum distortion to attain the critical\hyph privacy region.
On the other hand, the condition $\delta_{\rho} > \delta_{\sigma}$ implies that, at low rates, the pure forgery strategy offers better privacy protection than the pure suppression strategy does.
Therefore, the conclusion that follows from Corollary~\ref{Corollary:PureStrategies}
is that, together with the quantity $\oD(q\,\|\,p)$, the arithmetic and geometric mean
of the ratios $\frac{q_1}{p_1}$ and $\frac{q_n}{p_n}$ determine which strategy to choose.

Another interesting remark is the duality of these two ratios $\frac{q_1}{p_1}$ and $\frac{q_n}{p_n}$.
The former characterizes the minimum rate for the pure suppression strategy to reach the critical\hyph privacy region and,
 at the same time, it establishes the privacy gain at low forgery rates.
Conversely, the latter ratio defines the critical rate of the pure forgery strategy and determines the relative decrement in privacy risk at low suppression rates.

Lastly, we would like to establish a connection between our work and that of~\cite{Rebollo10IT,Parra12DKE},
where the \emph{pure} forgery and suppression strategies are investigated.
Denote by $\cR_{\textnormal{F}}$ the function derived in~\cite{Rebollo10IT} modeling the trade\hyph off between forgery rate and privacy \emph{risk},
the latter being measured as the KL divergence between the user's apparent profile and the population's distribution.
Define $\rho'$ as the ratio of forged ratings to \emph{total} number of ratings.
Accordingly, it can be shown that $\rho'=\frac{\rho}{1+ \rho}$ and that
$\cR(\rho,0) = \cR_{\textnormal{F}} (\rho').$
On the other hand, denote by $\cP_{\textnormal{S}}$ the function in~\cite{Parra12DKE} characterizing the trade\hyph off between suppression rate and privacy \emph{gain}.
In this case, privacy is measured as the Shannon's entropy of the user's apparent profile.
Under the assumption that the population's profile is uniform,
it can be proven that $\cR(0,\sigma) = \log n - \cP_{\textnormal{S}}(\sigma).$
In short, our formulation of the problem of optimal forgery and suppression of ratings encompasses, as particular cases, the cited works.

%%%%%%%%%%%%%%%%%%%%%%%%%%%%%%%%%%%%%%%%%%%%%%%%%%%%%%%%%%%%%%%%%%%%%%%%%%%%%%%%%%%%%%%%%%%%%%%%%%%%%%%%%%%%%%%%%%%%%%%%%%%%%%%%%%
%%%%%%%%%%%%%%%%%%%%%%%%%%%%%%%%%%%%%%%%%%%%%%%%%%%%%%%%%%%%%%%%%%%%%%%%%%%%%%%%%%%%%%%%%%%%%%%%%%%%%%%%%%%%%%%%%%%%%%%%%%%%%%%%%
%%%%%%%%%%%%%%%%%%%%%%%%%%%%%%%%%%%%%%%%%%%%%%%%%%%%%%%%%%%%%%%%%%%%%%%%%%%%%%%%%%%%%%%%%%%%%%%%%%%%%%%%%%%%%%%%%%%%%%%%%%%%%%%%%%
% 4.5 NUMERICAL EXAMPLE
%%%%%%%%%%%%%%%%%%%%%%%%%%%%%%%%%%%%%%%%%%%%%%%%%%%%%%%%%%%%%%%%%%%%%%%%%%%%%%%%%%%%%%%%%%%%%%%%%%%%%%%%%%%%%%%%%%%%%%%%%%%%%%%%%%
%%%%%%%%%%%%%%%%%%%%%%%%%%%%%%%%%%%%%%%%%%%%%%%%%%%%%%%%%%%%%%%%%%%%%%%%%%%%%%%%%%%%%%%%%%%%%%%%%%%%%%%%%%%%%%%%%%%%%%%%%%%%%%%%%%
%%%%%%%%%%%%%%%%%%%%%%%%%%%%%%%%%%%%%%%%%%%%%%%%%%%%%%%%%%%%%%%%%%%%%%%%%%%%%%%%%%%%%%%%%%%%%%%%%%%%%%%%%%%%%%%%%%%%%%%%%%%%%%%%%%
\subsection{Numerical Example}
\label{sec:Theory:Example}
\noindent
This subsection presents a numerical example that illustrates the theoretical analysis conducted in the previous subsections.
Later on in Sec.~\ref{sec:Experiments} we shall evaluate the effectiveness of our approach in a real scenario, namely in the movie recommendation system \emph{Movielens}.
In our numerical example we assume $n = 3$ categories of interests.
Although the example shown here is synthetic,
these three categories could very well represent interests across topics such as technology, sports and beauty.
\begin{figure}
\centering
\includegraphics[scale=0.57]{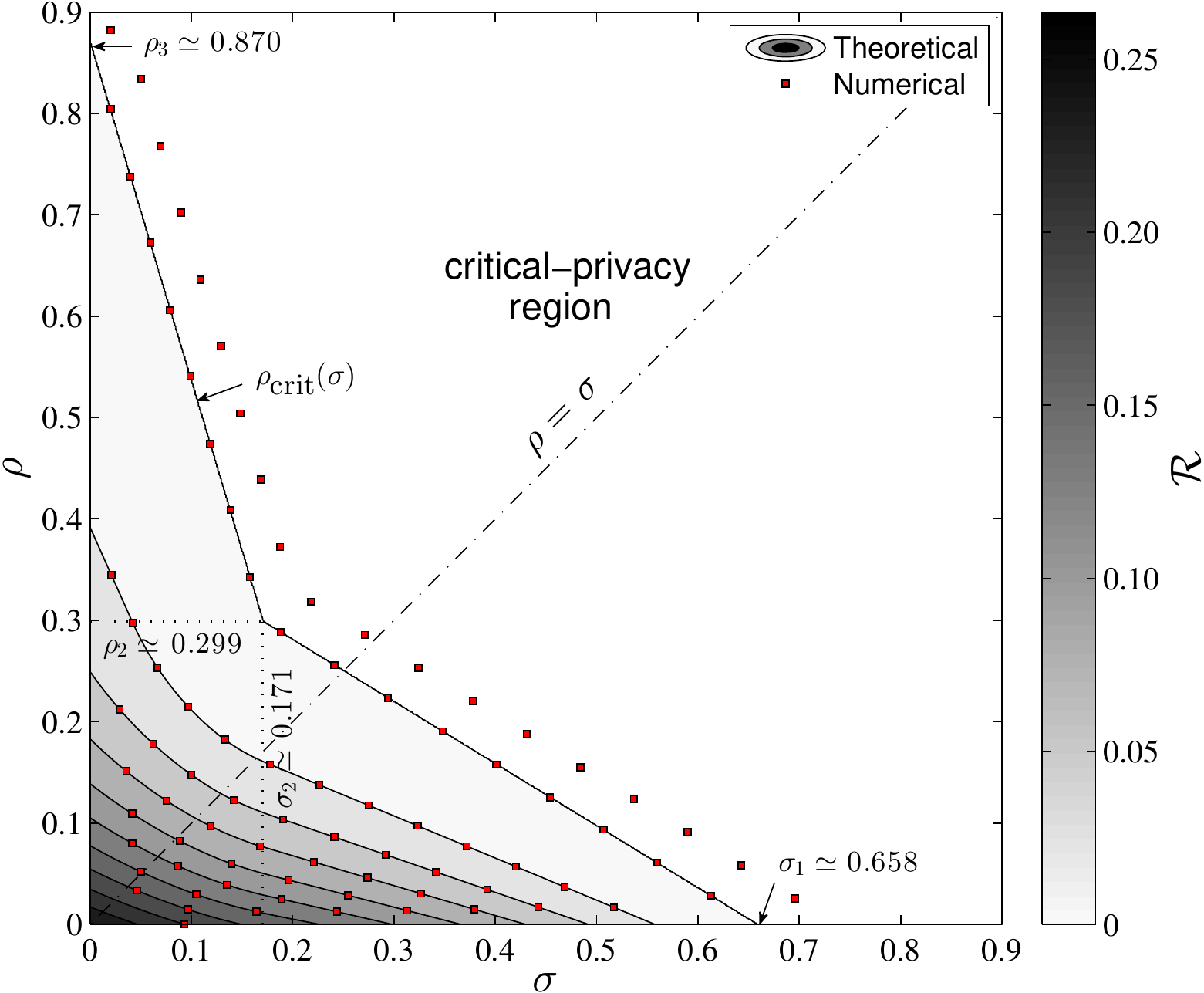}
\caption{Contour lines of the privacy\hyph forgery\hyph suppression function, the corresponding forgery and suppression thresholds, and the critical and noncritical privacy regions.}
\label{fig:trade-off}
\end{figure}
Accordingly, we suppose that the user's rating distribution is
\begin{equation*}
q=(0.130, 0.440, 0.430),
\end{equation*}
and the population's,
\begin{equation*}
p=(0.380, 0.390, 0.230).
\end{equation*}
Note that these distributions satisfy the positivity and labeling assumptions~\eqref{eqn:PositivityAssumption},~\eqref{eqn:LabelingAssumption}.

From Sec.~\ref{sec:Theory:ClosedFormSolution},
we easily obtain the forgery thresholds $\rho_1 =0$, $\rho_2 \simeq 0.299$ and $\rho_3 \simeq 0.870$ on the one hand,
and on the other the suppression thresholds $\sigma_3 = 0$, $\sigma_2 \simeq 0.171$ and $\sigma_1 \simeq 0.658$.
The thresholds $\rho_3$ and $\sigma_1$ are the critical rates of the pure strategies.
If we are to reach the critical\hyph privacy region and do not have any preference for either forgery or suppression,
the fact that $\rho_3 > \sigma_1$ leads us to opt for suppression as pure strategy.
However, the geometric mean of $\frac{q_1}{p_1}$ and $\frac{q_3}{p_3}$ is approximately $0.799$,
which is lower than $2^{\oD(q\,\|\,p)} \simeq 1.20$.
On account of Corollary~\ref{Corollary:PureStrategies}, this means that the pure forgery strategy contributes to a greater reduction in privacy risk at low rates than suppression does.
In fact, the gradient of the privacy\hyph forgery\hyph suppression function at the origin is $\nabla \cR(0,0)^{\textnormal{T}} \simeq (-1.81, -0.639)$,
by virtue of Proposition~\ref{Proposition:LowCase}.

Fig.~\ref{fig:trade-off} shows the contour lines of this function,
computed analytically from Theorem~\ref{Theorem:ClosedFormSolution} and numerically~\footnote{The numerical method chosen is the
interior\hyph point algorithm~\cite{Boyd04B} implemented by the Matlab R2012b function \texttt{fmincon}.}.
The region plotted in gray shades corresponds to the noncritical\hyph privacy region~$\bar{\cC}$.
The initial privacy risk is $\cR(0,0) \simeq 0.263$.
The white area represents the critical\hyph privacy region $\cC$,
where the apparent user profile coincides with the population's distribution and thus the privacy risk vanishes.
An interesting observation arising from Fig.~\ref{fig:trade-off} is the synergistic effect of combining forgery and suppression.
Just as an example, in the case when $\rho = \rho_2$ and $\sigma = \sigma_2$,
the sum of these two distortion measures is lower than the critical rates of the pure strategies.

\begin{figure*}[tbp]%
\centering\hspace*{\fill}
\subfigure[$\rho= 0.050$, $\sigma=0.100$, $\rho/\rhocrit(\sigma) \simeq 0.093$, $\cR(\rho,\sigma)\simeq 0.131$, $\cR(\rho,\sigma)/\cR_0 \simeq 0.498$,
$r^*=(0.050,0,0)$, $s^*=(0,0,0.100)$, $t^* \simeq (0.189, 0.463, 0.347)$.]%
{\includegraphics[scale=\FigScaleMultipleBigger]{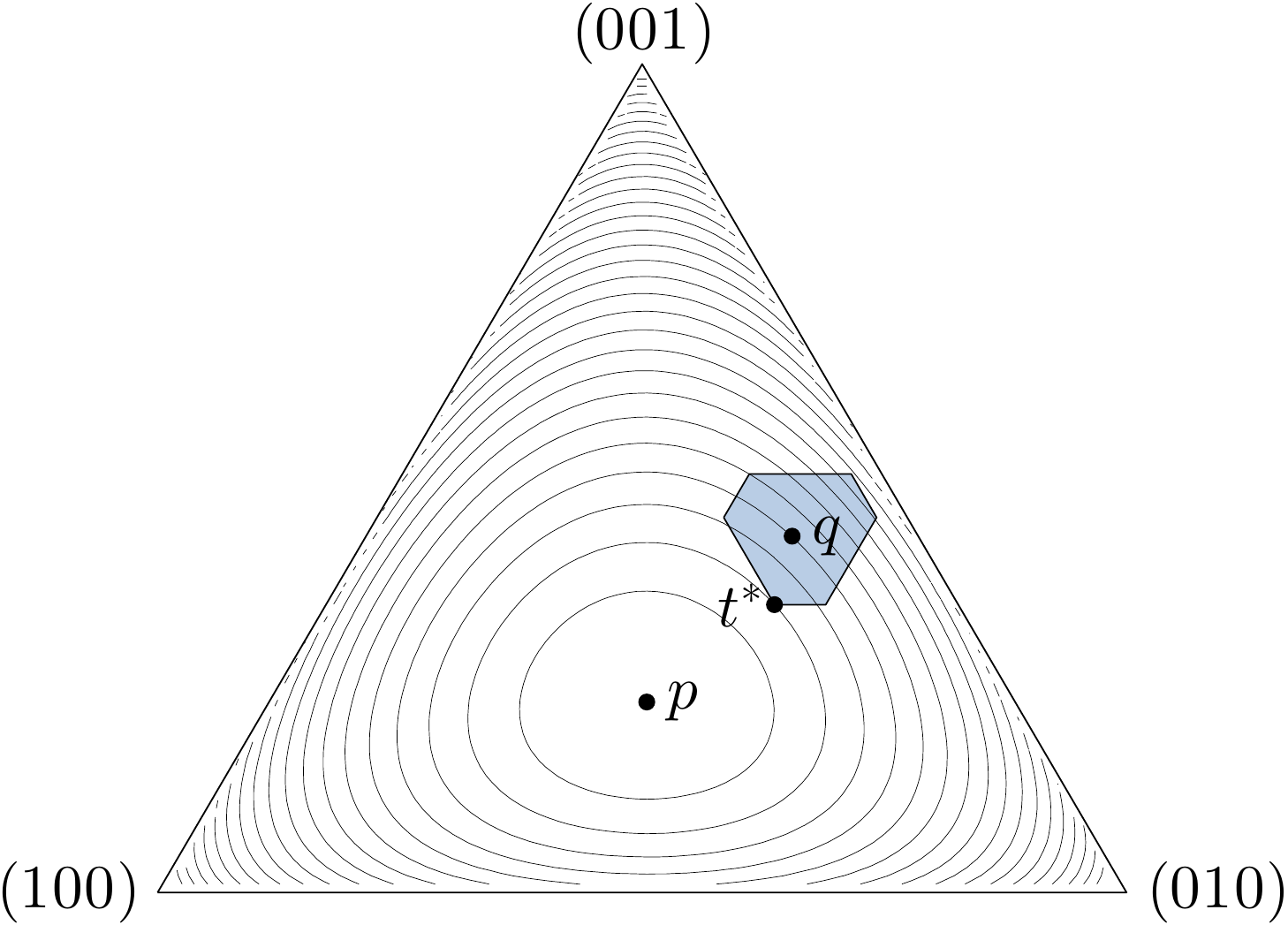}%
\label{fig:SubSimplex1}}\hfill
\subfigure[$\rho = 0.100$, $\sigma=0.200$, $\rho/\rhocrit(\sigma) \simeq 0.356$, $\cR(\rho,\sigma)\simeq 0.050$, $\cR(\rho,\sigma)/\cR_0 \simeq 0.190$,
$r^*=(0.100,0,0)$, $s^*\simeq (0,0.019,0.181)$, $t^*\simeq (0.256, 0.468, 0.276)$.]%
{\includegraphics[scale=\FigScaleMultipleBigger]{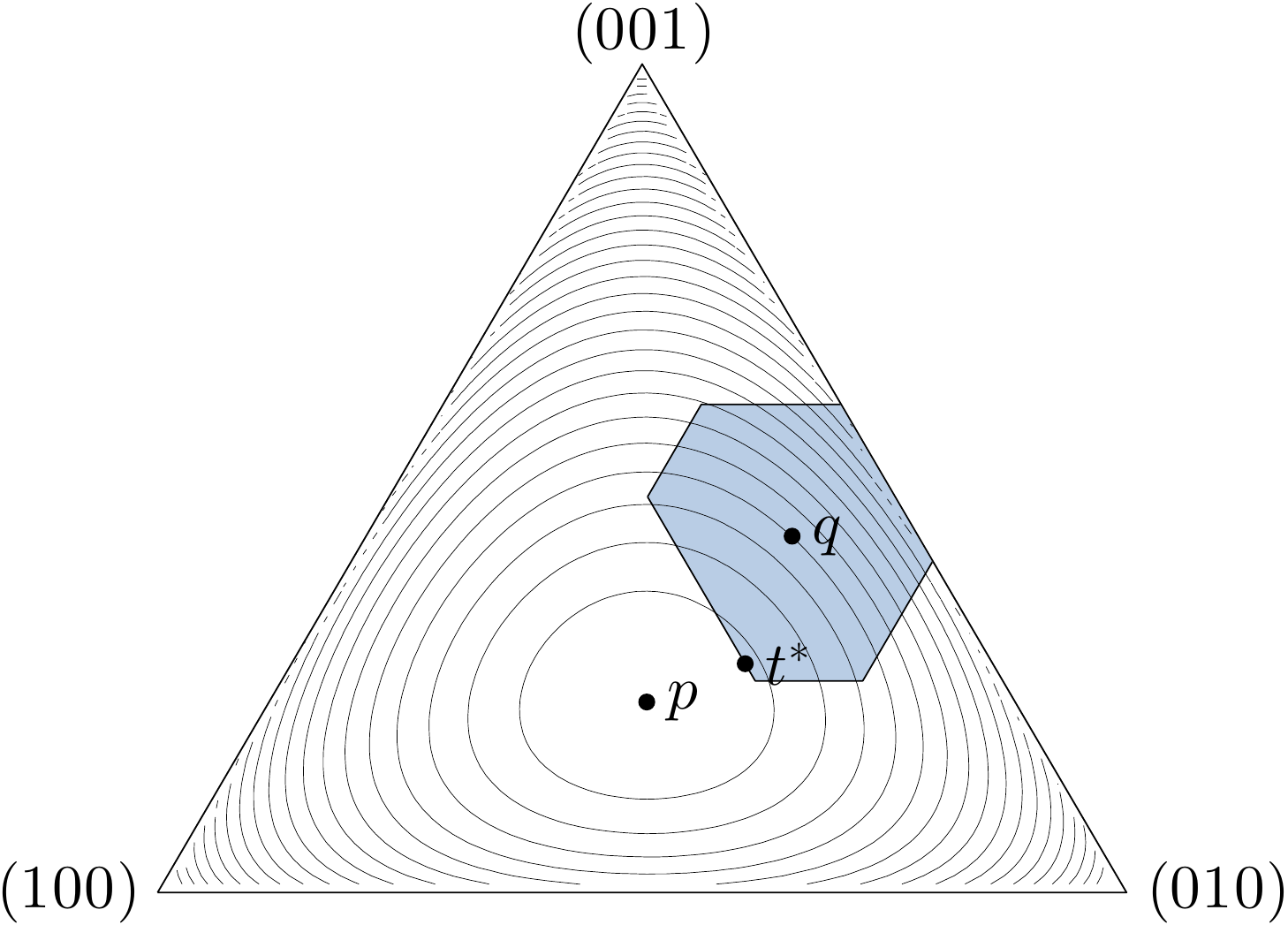}%
\label{fig:SubSimplex2}}\hspace*{\fill}
\\
\hspace*{\fill}
\subfigure[$\rho \simeq 0.219$, $\sigma=0.300$, $\rho/\rhocrit(\sigma) =1$, $\cR(\rho,\sigma)=0$, $\cR(\rho,\sigma)/\cR_0 =0$,
$r^*\simeq (0.219,0,0)$, $s^*\simeq (0,0.081,0.219)$, $t^*=p$.]%
{\includegraphics[scale=\FigScaleMultipleBigger]{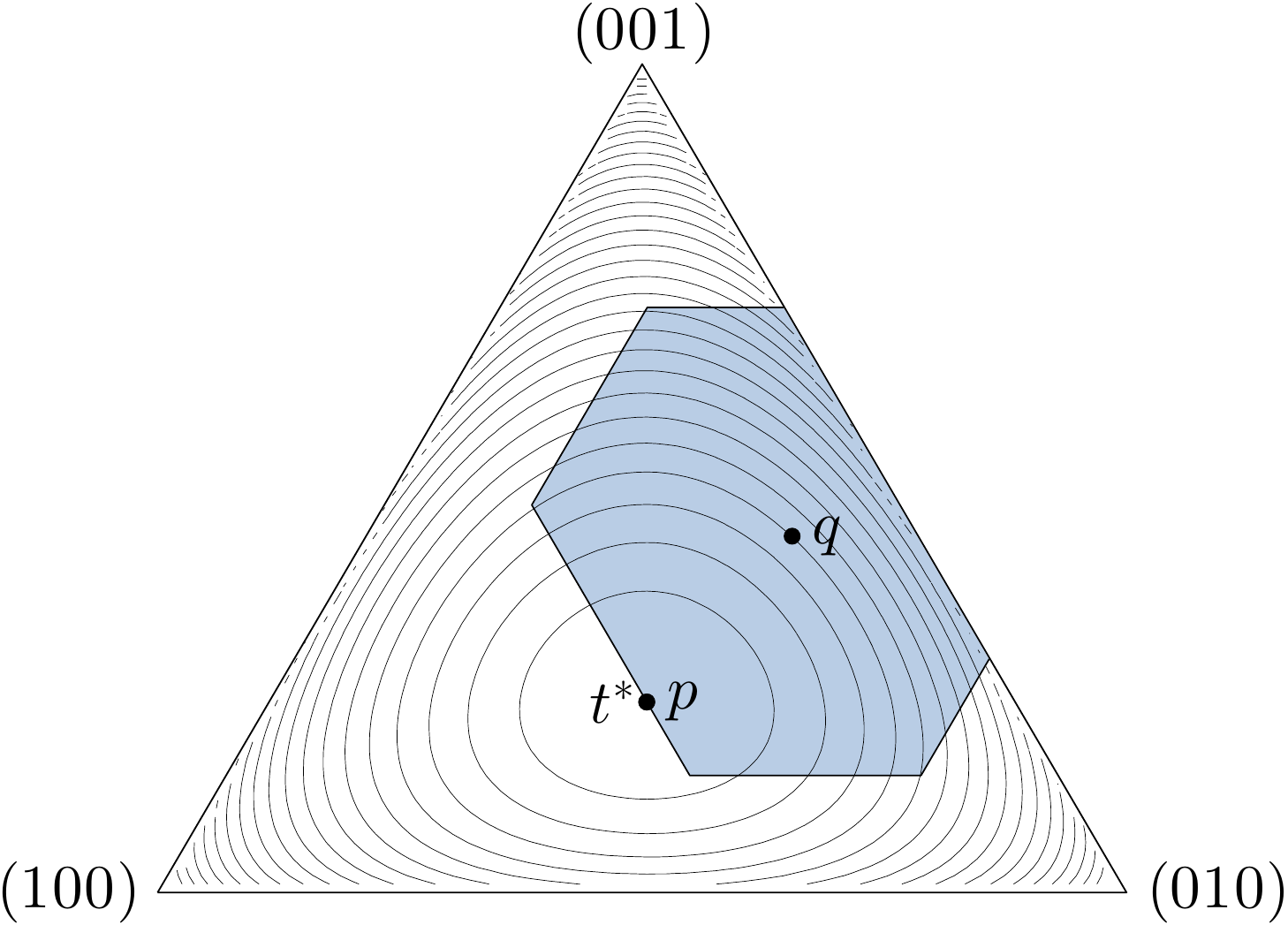}%
\label{fig:SubSimplex3}}\hfill
\subfigure[$\rho= 0.300$, $\sigma=0.300$, $\rho/\rhocrit(\sigma) \simeq 1.368$, $\cR(\rho,\sigma)=0$, $\cR(\rho,\sigma)/\cR_0 =0$,
$r^*\simeq (0.260,0.021,0.019)$, $s^*=(0.010,0.071,0.219)$, $t^*=p$.]%
{\includegraphics[scale=\FigScaleMultipleBigger]{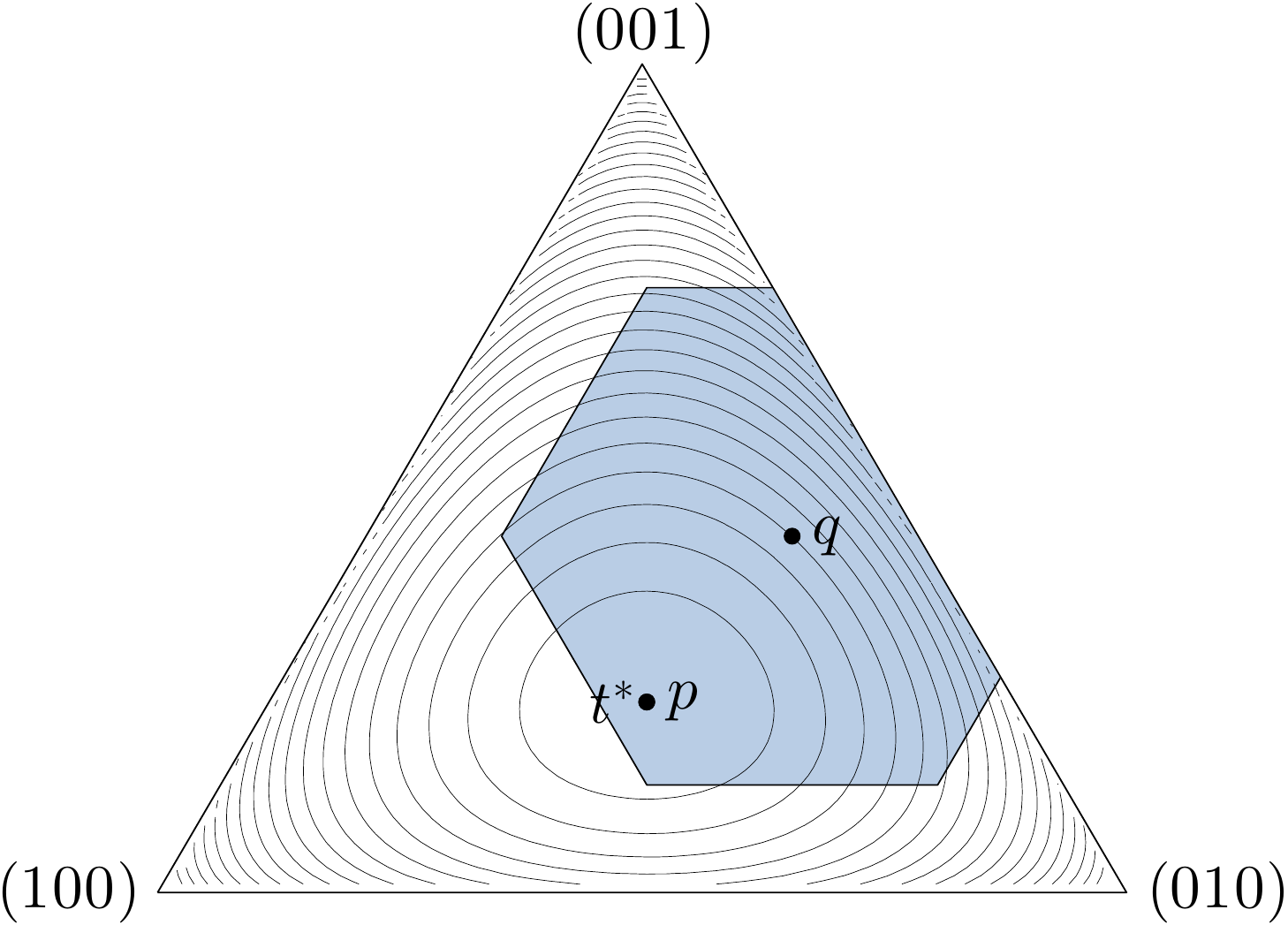}%
\label{fig:SubSimplex4}}\hspace*{\fill}
\caption{Probability simplices showing, for several interesting values of $\rho$ and $\sigma$, the user's actual profile $q=(0.130, 0.440, 0.430)$, the population's distribution $p=(0.380, 0.390, 0.230)$, the optimal apparent distribution $t^*$ and the set of feasible apparent distributions.}
\label{fig:Simplices}
\end{figure*}

Next, we examine the optimal apparent rating distribution for different values of $\rho$ and $\sigma$.
For this purpose, the user's genuine distribution $q$, the population's distribution $p$ and the optimal apparent distribution $t^*$ are depicted
in the probability simplices shown in Fig.~\ref{fig:Simplices}.
In each simplex, we also represent the contour lines of the KL divergence $\oD(\cdot\,\|\,p)$ between every distribution in the simplex and $p$.
Further, we plot the set of feasible apparent user distributions, not necessarily optimal, for four different combinations of $\rho$ and $\sigma$;
in any of these cases, the set takes the form of a hexagon.
Having said this, now we turn our attention to Fig.~\ref{fig:SubSimplex1}.
In this case, the optimal forgery and suppression strategies have $i=n-j+1=1$ nonzero component,
since $\rho \in [0,\rho_2]$ and $\sigma \in [0, \sigma_2]$.
This places the solution $t^*$ at one vertex of the hexagon.
A remarkable fact is that, for these rates, the privacy risk is approximately halved.
In the end, consistently with Proposition~\ref{Proposition:DecrementFactors},
the forgery and the suppression relative decrement factors are $\delta_{\rho} \simeq  6.87>1$ and $\delta_{\sigma} \simeq  2.42>0$.

In the case shown in Fig.~\ref{fig:SubSimplex2}, $r^*$ still has $i=1$ nonzero components,
while $s^*$ contains $n-j+1=2$ nonzero components.
Geometrically, the optimal apparent distribution lies at one edge of the feasible region.
This lowers privacy risk to a 19\% of its initial value.
The case in which $(\rho,\sigma)=(\rhocrit (\sigma),\sigma)$ is depicted in Fig.~\ref{fig:SubSimplex3}.
Here, the number of nonzero components of $r^*$ and $s^*$ remains the same as in the previous case,
but the privacy risk becomes zero.
The last case, illustrated in Fig.~\ref{fig:SubSimplex4}, does not have any practical application,
as $\cR(\rho,\sigma) = 0$ for any $(\rho,\sigma) \in \partial{\cC}$.
In this figure we can observe that the solution $t^*$ is placed in the interior of the hexagon,
and that the orthogonality principle of the strategies $r^*$ and $s^*$ stated in Corollary~\ref{Corollary:OrthoCont} is not satisfied.

%%%%%%%%%%%%%%%%%%%%%%%%%%%%%%%%%%%%%%%%%%%%%%%%%%%%%%%%%%%%%%%%%%%%%%%%%%%%%%%%%%%%%%%%%%%%%%%%%%%%%%%%%%%%%%%%%%%%%%%%%%%%%%%%%
%%%%%%%%%%%%%%%%%%%%%%%%%%%%%%%%%%%%%%%%%%%%%%%%%%%%%%%%%%%%%%%%%%%%%%%%%%%%%%%%%%%%%%%%%%%%%%%%%%%%%%%%%%%%%%%%%%%%%%%%%%%%%%%%%%
% 5 Experimental Evaluation
%%%%%%%%%%%%%%%%%%%%%%%%%%%%%%%%%%%%%%%%%%%%%%%%%%%%%%%%%%%%%%%%%%%%%%%%%%%%%%%%%%%%%%%%%%%%%%%%%%%%%%%%%%%%%%%%%%%%%%%%%%%%%%%%%%
%%%%%%%%%%%%%%%%%%%%%%%%%%%%%%%%%%%%%%%%%%%%%%%%%%%%%%%%%%%%%%%%%%%%%%%%%%%%%%%%%%%%%%%%%%%%%%%%%%%%%%%%%%%%%%%%%%%%%%%%%%%%%%%%%%
\section{Experimental Evaluation}
\label{sec:Experiments}
\noindent
In this section we evaluate the extent to which the forgery and the suppression of ratings could enhance user privacy in a real\hyph world recommendation system.
The system chosen to conduct this evaluation is \emph{Movielens}, a popular movie recommender developed by the GroupLens Research Lab~\cite{Grouplens} at the University of Minnesota.
As many other recommenders, \emph{Movielens} allows users to both rate and tag movies according to their preferences.
These preferences are then exploited by the recommender to suggest movies that users have not watched yet.

%%%%%%%%%%%%%%%%%%%%%%%%%%%%%%%%%%%%%%%%%%%%%%%%%%%%%%%%%%%%%%%%%%%%%%%%%%%%%%%%%%%%%%%%%%%%%%%%%%%%%%%%%%%%%%%%%%%%%%%%%%%%%%%%%%
%6.1 Data set
%%%%%%%%%%%%%%%%%%%%%%%%%%%%%%%%%%%%%%%%%%%%%%%%%%%%%%%%%%%%%%%%%%%%%%%%%%%%%%%%%%%%%%%%%%%%%%%%%%%%%%%%%%%%%%%%%%%%%%%%%%%%%%%%%%
\subsection{Data set}
\label{sec:Experiments:DataSet}
\noindent
The data set that we used to assess our data\hyph perturbative mechanism is the \emph{Movielens 10M} data set~\cite{MovielensDataSet11},
which contains 10\,000\,054 ratings and 95\,580 tags.
The ratings and tags included in this data set were assigned to 10\,681 movies by 71\,567 users.
The data are organized in the form of quadruples (\emph{username}, \emph{movie}, \emph{rating}, \emph{time}),
each one representing the action of a user rating a movie at a certain time.
Usernames have been replaced with numbers in an attempt to anonymize the data set.

For our purposes of experimentation,
we just needed the data fields \emph{username} and \emph{movie}, together with the categories each movie belongs to.
\emph{Movielens} contemplates $n=19$ categories or movies genres,
listed in alphabetical order as follows: \emph{action}, \emph{adventure}, \emph{animation}, \emph{children's}, \emph{comedy}, \emph{crime},
\emph{documentary}, \emph{drama}, \emph{fantasy}, \emph{film\hyph noir}, \emph{horror}, \emph{IMAX}, \emph{musical}, \emph{mystery},
\emph{romance}, \emph{sci\hyph fi}, \emph{thriller}, \emph{war} and \emph{western}.
As we shall see later in Sec.~\ref{sec:Experiments:Results},
for each particular user, we shall have to rearrange those categories in such a way that the labeling assumption~\eqref{eqn:LabelingAssumption} is satisfied.

In our data set, all users rated, at least, 20 movies.
This was the minimum number of ratings for the recommender to start working~\footnote{Nowadays, the algorithm implemented by \emph{Movielens} requires only 15 ratings to start generating predictions.}.
After the elimination of those users who exclusively tagged movies, the total number of users reduced to 69\,878.
Despite the large number of users, we found that only 4\,099 satisfied the positivity assumption~\eqref{eqn:PositivityAssumption}.
Considering that this small group of users represents just the 5.8\% of the total number of users,
we can assume that the application of our technique will have a negligible effect on the population's profile~$p$,
as supposed in Sec.~\ref{sec:FSTechnique:Formulation}.

\begin{table}
\caption{Category index of the particular user examined in our experiments.
The categories of \emph{Movielens} have been sorted and indexed in order to satisfy the labeling assumption~\eqref{eqn:LabelingAssumption}.}
\label{tab:categoriesML}
\footnotesize
\begin{tabularx}{\linewidth}{*3{c >{\centering\arraybackslash}X}}
    \toprule
    Index & Category name & Index & Category name & Index & Category name \\ \cmidrule(lr){1-2} \cmidrule(lr){3-4} \cmidrule(lr){5-6}
    1     & animation     & 7     & sci-fi        & 13    & war       \\
    2     & action        & 8     & comedy        & 14    & mystery       \\
    3     & film-noir     & 9     & thriller      & 15   & musical       \\
    4     & children's    & 10    & fantasy       & 16    & romance          \\
    5     & adventure     & 11    & horror        & 17    & IMAX         \\
    6     & crime         & 12    & western       & 18    & drama   \\
          &               &      &       & 19      &  documentary\\ \bottomrule
\end{tabularx}
\end{table}

%%%%%%%%%%%%%%%%%%%%%%%%%%%%%%%%%%%%%%%%%%%%%%%%%%%%%%%%%%%%%%%%%%%%%%%%%%%%%%%%%%%%%%%%%%%%%%%%%%%%%%%%%%%%%%%%%%%%%%%%%%%%%%%%%%
%6.2 Results
%%%%%%%%%%%%%%%%%%%%%%%%%%%%%%%%%%%%%%%%%%%%%%%%%%%%%%%%%%%%%%%%%%%%%%%%%%%%%%%%%%%%%%%%%%%%%%%%%%%%%%%%%%%%%%%%%%%%%%%%%%%%%%%%%%
\subsection{Results}
\label{sec:Experiments:Results}
\noindent
In this subsection we examine how the forgery and the suppression of ratings may help users of \emph{Movielens} to enhance their privacy.
With this aim, first, we analyze the effect of the perturbation of ratings on the privacy protection of a particular user from our data set.
Secondly, we consider the entire set of 4\,099 users and assess the relative reduction in privacy risk when these users apply the same forgery and suppression rates.
Lastly, we investigate the forgery and the suppression strategies separately, and draw some conclusions about these two pure strategies.

To conduct our first experiments, we choose a particular user from our data set~\footnote{The user considered in this first series of experiments
is identified by the number \texttt{3301} in~\cite{MovielensDataSet11}.}.
Before perturbing the movie rating history of this user,
it is necessary that the components of the user's profile $q$ and the population's distribution $p$ be rearranged to satisfy the labeling assumption~\eqref{eqn:LabelingAssumption}.
Table~\ref{tab:categoriesML} shows how movie categories have been sorted, and then indexed from 1 to $n$, to fulfill the assumption above.
We would like to note that the index provided in this table does not have to coincide with the index of other users in our data set.

\begin{figure*}[!tb]%
\centering\hspace*{\fill}
\subfigure[]
{\includegraphics[scale=0.50]{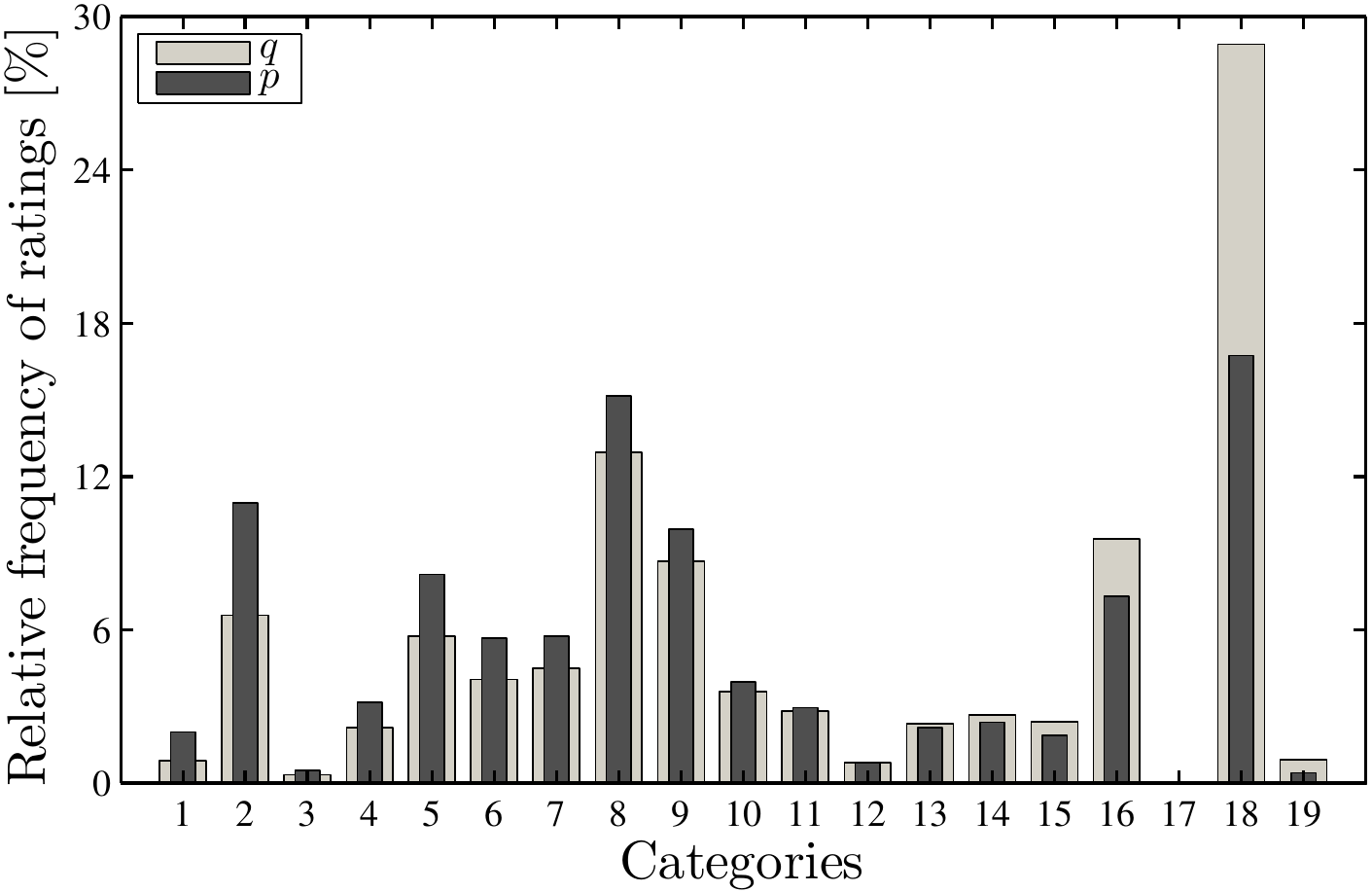}%
\label{fig:FSQvsP}}
\hspace*{\fill}
\\
\hspace*{\fill}
\subfigure[]
{\includegraphics[scale=0.50]{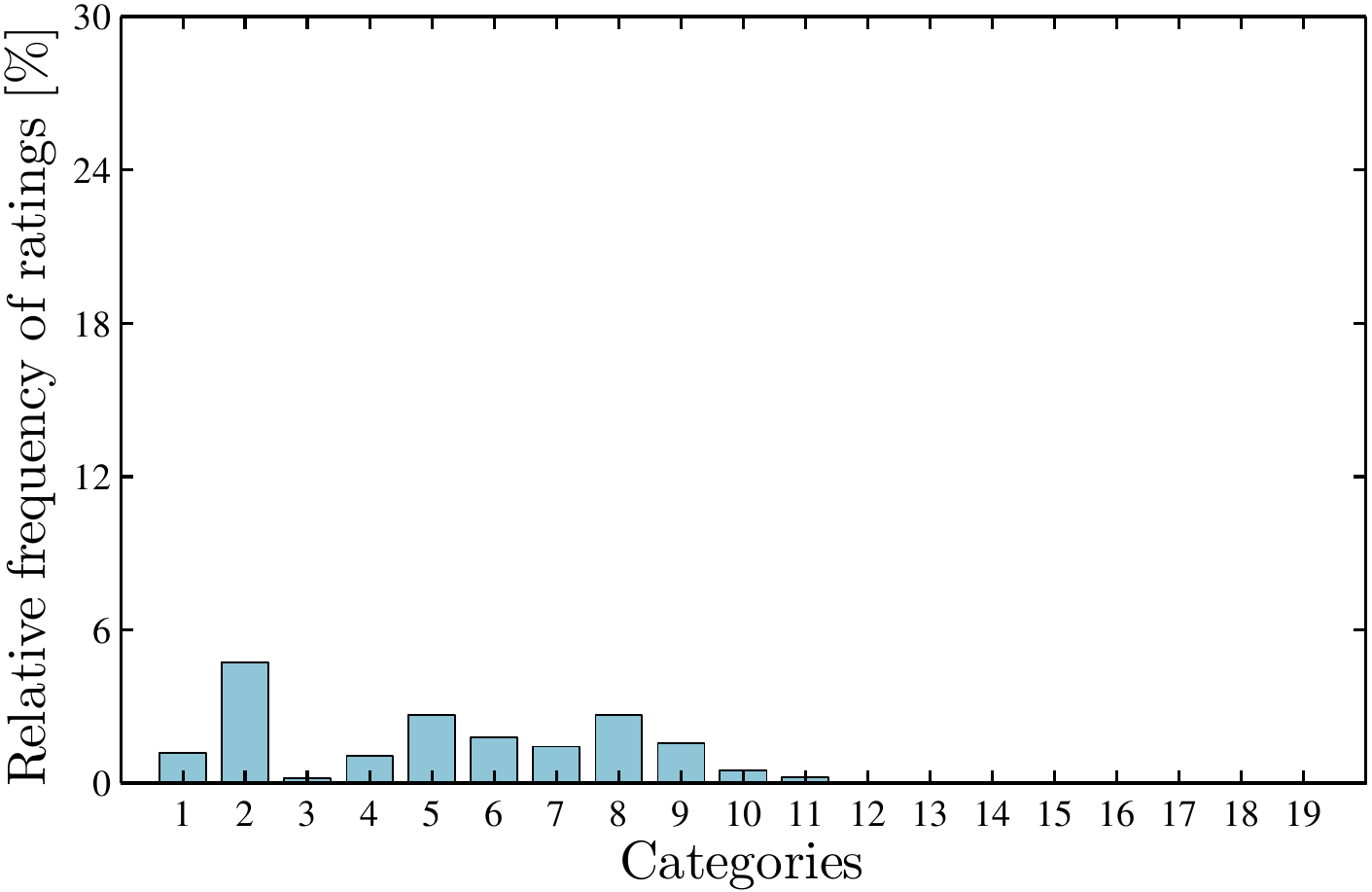}%
\label{fig:FSExampleR}}\hfill
\subfigure[]
{\includegraphics[scale=0.50]{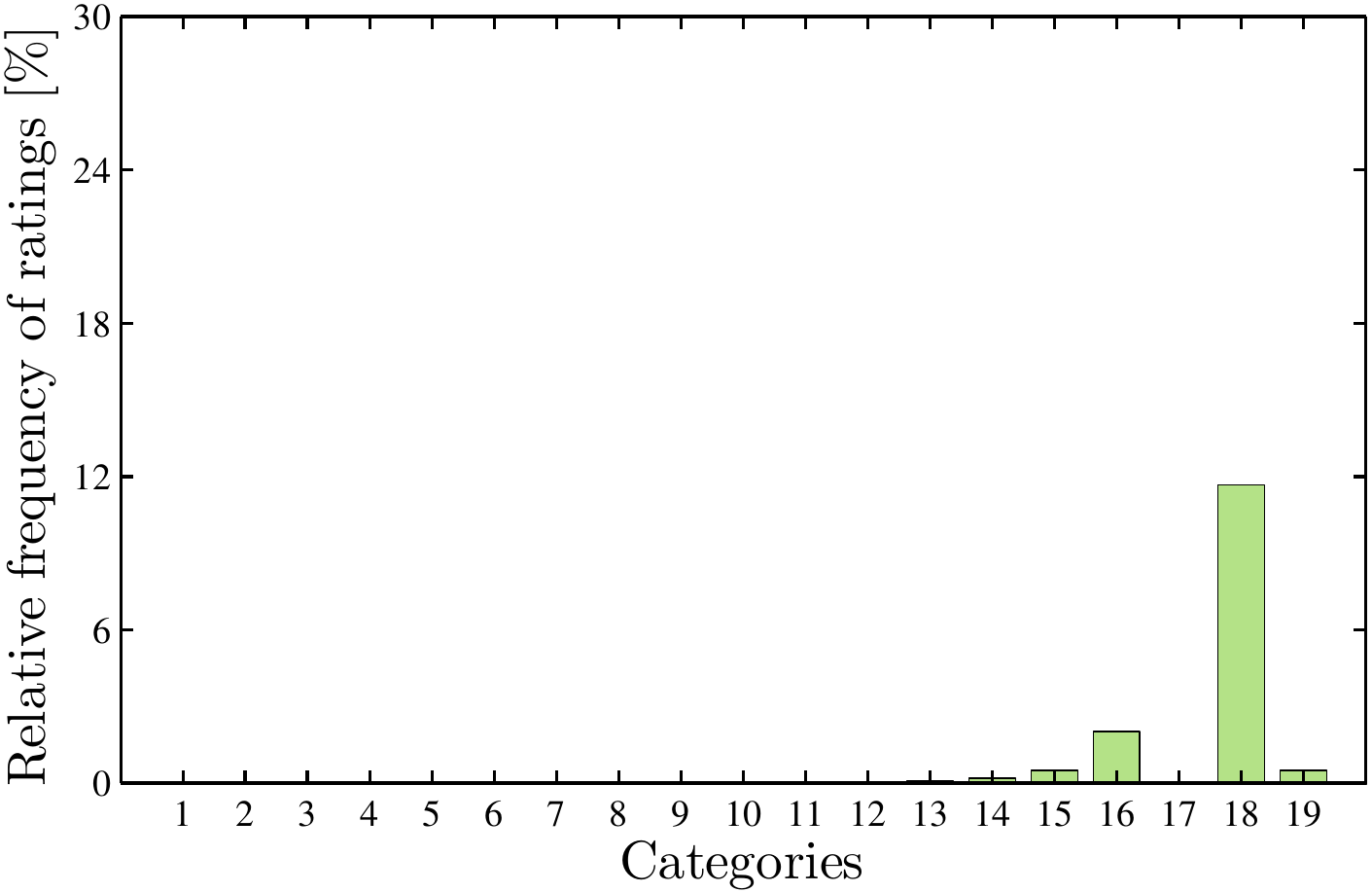}%
\label{fig:FSExampleS}}\hspace*{\fill}
\caption{In this figure we represent~(a) the item distribution $q$ of a particular user as well as the population's item distribution $p$.
In addition, we plot~(b) the optimal forgery strategy $r^*$ and~(c) the optimal suppression strategy $s^*$ that the user in question should adopt when they specify $\sigma=0.150$ and $\rho = \rhocrit(\sigma) \simeq 0.180$.}
\label{fig:FSExampleStrateg}
\end{figure*}

Fig.~\ref{fig:FSQvsP} depicts the user profile and the population profile, the latter being computed by averaging across the 69\,878 users.
From this figure we note that the user's interest far exceeds the population's in categories such as \emph{musical}, \emph{romance}, \emph{IMAX}, \emph{drama} and \emph{documentary}.
More precisely, such ratios $\tfrac{q_k}{p_k}$ yield
$$\left(\frac{q_k}{p_k}\right)_{k=15,\ldots, 19} \simeq (1.300, 1.306, 1.451, 1.728, 2.292).$$
In this figure, we also observe that the user's interest and the population's in the category 17 are nearly zero,
namely $q_{17} \simeq 0.0005$ and $p_{17} \simeq 0.0003$.

On the other hand, Fig.~\ref{fig:FSQvsP} indicates that the user shows little interest, compared to the population's preferences,
in categories such as \emph{animation}, \emph{action}, \emph{film\hyph noir} or \emph{children's}, to name just a few.
Specifically, the first six smallest ratios $\tfrac{q_k}{p_k}$ yield
$$\left(\frac{q_k}{p_k}\right)_{k=1,\ldots, 6} \simeq ( 0.444, 0.599, 0.651, 0.691, 0.705, 0.714).$$

%%%%%%%%%%%%%%%%%%%%%%%%%%%%%%%%%%%%%%% Q, P, R* and S*  %%%%%%%%%%%%%%%%%%%%%%%%%%%%%%%%%%
%%%%%%%%%%%%%%%%%%%%%%%%%%%%%%%%%%%%%%%%%%%%%%%%%%%%%%%%%%%%%%%%%%%%%%%%%%%%%%%%%%

Figs.~\ref{fig:FSExampleR} and~\ref{fig:FSExampleS} show the optimal forgery and suppression strategies that this particular user should apply,
in the case when $\sigma=0.150$ and $\rhocrit(\sigma) \simeq 0.180$.
The solutions plotted in these figures are consistent with our two previous observations---the optimal forgery strategy recommends that the user submit false ratings to
movies falling into the categories where the ratio $\tfrac{q_k}{p_k}$ is low;
and the optimal suppression strategy suggests that the user refrain from rating movies belonging to categories where the ratio~$\tfrac{q_k}{p_k}$ is high.
Just as an example, the fact that $s^*_{17} \simeq 0.0001$ means that the user at hand should eliminate one in five ratings to movies classified as \emph{IMAX}.

%%%%%%%%%%%%%%%%%%%%%%%%%%%%%%%%%%%%%%% Trade-off particular user  %%%%%%%%%%%%%%%%%%%%%%%%%%%%%
\begin{figure}[!hbt]
\centering
\includegraphics[scale=0.52]{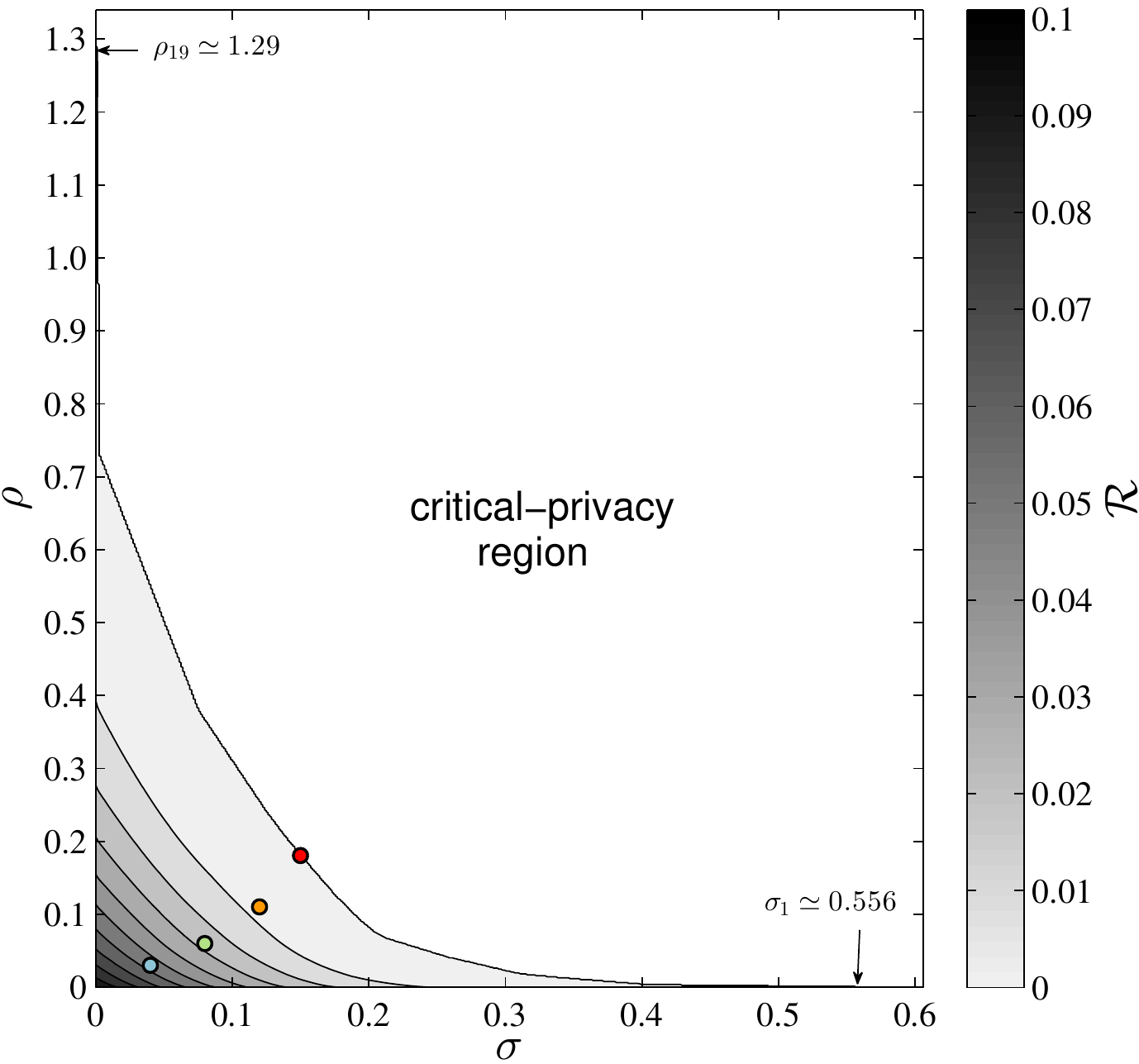}
\caption{Optimal trade\hyph off surface among privacy, forgery rate and suppression rate for one particular user in our data set.
The four points shown in this figure correspond to the pairs of values $(\rho,\sigma)$ that we used to show the proportionality relationship between $t^*$ and $p$ in Fig.~\ref{fig:ProportionalityML}.}
\label{fig:FSTradeOff_ML}
\end{figure}
%%%%%%%%%%%%%%%%%%%%%%%%%%%%%%%%%%%%%%%%%%%%%%%%%%%%%%%%%%%%%%%%%%%%%%%%%%%%%%%%%%

The optimal trade\hyph off surface among privacy, forgery rate and suppression rate is represented in Fig.~\ref{fig:FSTradeOff_ML}.
In this figure we plot the contour levels of the function $\cR(\rho,\sigma)$, which we computed theoretically.
The initial privacy risk is $\cR(0,0)\simeq 0.101$ and the arithmetic mean between the ratios $\frac{q_{1}}{p_{1}}$ and $\frac{q_{19}}{p_{19}}$ yields approximately 1.37.
Since the mean is higher than 1, Corollary~\ref{Corollary:PureStrategies} tells us that the user should opt for suppression as pure strategy, in lieu of forgery.
This is under the assumption that they wish to achieve the minimum privacy risk and do not have any preference for any of the pure strategies.
Nevertheless, the fact that $\delta_{\rho} \simeq 12.6 > \delta_{\sigma} \simeq 10.9$ leads us to choose forgery as pure strategy for $\rho,\sigma \simeq 0$.
When both strategies are combined, note that a forgery and suppression rate of just 0.1\% leads to a relative reduction in privacy risk of 2.35\%,
on account of the first\hyph order Taylor approximation derived in Sec.~\ref{sec:Theory:LowRates}.

\begin{figure*}[!tb]%
\centering\hspace*{13pt}
\subfigure[$\rho= 0.03$, {$\rho\in [\rho_6, \rho_7]$}, $\sigma=0.04$, {$\sigma\in [\sigma_{18}, \sigma_{17}]$}, $\rho/\rhocrit(\sigma) \simeq 0.055$,
$\cR(\rho,\sigma)\simeq 0.055$, $\cR(\rho,\sigma)/\cR_0 \simeq 0.549$.]%
{\includegraphics[scale=\FigScaleMultipleBiggest]{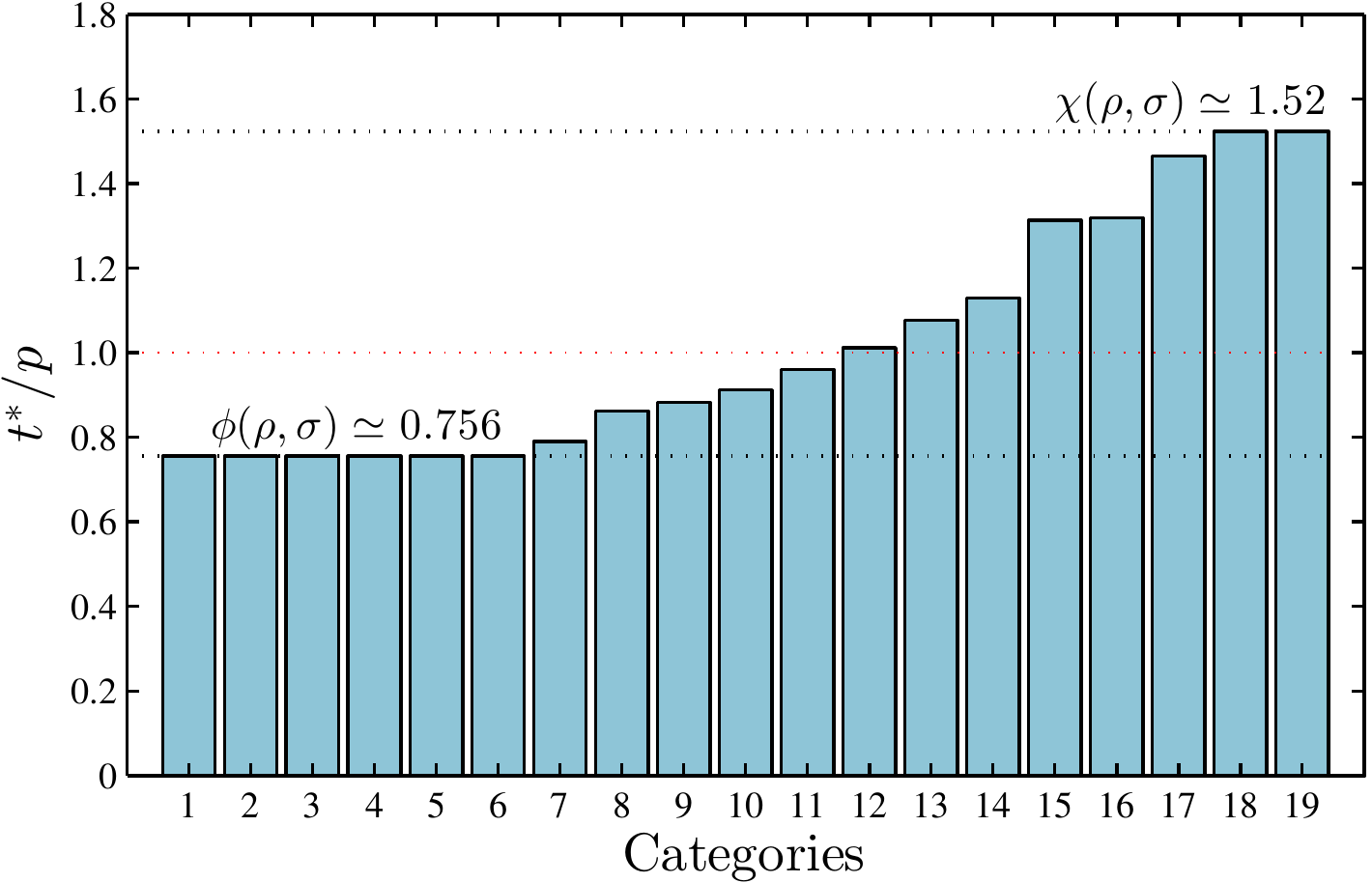}%
\label{fig:SubPropor1}}\hfill
\subfigure[$\rho= 0.06$, {$\rho\in [\rho_{7}, \rho_{8}]$}, $\sigma=0.08$, {$\sigma\in [\sigma_{15}, \sigma_{14}]$}, $\rho/\rhocrit(\sigma) \simeq 0.164$,
$\cR(\rho,\sigma)\simeq 0.026$, $\cR(\rho,\sigma)/\cR_0 \simeq 0.259$.]%
{\includegraphics[scale=\FigScaleMultipleBiggest]{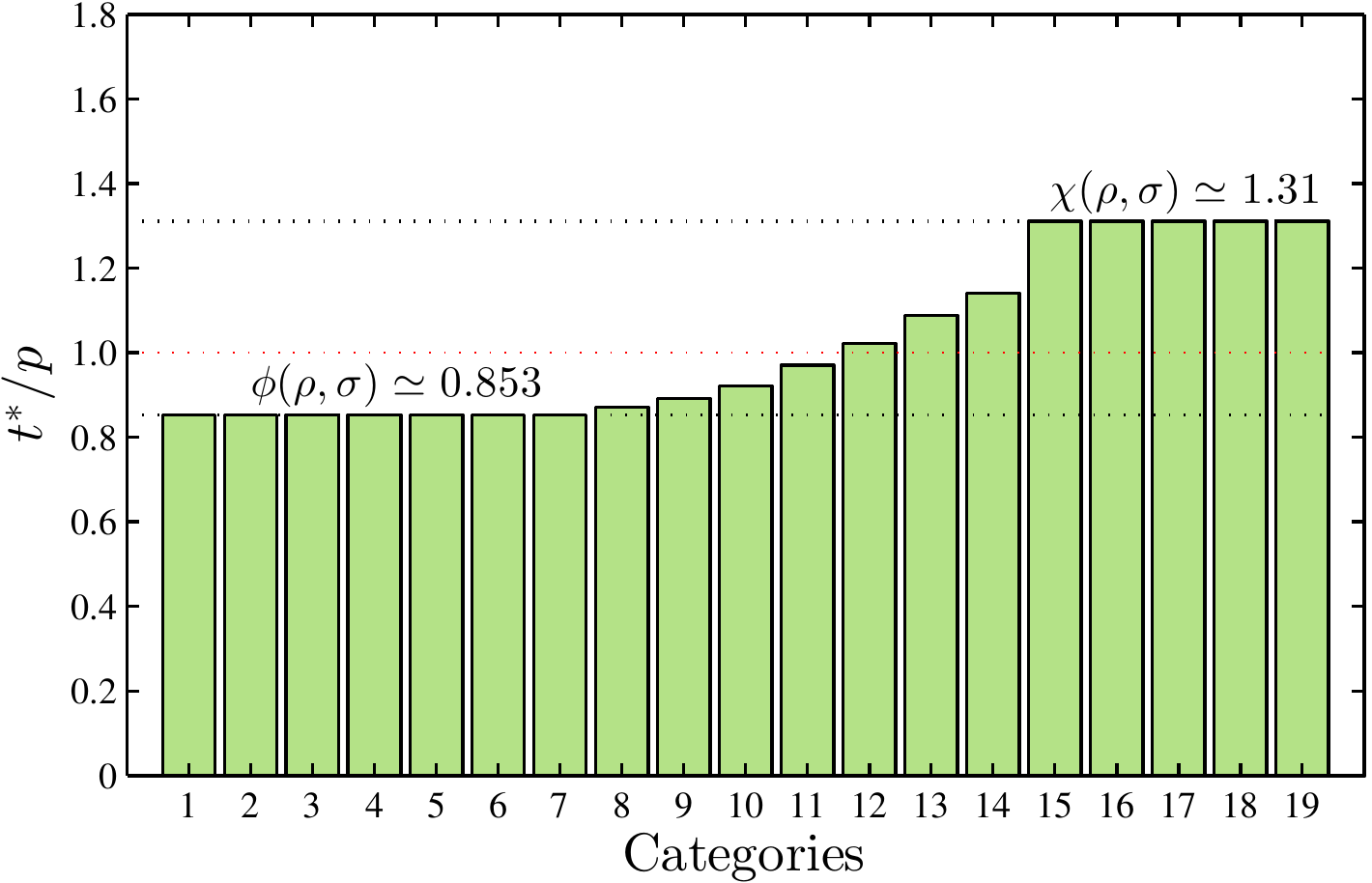}%
\label{fig:SubPropor2}}\hspace*{\fill}
\\
\hspace*{13pt}
\subfigure[$\rho= 0.11$, {$\rho\in [\rho_{10}, \rho_{11}]$}, $\sigma=0.12$, {$\sigma\in [\sigma_{15}, \sigma_{14}]$}, $\rho/\rhocrit(\sigma) \simeq 0.434$,
$\cR(\rho,\sigma)\simeq 0.006$, $\cR(\rho,\sigma)/\cR_0 \simeq 0.061$.]%
{\includegraphics[scale=\FigScaleMultipleBiggest]{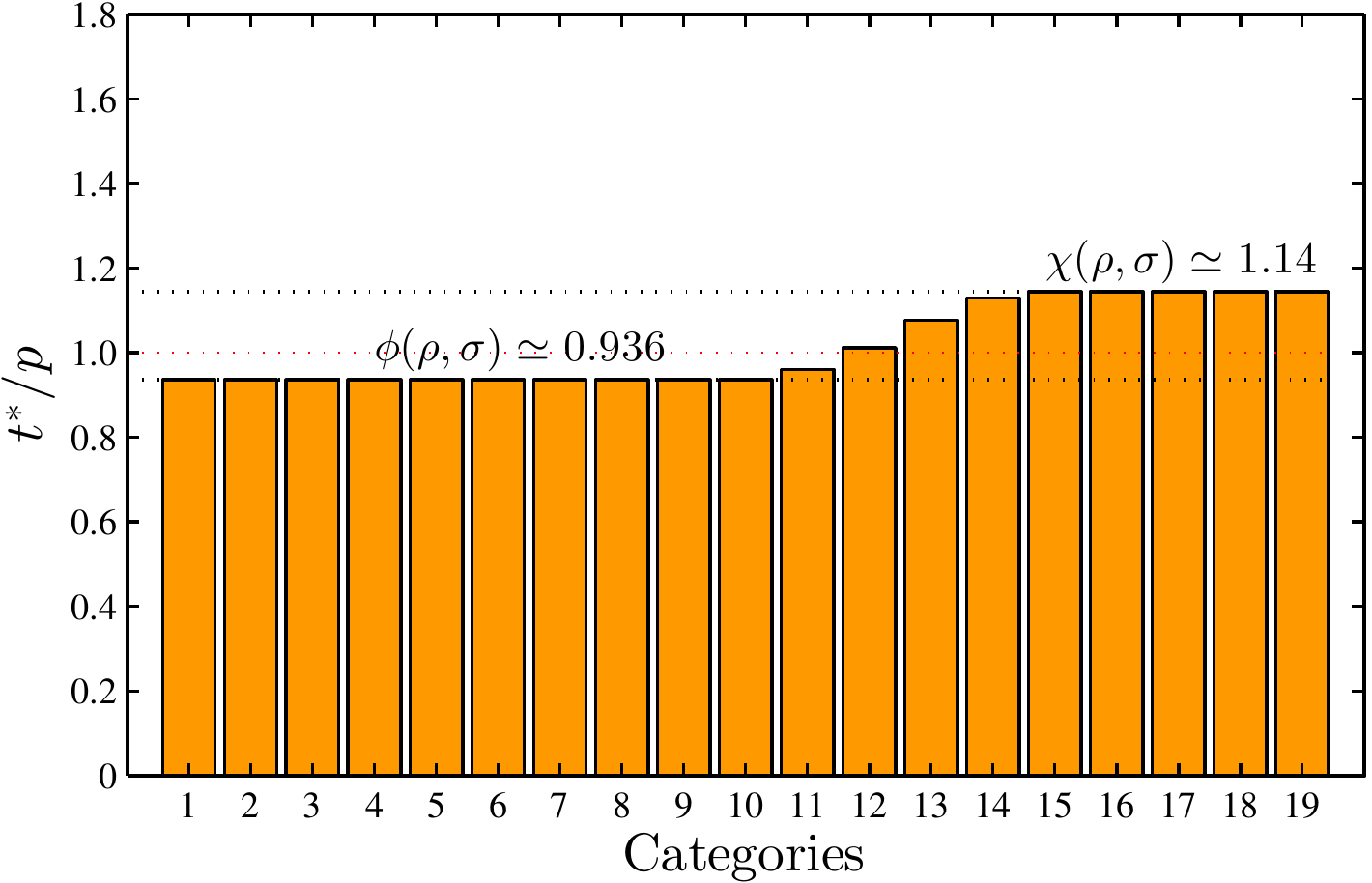}%
\label{fig:SubPropor3}}\hfill
\subfigure[$\rho \simeq 0.180$, {$\rho\in [\rho_{12}, \rho_{13}]$}, $\sigma=0.15$, {$\sigma\in [\sigma_{13}, \sigma_{12}]$}, $\rho/\rhocrit(\sigma) = 1$,
$\cR(\rho,\sigma)= 0$, $\cR(\rho,\sigma)/\cR_0 = 0$.]%
{\includegraphics[scale=\FigScaleMultipleBiggest]{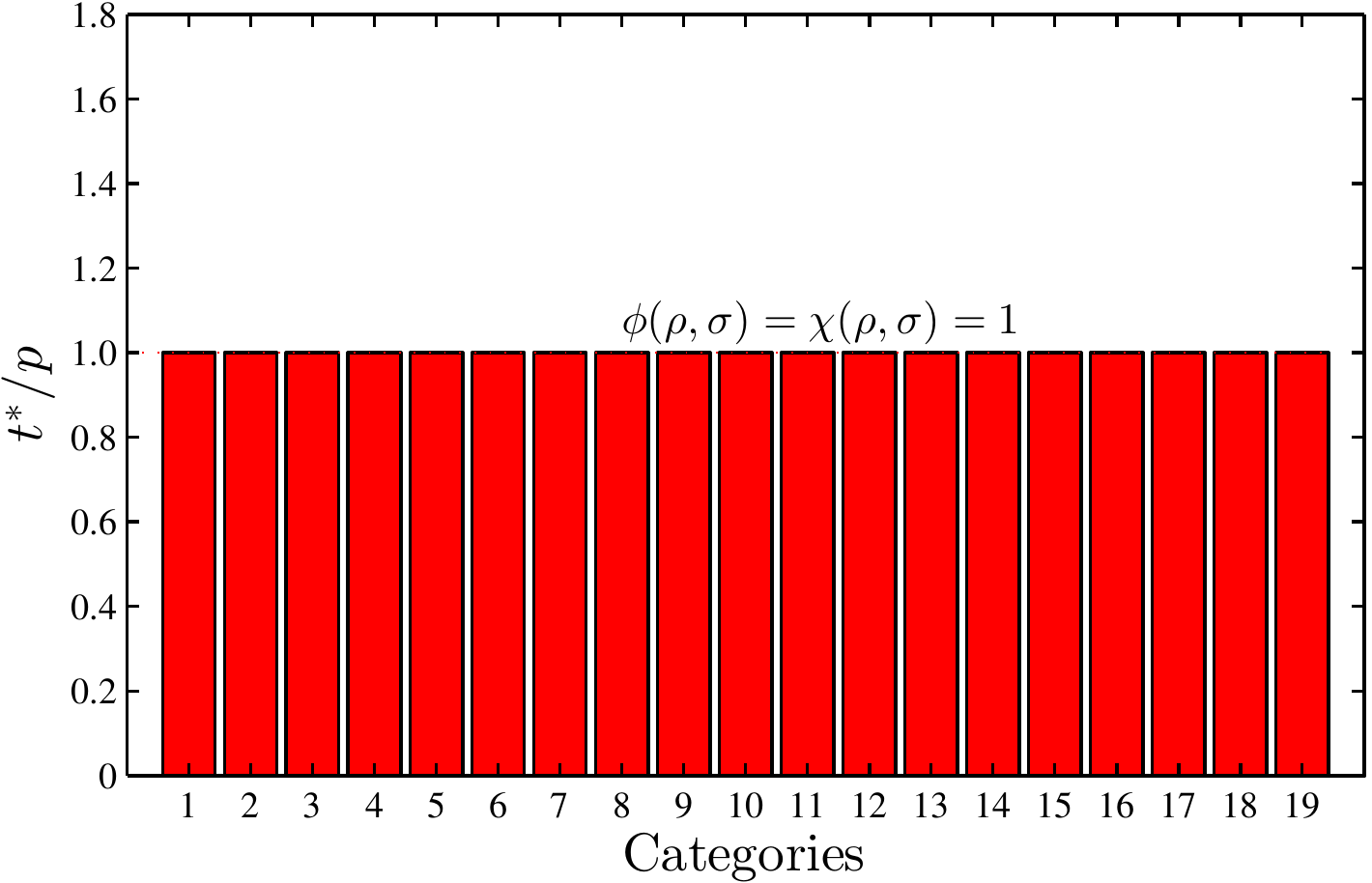}%
\label{fig:SubPropor4}}\hspace*{\fill}
\caption{Proportionality relationship between, on the one hand, the optimal apparent item distribution $t^*$ of the user identified as \texttt{3301} in our data set,
and on the other, the population's item distribution~$p$.}
\label{fig:ProportionalityML}
\end{figure*}

In Fig.~\ref{fig:FSTradeOff_ML} we have also plotted 4 points, which correspond to the following pairs of values $(\rho,\sigma)$:
$(0.03, 0.04)$, $(0.06, 0.08)$, $(0.11, 0.12)$ and $(0.18, 0.15)$.
For each of these pairs, we have represented the quotient $\frac{t^*_k}{p_k}$ in Fig.~\ref{fig:ProportionalityML}.
The aim is to show how the optimal apparent profile becomes proportional to the population's distribution, as the user approaches the critical\hyph privacy region.
Fig.~\ref{fig:SubPropor1} considers the first pair of values.
Here, $\rho$ and $\sigma$ fall into the intervals $[\rho_6, \rho_7]$ and $[\sigma_{18}, \sigma_{17}]$, respectively.
Consistently with Proposition~\ref{Proposition:Proportionality},
we check that $\frac{t^*_1}{p_1} = \cdots = \frac{t^*_6}{p_6} \simeq 0.756$
and that $\frac{t^*_{18}}{p_{18}} = \frac{t^*_{19}}{p_{19}} \simeq 1.52$.

In Fig.~\ref{fig:SubPropor2} we double the rates of forgery and suppression.
On the one hand, this leads to $\frac{t^*_1}{p_1} = \cdots = \frac{t^*_7}{p_7}$.
On the other, the fact that $\sigma \in [\sigma_{15}, \sigma_{14}]$ implies that
$\frac{t^*_{15}}{p_{15}} = \cdots = \frac{t^*_{19}}{p_{19}}$.
It is also interesting to note that, for these relatively small values of $\rho$ and $\sigma$, the final privacy risk is 26\% of the initial value $\oD(q\,\|\,p)$.

As $\rho$ and $\sigma$ increase, so does the function $\phi$. The contrary happens with the function $\chi$, which decreases with both rates.
In Fig.~\ref{fig:SubPropor3}, for example, the proportionality relationship between $t^*$ and $p$ holds for all except 4 categories.
The last pair $(\rho,\sigma) \simeq (0.18,0.15)$ lies at the boundary of $\cC$, as shown in Fig.~\ref{fig:FSTradeOff_ML}.
This implies that $\frac{t^*}{p}=1$ and therefore that $\cR(\rho,\sigma)=0$, as captured in Fig.~\ref{fig:SubPropor4}.

%%%%%%%%%%%%%%%%%%%%%%%%%%%%%%%%%%%%% PROPORTIONALITY  %%%%%%%%%%%%%%%%%%%%%%%%%%%%%%%%%
%%%%%%%%%%%%%%%%%%%%%%%%%%%%%%%%%%%%%%%%%%%%%%%%%%%%%%%%%%%%%%%%%%%%%%%%%%%%%%%%%%

\begin{figure*}[!htb]%
\centering\hspace*{13pt}
\subfigure[10\textsuperscript{th} percentile.]%
{\includegraphics[scale=\FigScaleMultipleBigger]{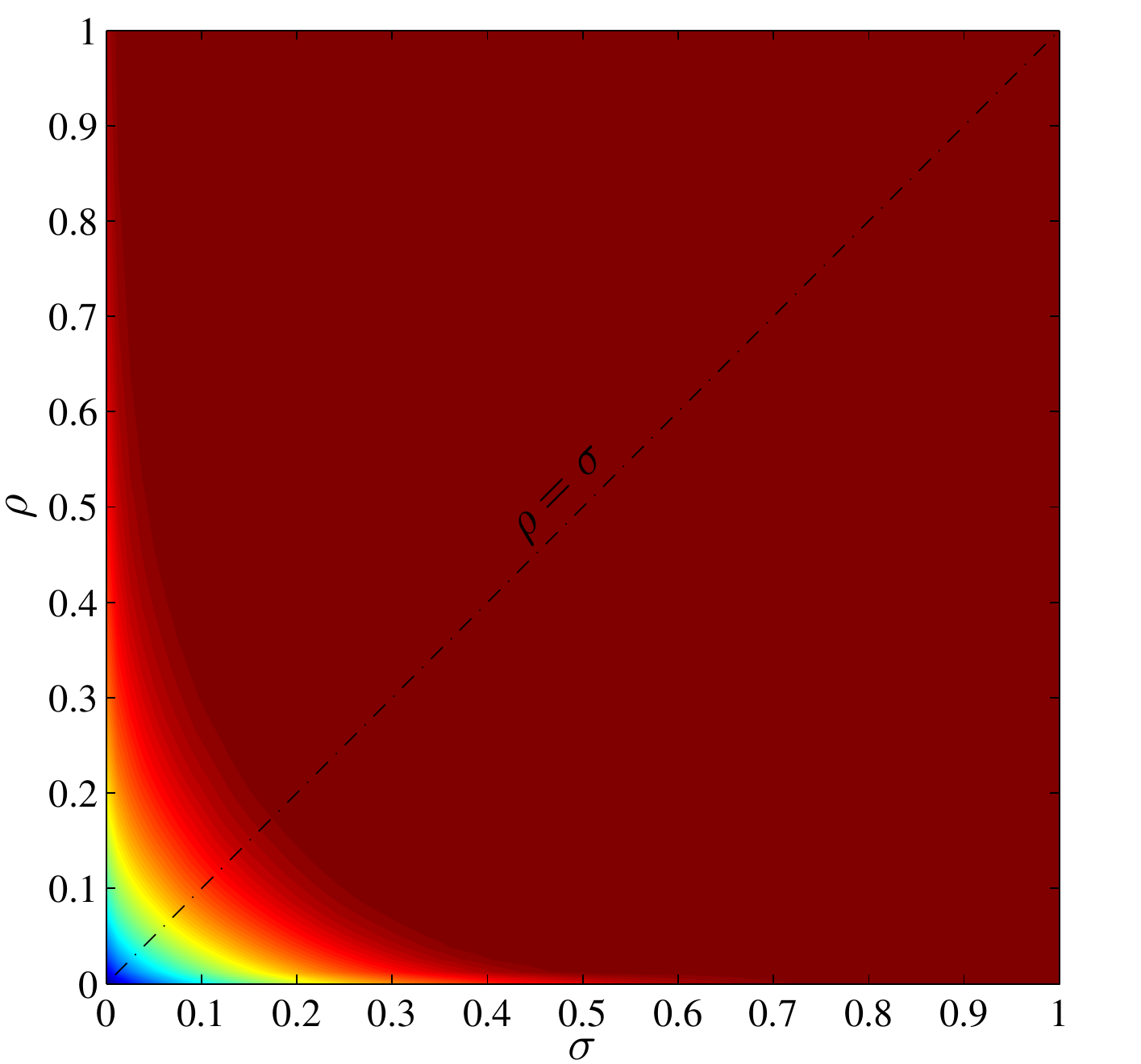}%
\label{fig:FSPercentile10}}\hfill
\subfigure[50\textsuperscript{th} percentile.]%
{\includegraphics[scale=\FigScaleMultipleBigger]{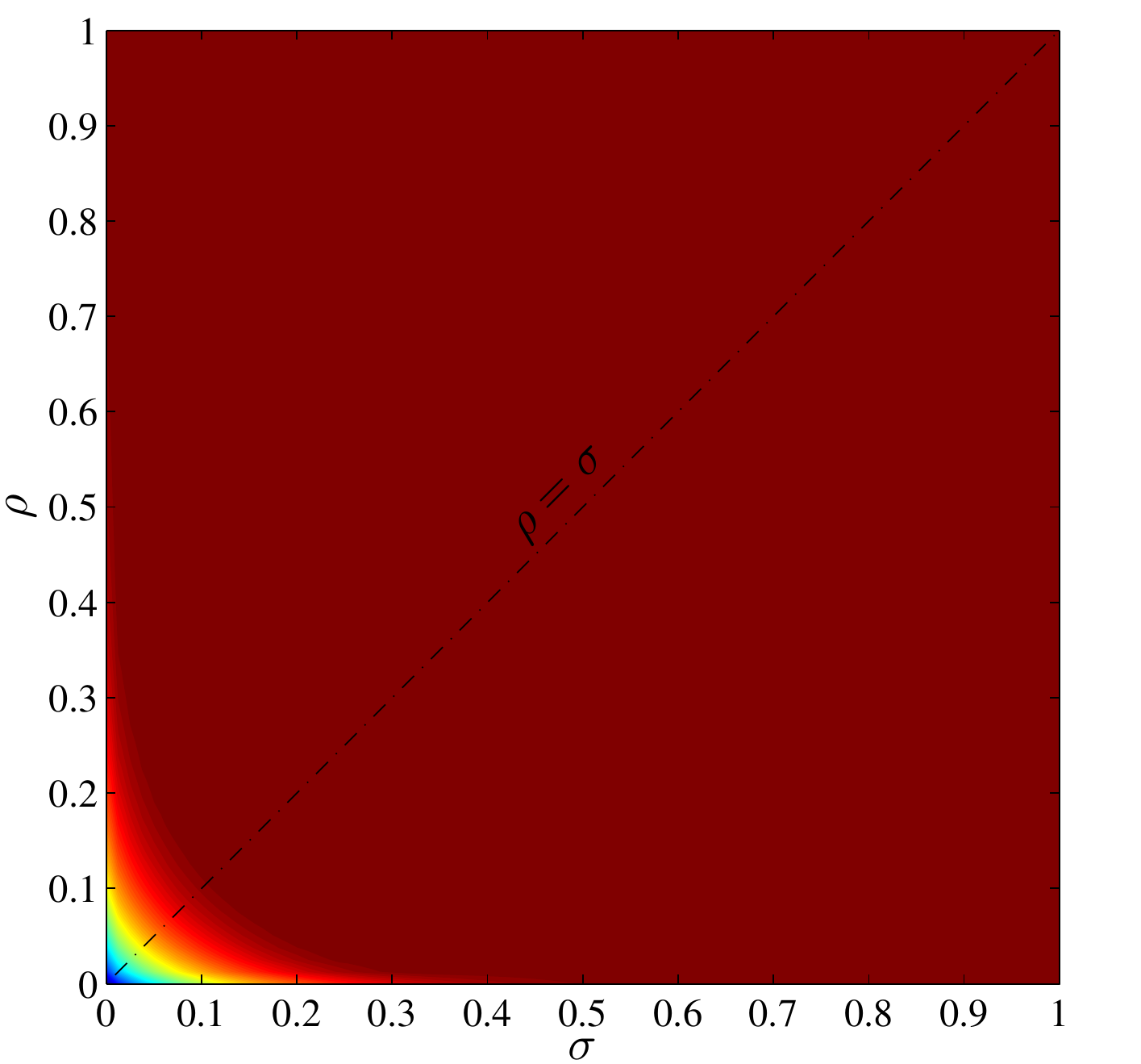}%
\label{fig:FSPercentile50}}\hspace*{1pt}
\\
\hspace*{13pt}
\subfigure[90\textsuperscript{th} percentile.]%
{\includegraphics[scale=\FigScaleMultipleBigger]{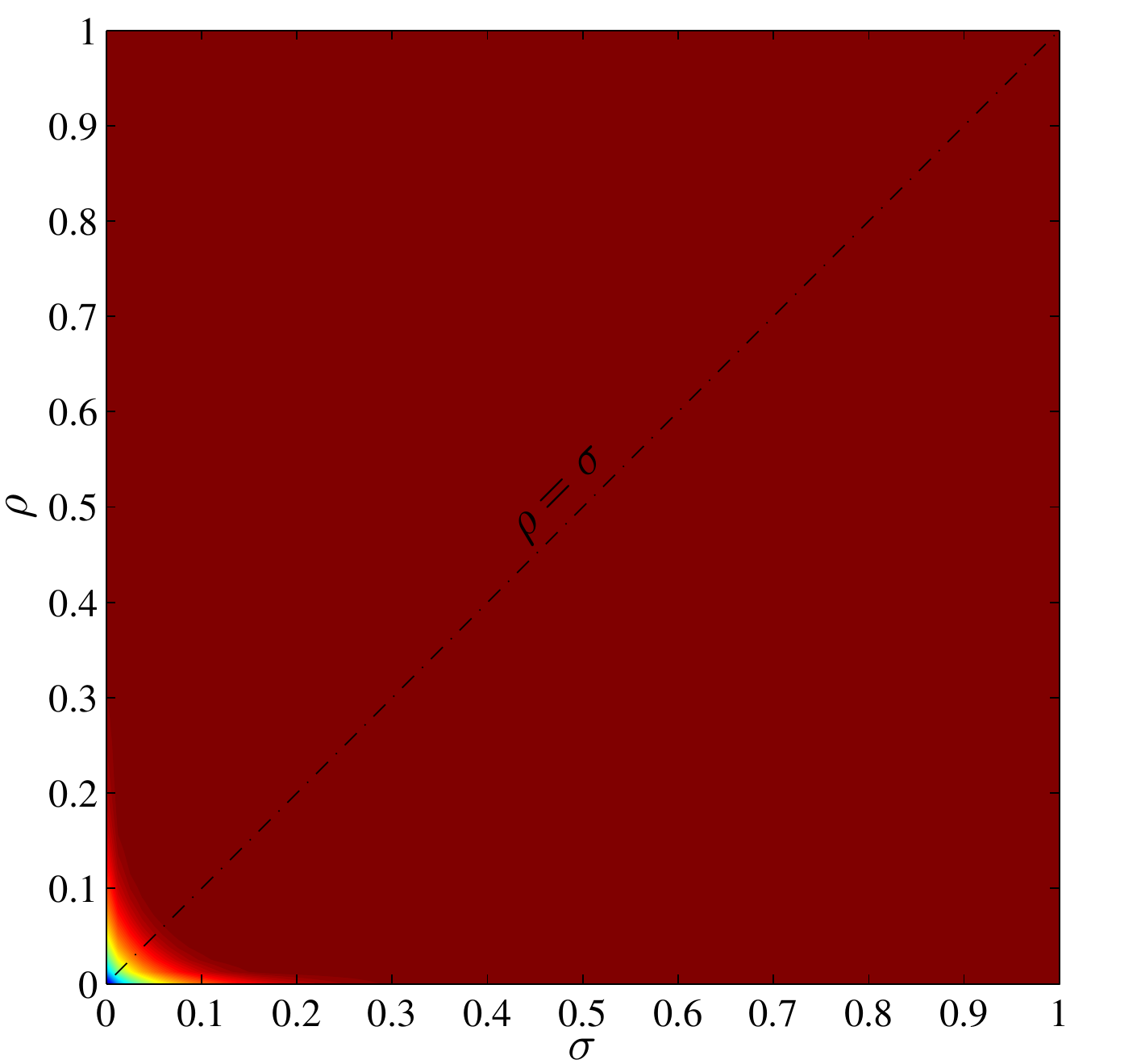}%
\label{fig:FSPercentile90}}
\subfigure
{\includegraphics[scale=\FigScaleMultipleBigger]{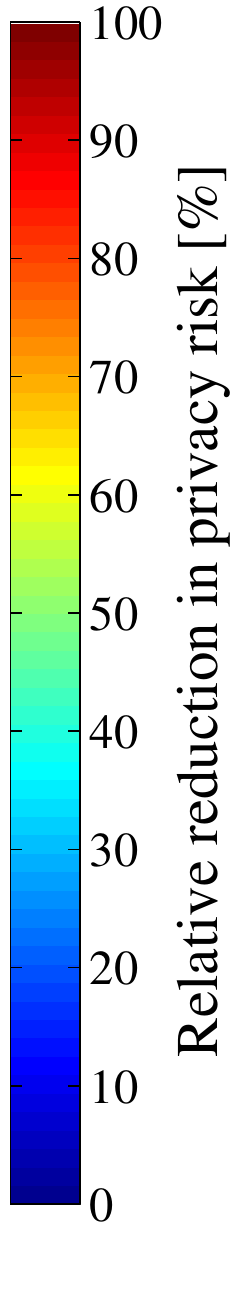}}\hspace*{1pt}
\label{fig:FSPercentileScale}
\caption{We assume that the 4\,099 users satisfying the positivity assumption~\eqref{eqn:PositivityAssumption}
protect their privacy by using a common forgery rate and a common suppression rate.
Under this assumption, we plot some percentiles surfaces of relative reduction in privacy risk, against these two common rates.}
\label{fig:FSPercentile}
\end{figure*}

Having examined the case of a specific user,
in our next series of experiments we evaluate the privacy\hyph protection level that users can achieve if they are disposed to forge and eliminate a fraction of their ratings.
For simplicity, we suppose that all users satisfying the positivity assumption~\eqref{eqn:PositivityAssumption} apply a common forgery rate and a common suppression rate.
Fig.~\ref{fig:FSPercentile} depicts the contours of the 10\textsuperscript{th}, 50\textsuperscript{th} and 90\textsuperscript{th}
percentile surfaces of relative reduction in privacy risk, for different values of $\rho$ and~$\sigma$.
Two conclusions can be drawn from this figure.
\begin{itemize}
  \item [$\bullet$] First, for relatively small values of $\rho$ and $\sigma$ (lower than 15\%), a vast majority of users lowered privacy risk significantly.
In quantitative terms, we observe in Fig.~\ref{fig:FSPercentile10} that, for $\rho =\sigma = 0.05$, the 10\% of users adhered to our technique
obtained a reduction in privacy risk by at least 52.4\%.
For those same rates of forgery and suppression rates, the 50\textsuperscript{th} and 90\textsuperscript{th} percentiles are 73.9\% and 94.8\%.
For higher rates, e.g., $\rho=\sigma=0.15$, Fig.~\ref{fig:FSPercentile50} highlights that half of users experienced a reduction in privacy risk
less than or equal to 100\%.
  \item [$\bullet$] Secondly, the three percentile surfaces exhibit a certain symmetry with respect to the line $\rho=\sigma$.
If this symmetry were exact, the exchange of the rates of forgery and suppression would not have any impact on the resulting privacy\hyph protection achieved.
However, this is not the case. For example, Fig.~\ref{fig:FSPercentile10} shows a lower reduction in privacy risk for $\rho<\sigma$,
particularly accentuated when~$\sigma\simeq 0$. The reason for this may be found in the fact that, for most users, $\rho_n$ is greater than $\sigma_1$.
We shall elaborate more on this later on when we consider forgery and suppression as pure strategies.
\end{itemize}

%%%%%%%%%%%%%%%%%%%%%%%%%%%%%%%%%%%%%% PERCENTILES  %%%%%%%%%%%%%%%%%%%%%%%%%%%%%%%%%%
%%%%%%%%%%%%%%%%%%%%%%%%%%%%%%%%%%%%%%%%%%%%%%%%%%%%%%%%%%%%%%%%%%%%%%%%%%%%%%%%%

\begin{figure*}[htb]%
\centering\hspace*{\fill}
\subfigure%
{\includegraphics[scale=\FigScaleMultipleBiggest]{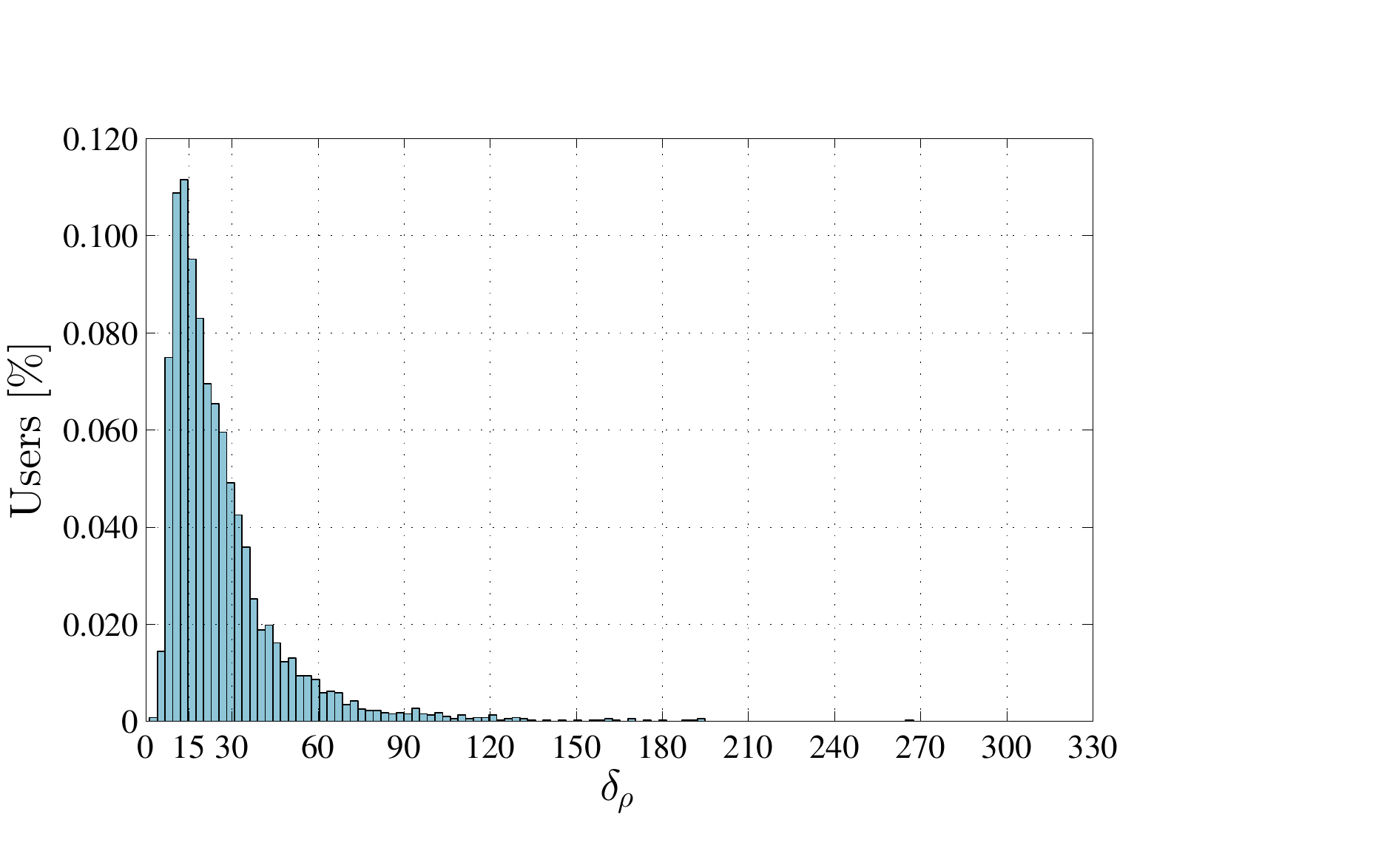}%
\label{fig:deltarho}}\hfill
\subfigure%
{\includegraphics[scale=\FigScaleMultipleBiggest]{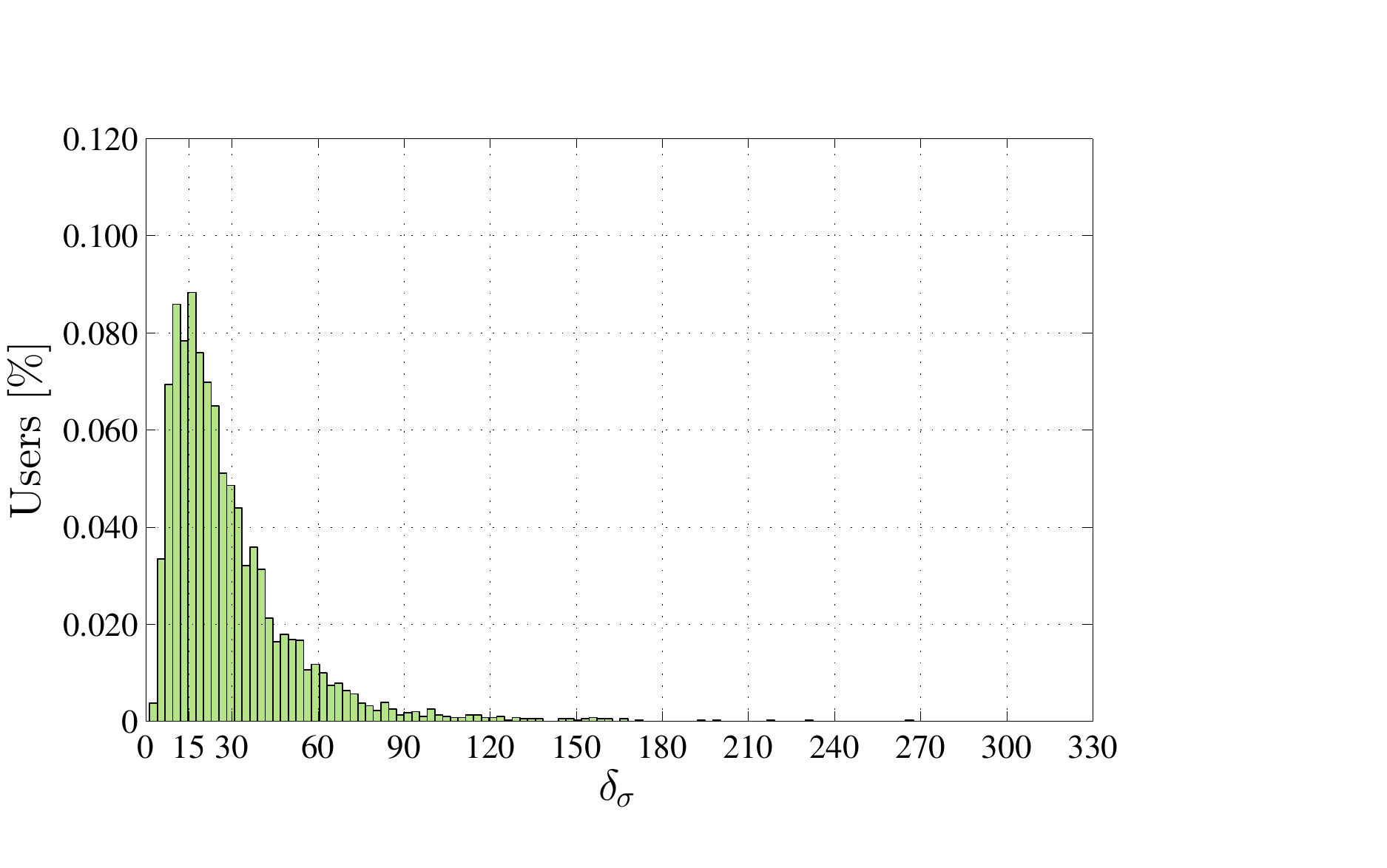}%
\label{fig:deltasigma}}\hspace*{\fill}
\caption{Probability distribution of the relative decrement factors of forgery and suppression.}
\label{fig:deltarhosigma}
\end{figure*}

Next, we analyze the privacy protection provided by our technique for $\rho,\sigma \simeq 0$.
In the theoretical analysis conducted in Sec.~\ref{sec:Theory:LowRates} we derived an expression for the relative reduction in privacy risk at low rates.
Particularly, said expression was in terms of two factors, namely $\delta_{\rho}$ and $\delta_{\sigma}$.
In Fig.~\ref{fig:deltarhosigma} we show the probability distribution of these factors.
Consistently with Proposition~\ref{Proposition:DecrementFactors},
the minimum values of these factors are $\delta_{\rho} \simeq  3.12>1$ and $\delta_{\sigma} \simeq 2.30 >0.$
The maximum values attained by these forgery and the suppression factors are approximately $324.98$ and $266.13$.
On the other hand, in favour of suppression is the fact that the percentage of users with $\delta_{\rho} \geqslant 30$ is lower than those users with $\delta_{\sigma} \geqslant 30$.
More precisely, these percentages yield 26.8\% and 33.1\%, respectively.
In the end, an eye\hyph opening finding is that $\delta_{\rho}>\delta_{\sigma}$ in 43.45\% of users, which suggests introducing a suppression rate higher than that of forgery,
at least at low rates.

%%%%%%%%%%%%%%%%%%%%%%%%%%%%%%% DELTA RHO AND DELTA SIGMA %%%%%%%%%%%%%%%%%%%%%%%%%%%%%%%%
%% Opción 1
%%%%%%%%%%%%%%%%%%%%%%%%%%%%%%%%%%%%%%%%%%%%%%%%%%%%%%%%%%%%%%%%%%%%%%%%%%%%%%%%%%

\begin{figure*}[!tb]%
\centering\hspace*{\fill}
\subfigure%
{\includegraphics[scale=\FigScaleMultipleBiggest]{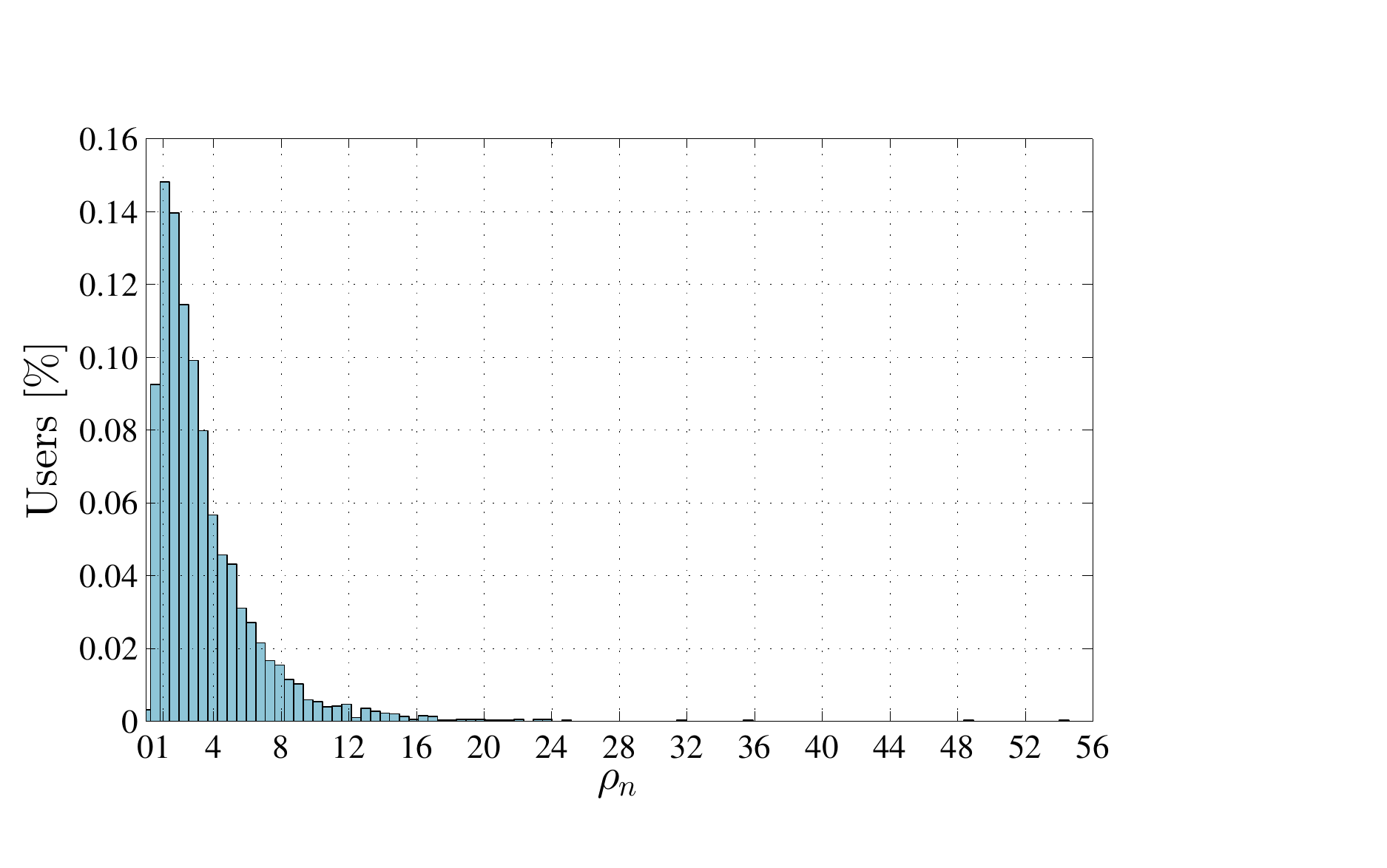}%
\label{fig:RhoN}}\hfill
\subfigure%
{\includegraphics[scale=\FigScaleMultipleBiggest]{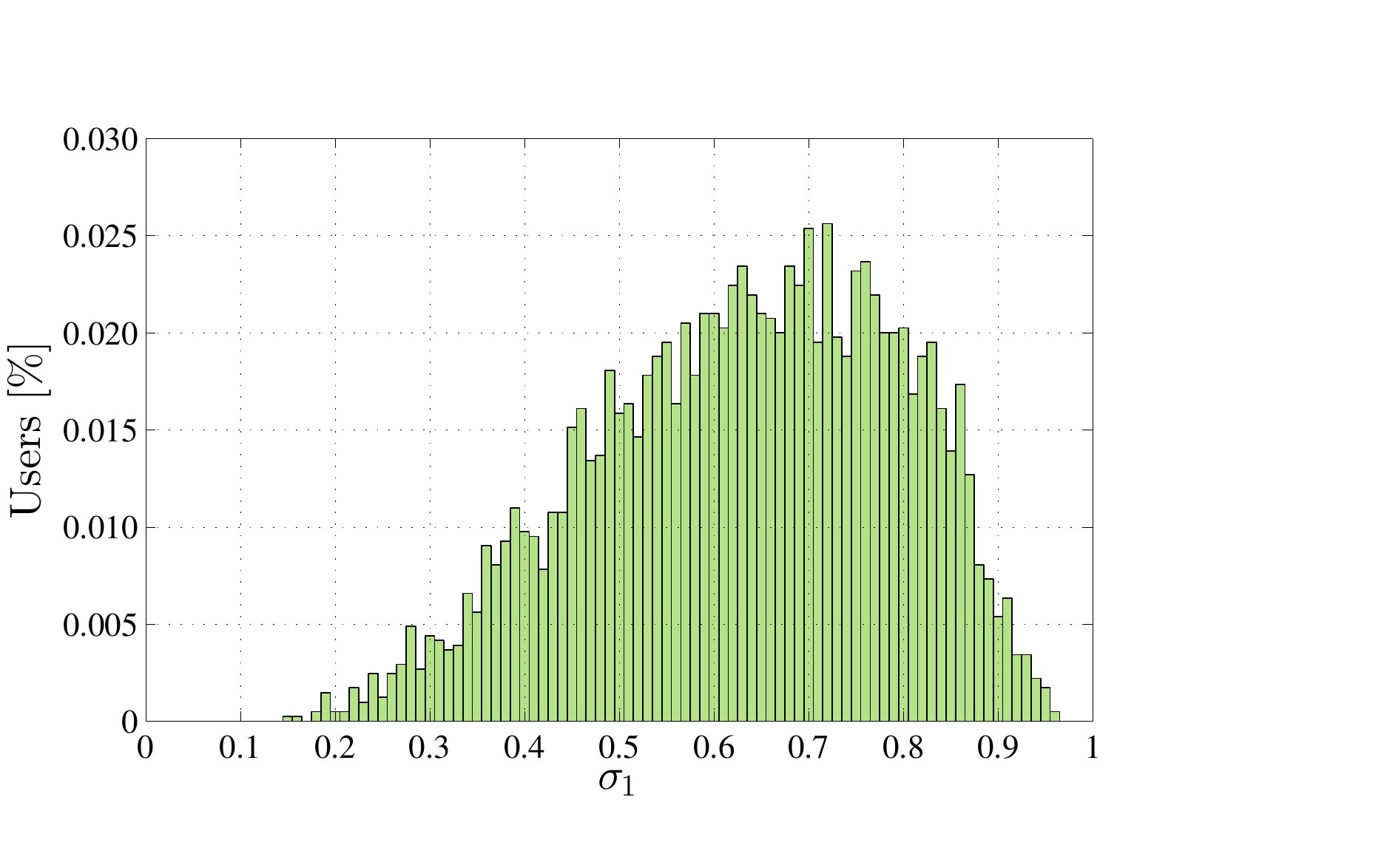}%
\label{fig:Sigma1}}\hspace*{\fill}
\caption{Probability distribution of the critical forgery and suppression rates.}
\label{fig:RhoNSigma1}
\end{figure*}

After analyzing the forgery and the suppression of ratings as a mixed strategy,
our last experimental results contemplate the application of forgery and suppression as pure strategies.
In Fig.~\ref{fig:RhoNSigma1} we illustrate the probability distribution of the critical rates $\rho_n$ and $\sigma_1$.
The critical forgery rate ranges approximately from $0.171$ to $54.18$, and its average is $3.45$.
The critical suppression rate, on the other hand, goes from $0.153$ to $0.963$, and its average is $0.632$.
These figures indicate that, on average, a user will have either to refrain from rating an item six out of ten times,
or submit nearly 3.45 false ratings per each original rating.
This is, of course, when the user wishes to reach the critical\hyph privacy region.
Bearing these figures in mind, it is not surprising then that 95.3\% of the users in our data set would opt for suppression as pure strategy,
as it comes at the cost of a lower impact on utility.

%%%%%%%%%%%%%%%%%%%%%%%%%%%%%%%%%%%%%% RhoN AND Sigma1 %%%%%%%%%%%%%%%%%%%%%%%%%%%%%%%%%
% Opción 1
%%%%%%%%%%%%%%%%%%%%%%%%%%%%%%%%%%%%%%%%%%%%%%%%%%%%%%%%%%%%%%%%%%%%%%%%%%%%%%%%%%

%%%%%%%%%%%%%%%%%%%%%%%%%%%%%%%%%%%%%%%%%%%%%%%%%%%%%%%%%%%%%%%%%%%%%%%%%%%%%%%%%%%%%%%%%%%%%%%%%%%%%%%%%%%%%%%%%%%%%%%%%%%%%%%%%%
%%%%%%%%%%%%%%%%%%%%%%%%%%%%%%%%%%%%%%%%%%%%%%%%%%%%%%%%%%%%%%%%%%%%%%%%%%%%%%%%%%%%%%%%%%%%%%%%%%%%%%%%%%%%%%%%%%%%%%%%%%%%%%%%%
%%%%%%%%%%%%%%%%%%%%%%%%%%%%%%%%%%%%%%%%%%%%%%%%%%%%%%%%%%%%%%%%%%%%%%%%%%%%%%%%%%%%%%%%%%%%%%%%%%%%%%%%%%%%%%%%%%%%%%%%%%%%%%%%%%
%6 CONCLUSION
%%%%%%%%%%%%%%%%%%%%%%%%%%%%%%%%%%%%%%%%%%%%%%%%%%%%%%%%%%%%%%%%%%%%%%%%%%%%%%%%%%%%%%%%%%%%%%%%%%%%%%%%%%%%%%%%%%%%%%%%%%%%%%%%%%
%%%%%%%%%%%%%%%%%%%%%%%%%%%%%%%%%%%%%%%%%%%%%%%%%%%%%%%%%%%%%%%%%%%%%%%%%%%%%%%%%%%%%%%%%%%%%%%%%%%%%%%%%%%%%%%%%%%%%%%%%%%%%%%%%%
%%%%%%%%%%%%%%%%%%%%%%%%%%%%%%%%%%%%%%%%%%%%%%%%%%%%%%%%%%%%%%%%%%%%%%%%%%%%%%%%%%%%%%%%%%%%%%%%%%%%%%%%%%%%%%%%%%%%%%%%%%%%%%%%%%
\section{Conclusion}
\label{sec:Conclusion}
\noindent
In the literature of recommendation systems there exists a variety of approaches aimed at protecting user privacy.
Among these approaches, the forgery and the suppression of ratings emerge as a technique that may hinder attackers in their efforts to accurately profile users on the basis of the items they rate.
Our technique
does not require that users trust neither the recommender nor the network operator,
it is simple in terms of infrastructure requirements,
and it can be used in combination with other approaches providing soft privacy.
However, as any data\hyph perturbative approach, our privacy\hyph enhancing technology comes at the expense of a loss in data utility,
in particular a degradation of the quality of the recommender's predictions.
Put another way, it poses a trade\hyph off between privacy and utility.

The objective of this paper is to investigate mathematically said trade\hyph off.
For this purpose, first we propose a quantitative measure of both privacy and utility.
We quantify privacy risk as the KL divergence between the user's rating distribution and the population's,
and measure utility as the fraction of ratings the user is willing to forge and suppress.
With these two quantities, we formulate a multiobjective optimization problem characterizing the trade\hyph off between privacy risk on the one hand,
and on the other forgery rate and suppression rate.

Our theoretical analysis provides a closed\hyph form solution to this problem and characterizes the optimal trade\hyph off surface between privacy and utility.
The solution is confined to the closure of the noncritical\hyph privacy region.
The interior of the critical\hyph privacy region is of no interest as the privacy risk attains its minimum value at the boundary of $\bar{\cC}$.
In the region of interest, our analysis finds that the optimal forgery and suppression strategies are orthogonal.
In addition, these two strategies follow an intuitive principle.
The forgery strategy recommends adding ratings to those categories where the user's interest is lower than the population's.
The suppression strategy suggests eliminating those ratings belonging to the categories where the user shows too much interest compared to the reference distribution.

Our theoretical study also examines how these optimal strategies perturb user profiles.
It is interesting to observe that the optimal apparent profile becomes proportional to the population's distribution in those categories with the lowest and highest ratios $\frac{q_k}{p_k}$.
Our analysis also includes the characterization of $\cR$ at low rates of forgery and suppression.
More accurately, we provide a first\hyph order Taylor approximation of the privacy\hyph utility trade\hyph off function,
from which we conclude that the ratios $\frac{q_1}{p_1}$ and $\frac{q_n}{p_n}$ determine, together with the quantity $\oD(q\,\|\,p)$,
the privacy risk at low rates.
An eye\hyph opening fact is that the relative decrement in privacy risk is greater than the forgery rate introduced.

Further, we consider the special case when forgery and suppression are not used in combination.
Under this consideration, we investigate which one is the most appropriate technique,
first, in terms of causing the minimum distortion to reach the critical\hyph privacy region,
and secondly, in terms of offering better privacy protection at low rates.
Our findings show that the arithmetic and geometric mean of the maximum and minimum ratios $\frac{q_k}{p_k}$
play a fundamental role in deciding the best technique to use.
Afterwards, our formulation and theoretical analysis are illustrated with a numerical example.

In the end, the last section is devoted to the experimental evaluation of our data\hyph perturbative mechanism in a real\hyph world recommendation system.
In particular, we examine how the application of the forgery and the suppression of ratings may preserve user privacy in \emph{Movielens}.
Among other results, we find that a large majority of users significantly reduce privacy risk for forgery and suppression rates of just 15\%.
In our data set, the probability distributions of the relative decrement factors indicate that, at low rates, forgery provides a higher reduction in privacy risk than suppression does.
By contrast, we observe that the suppression relative decrement factor is greater than that of forgery in 43.45\% of users.
Lastly, we consider the case when users must opt for either forgery or suppression;
and find that the latter is the best strategy to use in 95.3\% of users who wish to vanish privacy risk while causing the minimum distortion.

%%%%%%%%%%%%%%%%%%%%%%%%%%%%%%%%%%%%%%%%%%%%%%%%%%%%%%%%%%%%%%%%%%%%%%%%%%%%%%%%%%%%%%%%%%%%%%%%%%%%%%%%%%%%%%%%%%%%%%%%%%%%%%%%%%
%%%%%%%%%%%%%%%%%%%%%%%%%%%%%%%%%%%%%%%%%%%%%%%%%%%%%%%%%%%%%%%%%%%%%%%%%%%%%%%%%%%%%%%%%%%%%%%%%%%%%%%%%%%%%%%%%%%%%%%%%%%%%%%%%%
%REFERENCES
%%%%%%%%%%%%%%%%%%%%%%%%%%%%%%%%%%%%%%%%%%%%%%%%%%%%%%%%%%%%%%%%%%%%%%%%%%%%%%%%%%%%%%%%%%%%%%%%%%%%%%%%%%%%%%%%%%%%%%%%%%%%%%%%%%
%%%%%%%%%%%%%%%%%%%%%%%%%%%%%%%%%%%%%%%%%%%%%%%%%%%%%%%%%%%%%%%%%%%%%%%%%%%%%%%%%%%%%%%%%%%%%%%%%%%%%%%%%%%%%%%%%%%%%%%%%%%%%%%%%%
\bibliographystyle{IEEEtran}


\end{document}